\documentclass[fleqn,usenatbib]{mnras}
\usepackage{bm}
\usepackage{fix-cm}
\usepackage{newtxtext,newtxmath}
\usepackage[T1]{fontenc}

\DeclareRobustCommand{\VAN}[3]{#2}
\let\VANthebibliography\thebibliography
\def\thebibliography{\DeclareRobustCommand{\VAN}[3]{##3}\VANthebibliography}

\usepackage{graphicx}	% Including figure files
\usepackage{amsmath}	% Advanced maths commands
\usepackage{pdflscape}
\usepackage{subcaption}
\usepackage{pgffor}
\usepackage{textcomp}
\usepackage{soul}
\title[Radio Continuum Monitoring in the Coronet]{A dual-band centimetre continuum monitoring survey of Young Stellar Objects in the Coronet Cluster}

\author[J. Ram\'irez-Arellano et al.]{
Johanan Ram\'irez-Arellano,$^{1}$\thanks{E-mail: j.ramirez@irya.unam.mx}
Carlos Carrasco-Gonz\'alez,$^{1}$
Roberto Galv\'an-Madrid,$^{1}$
\newauthor
Hauyu Baobab Liu,$^{2,3}$
Jan Forbrich,$^{4}$
Arpan Ghosh$^{1}$,
\newauthor
Yenifer Angarita,$^{5}$
and Carlos G. Rom\'an-Zuñiga$^{6}$
\\
$^{1}$Universidad Nacional Aut\'onoma de M\'exico, Instituto de Radioastronom\'ia y Astrof\'isica, 58090 Morelia, Michoac\'an, M\'exico\\
$^{2}$Department of Physics, National Sun Yat-Sen University, No. 70, Lien-Hai Road, Kaohsiung City 80424, Taiwan, R.O.C.\\
$^{3}$Center of Astronomy and Gravitation, National Taiwan Normal University, Taipei 116, Taiwan, R.O.C.\\
$^{4}$Centre for Astrophysics Research, University of Hertfordshire, College Lane, Hatfield AL10 9AB, UK\\
$^{5}$Department of Physics and Astronomy, Chalmers University of Technology, 412 96 Gothenburg, Sweden \\
$^{6}$Universidad Nacional Aut\'onoma de M\'exico, Instituto de Astronom\'ia, AP 106, Ensenada 22800, BC, M\'exico\\
}
\date{Accepted 2026 April 28. Received 2026 April 28; in original form 2026 March 12}

\pubyear{\the\year{}}

\begin{document}
\label{firstpage}
\pagerange{\pageref{firstpage}--\pageref{lastpage}}
\maketitle

\begin{abstract}
We present sensitive ($\sim9\ \mu$Jy), sub-arcsecond resolution radio continuum observations at 9.0 GHz (3.3 cm) and 14.0 GHz (2.1 cm) obtained with the Karl G. Jansky Very Large Array (VLA) toward the nearby Coronet Cluster in Corona Australis ($d \approx 150$ pc). We monitored the region from March 2012 to February 2015 using all available VLA configurations, allowing us to construct deep \textit{X}- and \textit{Ku}-band maps at multiple angular resolutions.
We detected 20 radio sources, including 14 previously known Young Stellar Objects (YSOs), five sources possibly associated with shock emission, and one background galaxy. We resolved IRS~5, previously known to be a binary system, and identified IRS~7A and IRS~7B as multiple systems at centimetre wavelengths. The younger Class 0 and I YSOs exhibit spectral indices $\alpha_{\mathrm{pk}}$ ranging from $-0.4$ to $1.7$, while the more evolved Class II YSOs show flatter values between $0$ and $0.8$, consistent with free-free emission, with minor contributions from non-thermal emission.
The Class III source is only constrained by an upper limit.
Radio variability, measured as a fraction of the mean intensity peak, is found to be ubiquitous and independent of evolutionary stage.
Variability structure functions computed for nine sources indicate no preferred timescales for most of them. We also investigate spectral index variability for six sources and find significant variations in only one object. Finally, we analyse the extended radio emission toward IRS~7B, where some subcomponents exhibit negative spectral indices suggestive of non-thermal processes.

\end{abstract}


%%%%%%%%%%%%%%%%%%%%%%%%%%%%%%%%%%%%%%%%%%%%%%%%%%

%%%%%%%%%%%%%%%%% BODY OF PAPER %%%%%%%%%%%%%%%%%%

\section{Introduction}
\label{sec:intro}

Young Stellar Objects (YSOs) in star forming regions proceed through a series of evolutionary stages \citep{Lada1987,Andre1993}. Class 0 and I objects represent the earliest protostellar stages, where a recently formed disk is still embedded in a comparatively massive envelope. Class II objects contain a pre-main-sequence star at their centre, surrounded by a gas-rich disk mostly devoid of a surrounding envelope. Class III YSOs retain the gas-poor leftovers of a disk. 
\citep[recent reviews of particular evolutionary stages include, e.g.,][]{Tobin2024,Manara2023}.
Each stage has a progressively longer lifetime.  Class~0 and Class~I objects persist for $< 0.5$ Myr, whereas class II YSOs last for a few Myr \citep{Hernandez2007,Evans2009,Dunham2015}. 
During the earliest stages, accretion proceeds at higher rates, driving powerful winds and collimated outflows \citep{Frank2014}; as the system evolves, accretion gradually declines \citep{Hartmann2016}. 

Radio continuum emission traces different physical processes across the evolutionary stages of young stellar objects. It is well established that the youngest protostars exhibit free–free emission, which traces regions  where partially-ionized outflows originate \citep[e.g.,][]{Anglada1998}. Young YSOs may additionally display gyrosynchrotron emission; this non-thermal mechanism has been reported in YSOs since the early 1990s \citep[e.g.,][]{Phillips1993,Ray1997, Feigelson1998}. 
As evolution proceeds, thermal free–free emission becomes weaker and may arise from less powerful jets or photoevaporating disks \citep[e.g.,][]{Pascucci2012, Galvan2014, Rodriguez2014, Macias2016}.   
More evolved objects are often dominated by optically-thin non-thermal (gyrosynchrotron) emission, generated in their active magnetospheres \citep[e.g.,][]{Gudel2002, Forbrich2007,Liu2014}. Thanks to the high angular resolution and sensitivity of modern interferometers, radio emission can now be detected and characterized across all evolutionary stages \citep[e.g.,][]{Diaz-Marquez2024}.

The radio flux of thermal radio jets is directly linked to the amount of material being ejected \citep{Anglada2018}. 
Owing to the strong connection between accretion and the mechanisms responsible for launching winds and jets \citep[e.g.,][]{Ray2007}, thermal radio emission from ionized jets is expected  to be directly related to the accretion rate \citep{Rota25,Garuffi2025,Ghosh2026}. Therefore, by studying radio emission from ionized outflows, we gain indirect access to the accretion processes and the evolutionary history of protostars. It is therefore important to carry out radio censuses of star-forming regions targeting YSOs spanning all evolutionary stages. Since radio continuum emission from YSOs is intrinsically weak, deep observations are required. Multi-band data are necessary to study the radio spectrum and thus characterize the emission mechanisms of individual YSOs. Finally, radio surveys of YSOs are seldom capable of covering their samples over many epochs. For example, the Gould’s Belt Very Large Array Survey mapped large areas of nearby star-forming regions with a few observing epochs in each of them \citep[e.g.,][]{Ortiz-Leon2015, Dzib2013}.
 
In this paper, we present a sensitive ($9\ \mu\mathrm{Jy\ beam}^{-1}$), dual-band (9 and 14 GHz), and multi-epoch (39 epochs) radio study of the Coronet Cluster conducted with the Karl G. Jansky Very Large Array (JVLA). The Coronet Cluster, located within the Corona Australis star-forming region at a distance of about 150 pc \citep[][]{Galli2020}, hosts a  compact group of YSOs at different evolutionary stages. Our  field of view of $\sim5'$ covers most of the extent of the cluster. This region has been extensively studied at radio \citep[e.g.,][]{Choi2008,Miettinen2008,Liu2014}, far-infrared \citep{SiciliaAguilar2013}, (sub)millimetre \citep[e.g.,][]{Groppi2004,Cazzoletti2019,Pattle2025}, near- to mid-infrared \citep[e.g.,][]{Peterson11,Sandell2021,Esplin22}, and X-ray wavelengths \citep[e.g.,][]{ ForbrichyPreibisch2007,Forbrich2006}. 
A review of earlier observations of this regions was presented by \citet{NeuhauserForbrich2008}. 
In Sect.~\ref{sec:observations}, we describe the observations and data reduction. In Sect.~\ref{sec:results}, we present the source identification, spectral indices, radio variability analysis, and the comparison with infrared properties. In Sect.~\ref{sec:discussion}, we discuss the evolutionary state of the YSOs in the Coronet, the nature of their radio emission, and their variability.

\section{Observations and data reduction}
\label{sec:observations}

Radio continuum observations toward the Coronet Cluster in the Corona Australis star forming region were conducted with the National Radio Astronomy Observatory’s (NRAO) Karl G. Jansky Very Large Array (JVLA) at \textit{X}-band (9.0 GHz or 3.3 cm) and \textit{Ku}-band (14.0 GHz or 2.1 cm) from March 2012 to February 2015. The observations were carried out in all available VLA configurations, from the most extended (A configuration) to the most compact (D configuration), including the hybrid configurations (DnC, CnB, and BnA). A subset of \textit{X}-band data obtained between March and September 2012, mostly in the C configuration,  was previously reported by \citet{Liu2014} and \citet{Galvan2014}.  
A summary of the observations that we present in this work is given in Table \ref{tab:obs}. 

Data in each band were collected using a 2 GHz baseband in full-polarization mode. The total bandwidth was split into 16 spectral windows, each with a bandwidth of 128 MHz, further divided into 64 channels of 2 MHz each. Flux density, complex gain, and bandpass calibrations were performed by observing 3C286 or 3C48, J1924–2914, and J2355+4950, respectively. The pointing centre for the \textit{X}-band observations was $\alpha\mathrm{(J2000)=19^h01^m48^s.00},\ \delta\mathrm{(J2000)=-36^{\circ}57'59\farcs0}$. For the \textit{Ku}-band observations, two pointing centres were used: $\alpha\mathrm{(J2000)=19^h01^m52^s.32},\ \delta\mathrm{(J2000)=-36^{\circ}57'51\farcs9}$ and $\alpha\mathrm{(J2000)=19^h01^m55^s.67},\ \delta\mathrm{(J2000)=-36^{\circ}56'54\farcs2}$. Each \textit{X}-band epoch had a total on-source integration time of approximately 5 minutes, while each \textit{Ku}-band epoch had about 4.5 minutes per pointing.

Data calibration and imaging were performed using the Common Astronomy Software Applications (\texttt{CASA}) package, release 6.5.4.9 \citep[][]{Casa22}. 
First, each epoch was calibrated using the standard \texttt{CASA} procedures. The calibrated data were further improved through self-calibration. We used the task \texttt{gaincal} to perform phase calibration, followed by amplitude calibration; in both cases, using one solution per scan, we achieved signal-to-noise ratio improvements between 5\% and 200\%, depending on the epoch.

Once all the self-calibrated data were obtained, imaging was performed using the \texttt{CASA} task \texttt{tclean}: \textit{X}-band epochs were imaged with \textit{Briggs} weighting robust = 0.5, and \textit{Ku}-band epochs with robust = 0.0, providing a good balance between sensitivity and resolution in each epoch. The images of individual epochs were divided into five resolution groups, with Group 1 corresponding to the highest resolution and Group 5 to the lowest (see Table \ref{tab:obs}). The common beam of each group was computed using the Python package \texttt{radio-beam}\footnote{\href{https://radio-beam.readthedocs.io}{https://radio-beam.readthedocs.io}}, and the convolution of each epoch’s Point Spread Functions (PSFs) with the common beam was carried out using the \texttt{CASA} task \texttt{imsmooth}. Two-dimensional Gaussian fitting was performed using the \texttt{imfit} task on the images prior to primary beam (PB) correction. This correction was then applied by dividing by the mean value in the corresponding region of the PB response image, computed using the \texttt{imstat} task. 

Additionally, the visibilities of the individual epochs in each band were concatenated using the \texttt{CASA} task \texttt{concat}. Imaging of these concatenated data was performed with \texttt{tclean} using different weightings of the visibilities: natural, uniform, and \textit{Briggs} weighting with robust = –0.5, 0.0, 0.5, and 1.0, to explore the parameter space of synthesized beam (PSF) sizes. 
The final catalogue, presented in Table \ref{tab:survey}, was obtained from the deep, concatenated \textit{X}-band map generated with Robust = 0.0 (Fig. \ref{fig:coraus}, panel a). 
This image has a synthesized-beam FWHM of $1\farcs84 \times 0\farcs78$ ($\approx 280\times120$ au) and an rms noise of about $9\ \mu$Jy beam$^{-1}$, prior to primary beam correction. The \textit{Ku}-band deep map (Fig. \ref{fig:coraus}, panel b) has synthesized beam of $0\farcs59 \times 0\farcs25$ ($\approx 90\times40$ au), with an rms noise of $\sim8\ \mu$Jy beam$^{-1}$. For the joint analysis the \textit{Ku}-band deep map was convolved to match the PSF size of the \textit{X}-band map, using the \texttt{CASA} task \texttt{imsmooth}.
The corresponding catalogue for this band is presented in Table~\ref{tab:survey_ku}. The spectral index was computed using these deep maps with a common resolution. 

We report the flux density and peak intensity uncertainties for each source using standard error propagation. The total flux and intensity errors for each source are obtained by adding in quadrature the errors derived from the Gaussian fits (\texttt{imfit}) and the systematic flux calibration error, which is assumed to be 5\% of the measured flux density or intensity peak in both bands.

\subsection{Challenges}
\label{sec:challenges}

The generation of the \textit{Ku}-band deep continuum map presented a significant challenge due to flaring of IRS~5. 
This source is the first YSO known to have 
non-thermal (gyrosynchrotron) emission  \citep{Feigelson1998}. 
It has been reported to exhibit strong variability in its centimetre emission on timescales shorter than an individual observing block \citep{Liu2014}.  
An emission flare in one of the \textit{Ku}-band epochs produced intense image side-lobes (26$^{\mathrm{th}}$ epoch, see Table \ref{tab:obs}). To mitigate this problem, we constructed models of the emission of IRS~5 for each epoch using \texttt{tclean}. The model was then subtracted in the visibility plane in each of the \textit{Ku}-band epochs using the \texttt{CASA} task \texttt{uvsub}. Finally, we used the task \texttt{concat} to concatenate all the epochs with IRS 5 subtracted, but we still excluded the flare epoch entirely from the analysis. Besides the flaring event in the day $JD-2456137=590$, IRS 5 also had a milder flaring event in the day $JD-2456137=445$ (see Fig.~\ref{fig:panel1}), which did not introduce significant artifacts in the deep map. Therefore, this epoch was not excluded from the analysis. The final concatenated map is shown in Fig. \ref{fig:coraus}b.

\section{Results}
\label{sec:results}
\subsection{Source Identification}
\label{sec:sources}

\begin{figure*}
 \includegraphics[width=1\textwidth]{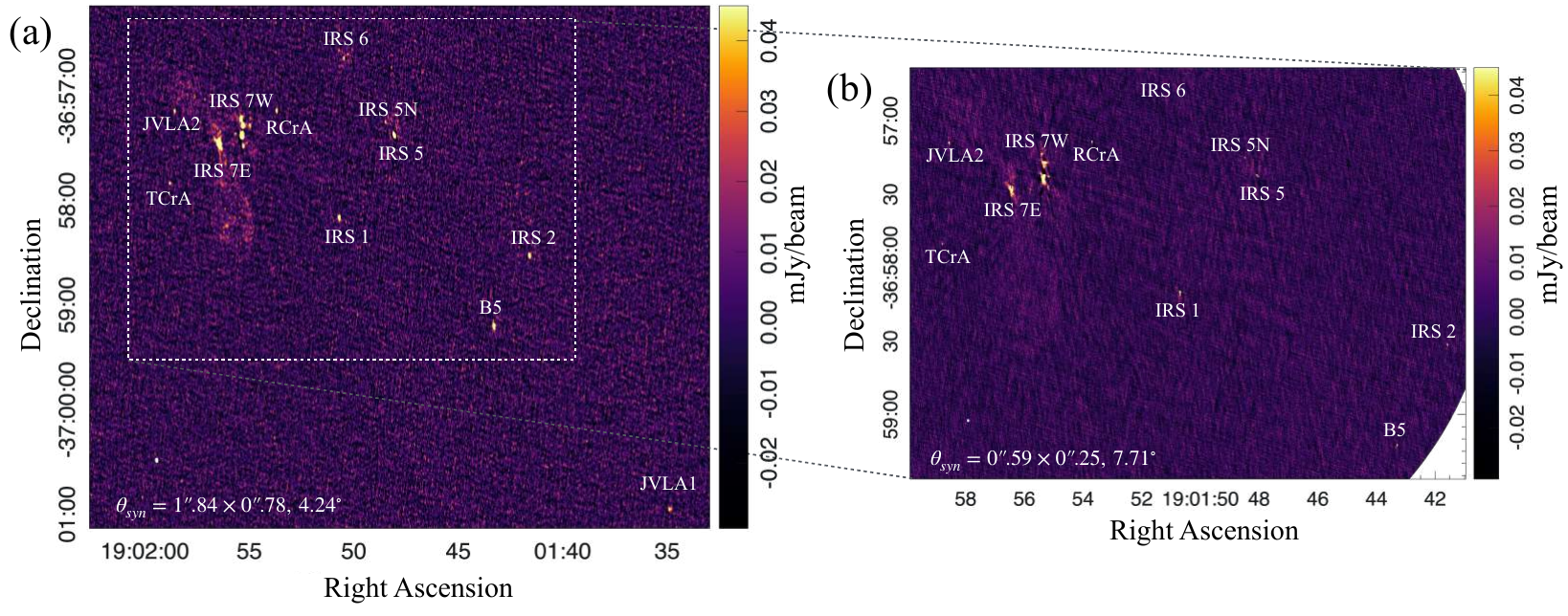}
 \caption{Deep continuum maps of the Coronet cluster at: (a) 9.0 GHz (3.3 cm), (b) 14.0 GHz (2.1 cm); imaged with Briggs weighting (robust = 0). Maps are shown prior to primary beam correction. The synthesized beam size for the 9 GHz map is $1''.84 \times 0''.78,\ \mathrm{PA}=4.24^\circ$ with an rms noise of $\sim9\ \mu$Jy beam$^{-1}$, while for the 14 GHz map it is $0''.59 \times 0''.25,\ \mathrm{PA}=7.71^\circ$, with an rms noise of $\sim8\ \mu$Jy beam$^{-1}$. The 9.0 GHz map covers a region of approximately 0.25 pc.}
    \label{fig:coraus}
\end{figure*}

Source identification was carried out using the \textit{X}-band deep map (Robust = 0.0, Fig. \ref{fig:coraus}a) prior to primary beam correction. We used the Python Blob Detector and Source Finder (\texttt{PyBDSF})\footnote{\href{https://pybdsf.readthedocs.io}{https://pybdsf.readthedocs.io}}. This package computes the mean and rms of the image, divides it into rms boxes, and uses these values to identify sources with peak intensities above a given threshold relative to the local rms noise ($\sigma$). For this work, we adopted a threshold of $2.5\sigma$, which enables accurate source detection after the visual inspection of our images. For each identified source, \texttt{PyBDSF} returns the position, which we used to define masks to perform Gaussian fits with the CASA \texttt{imfit} task.

At a nominal resolution of $\sim 180$ au in the $X$-band detection image, we identify 20 sources (see Table~\ref{tab:survey}), most of which were previously reported by \citet{Liu2014}, the source CrA-16 is marginally detected at a $3\sigma$ level. In the \textit{Ku}-band deep map (Fig. \ref{fig:coraus}b) we detected 16 of the 20 sources previously identified in the \textit{X}-band map, with no additional sources identified. The sources IRS~7AS and IRS~6 were detected exclusively in the \textit{X}-band. The source JVLA1 lies outside the  primary beam of the \textit{Ku}-band map. 

\begin{figure*}
 \includegraphics[width=0.75\textwidth]{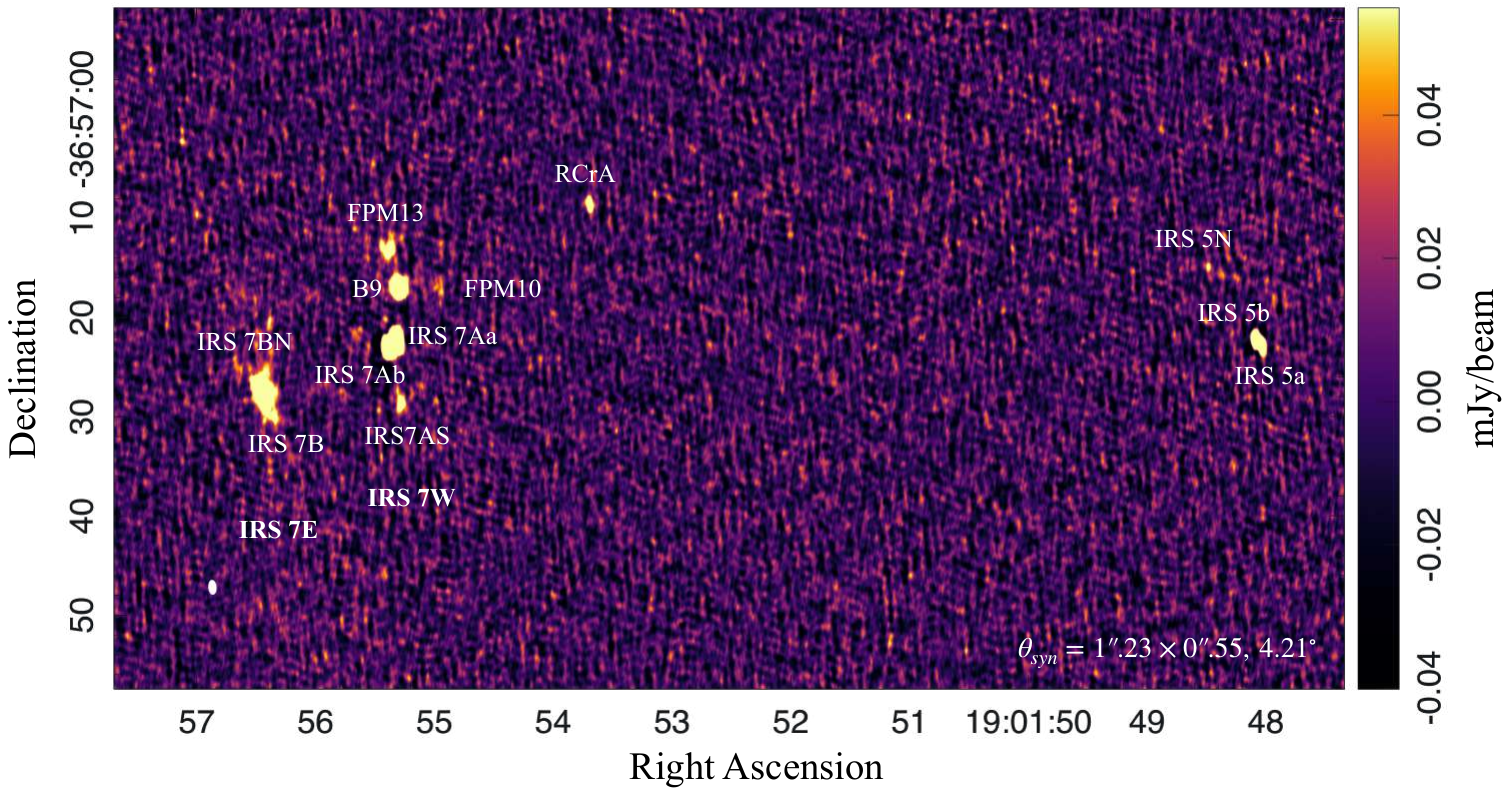}
 \caption{Zoom-in of the deep continuum map of the Coronet at 9.0 GHz (3.3 cm), imaged with Briggs weighting (robust = -0.5). The synthesized beam size is $1''.23 \times 0''.55,\ \mathrm{PA}=4.21^\circ$, with an rms noise of $\sim13\ \mu$Jy beam$^{-1}$ prior to primary beam correction.}
    \label{fig:coraus_zoom}
\end{figure*}

The source formerly known as IRS~7A is resolved into the components IRS~7Aa, IRS~7Ab, and IRS~7AS, revealing a newly identified multiple system. The components IRS~7Ab and IRS~7AS are newly detected at centimetre wavelengths. Similarly, IRS~7B is resolved into the known submillimetre binary system IRS~7Ba and IRS~7Bb \citep[e.g.,][]{Maureira2025, Hsieh24}, as well as a new centimetre source to the northeast, designated IRS~7BN. The source IRS~5, a previously known multiple system, is resolved into IRS~5a and IRS~5b, which are distinct from IRS~5N, located to the northeast (see Fig.~\ref{fig:coraus_zoom}). The source IRS~6 is a known multiple system with a separation of $0.75$ arcsec \citep[e.g.,][]{Nisini2005}  Although our \textit{Ku}-band observations reach an angular resolution of $0\farcs59 \times 0\farcs25$, the source exhibits faint emission and we were unable to resolve it spatially. Throughout the paper, we adopt the nomenclature IRS~7W and IRS~7E for IRS~7A and IRS~7B, respectively, in cases where the angular resolution is insufficient to resolve the individual components.

Sources in Tables \ref{tab:survey} and \ref{tab:survey_ku} are ordered by increasing evolutionary stage. For Class 0 YSOs, we used the bolometric temperature parameter $T_\mathrm{{bol}}$ to assign their evolutionary stage \citep[][]{Myers1993}. For Class~I and flat-spectrum sources, we used the bolometric temperature $T_\mathrm{bol}$ and the infrared spectral index $\alpha_{\mathrm{IR}}$ \citep[from][]{Hsieh24, Dunham2015, Esplin22}. In addition, we computed the colour index $K_s - [8.0]$ using $K_s$-band photometry from 2MASS \citep{Skrutskie2006} and Spitzer 8.0~$\mu$m photometry from \citet{Esplin22} to provide a more robust classification (see Sect.~\ref{sec:radvsir}). Our sample contains only one Class III YSO (JVLA1). The sources FPM10, FPM13, IRS~7Ab, IRS~7AS and IRS~7BN had not been assigned a YSO class in the literature, and we suggest that their centimetre continuum is due to emission in shocks (see Sec. \ref{sec:natureofsubcomps}). 

In Tables \ref{tab:survey} and \ref{tab:survey_ku}, we report the deconvolved source sizes. In general, most sources are at least marginally resolved, suggesting that their radio emission originates in structures with extensions of $\sim 100$ au, such as ionized winds. 

%\begin{landscape}
 \begin{table*}
  \caption{Coronet Cluster catalogue at 9.0 GHz or 3.3 cm.}
  \label{tab:survey}
  \begin{minipage}{\textwidth}
  \begin{tabular}{ccccccccc}
    \hline
    Source & R.A   & Decl.  & $I_{\rm pk}$ & $S_{\nu}$ & Maj. Axis FWHM & Min. Axis FWHM & P.A. &  Class (Ref.) \\
    Name & (J2000) & (J2000) & (mJy/beam) & (mJy) & (arcsec) & (arcsec) & (deg) &  \\
    \hline 
        IRS 5N&$19:01:48.479$&$-36:57:14.99$&$0.072\pm0.010$&$0.072\pm0.017$& \multicolumn{3}{|c|}{P.S}  &0 (1)\\
        IRS 7A&$19:01:55.333$&$-36:57:22.45$&$5.40\pm0.31$&$8.89\pm0.59$&$1.09\pm0.11$&$0.34\pm0.28$&$124\pm10$& Multiple \\
        IRS 7Aa&$19:01:55.323$&$-36:57:22.36$&$4.98\pm0.25$&$6.98\pm0.36$&$0.517\pm0.013$&$0.31\pm0.054$&$108\pm6$& 0 (1)\\
        IRS 7Ab &$19:01:55.395$&$-36:57:22.98$&$1.762\pm0.094$&$2.11\pm0.13$&$0.412\pm0.072$&$0.11\pm0.17$&$59\pm19$ & Shock? \\
        IRS 7AS &$19:01:55.282$&$-36:57:28.64$&$0.171\pm0.020$&$0.243\pm0.043$&$0.99\pm0.54$&$0.58\pm0.17$&$8\pm11$ &Shock?\\
        B9&$19:01:55.301$&$-36:57:16.95$&$1.300\pm0.070$&$1.91\pm0.11$&$0.785\pm0.051$&$0.50\pm0.26$&$84\pm21$& 0 (1)\\
        IRS 7E &$19:01:56.433$&$-36:57:27.91$&$1.022\pm0.066$&$2.96\pm0.22$& $2.59\pm0.20$&$1.020\pm0.080$&$20\pm3$ & Multiple\\
        IRS 7B &$19:01:56.423$&$-36:57:28.36$&$0.803\pm0.048$&$0.779\pm0.063$& \multicolumn{3}{|c|}{P.S} & 0 (1)\\
        IRS 7BN&$19:01:56.448$&$-36:57:27.42$&$0.339\pm0.030$&$2.42\pm0.24$&$3.02\pm0.29$&$1.35\pm0.14$&$26\pm5$ & Shock? \\
        JVLA2&$19:01:58.558$&$-36:57:08.76$&$0.186\pm0.018$&$0.220\pm0.034$&$\leq1.39$&$\leq0.39$& &I (1)\\
        IRS5 &$19:01:48.070$&$-36:57:22.38$&$1.046\pm0.059$&$1.284\pm0.088$& \multicolumn{3}{|c|}{P.S} & Multiple\\
        IRS 5a &$19:01:48.032$&$-36:57:22.83$&$0.532\pm0.035$&$0.475\pm0.048$& \multicolumn{3}{|c|}{P.S} &Flat (1)\\
        IRS 5b &$19:01:48.085$&$-36:57:22.21$&$1.122\pm0.060$&$0.868\pm0.057$& \multicolumn{3}{|c|}{P.S} & Flat (1)\\
        IRS 1&$19:01:50.692$&$-36:58:09.97$&$0.563\pm0.030$&$0.605\pm0.036$&$
        \leq0.68$&$\leq0.20$& &I (2)\\
        IRS 2&$19:01:41.587$&$-36:58:31.56$&$0.399\pm0.024$&$0.509\pm0.038$&$0.76\pm0.23$&$0.43\pm0.38$&$29\pm39$ & I (2)\\
        IRS 6&$19:01:50.472$&$-36:56:38.38$&$0.096\pm0.012$&$0.127\pm0.026$&$1.06\pm0.59$&$0.30\pm0.28$&$160\pm18$&II (2)\\
        TCrA&$19:01:58.785$&$-36:57:50.09$&$0.130\pm0.015$&$0.176\pm0.029$& \multicolumn{3}{|c|}{P.S}  &II (3)\\
        CrA-16&$19:01:33.876$&$-36:57:44.71$& &$\leq0.077$& \multicolumn{3}{|c|}{M.D} &II (4)\\
        RCrA&$19:01:53.689$&$-36:57:08.52$&$0.221\pm0.015$&$0.234\pm0.023$&$0.53\pm0.37$&$0.157\pm0.090$&$180\pm8.5$&II (3)\\
        JVLA1&$19:01:34.868$&$-37:00:56.93$&$0.53\pm0.10$&$1.76\pm0.42$&$1.74\pm0.85$&$1.7\pm1.3$&$25\pm88$ & III (3)\\
        FPM13&$19:01:55.388$&$-36:57:13.19$&$0.199\pm0.029$&$0.60\pm0.11$&$2.30\pm0.68$&$1.23\pm0.33$&$180\pm30$ & Shock? \\
        FPM10&$19:01:54.963$&$-36:57:17.04$&$0.087\pm0.015$&$0.203\pm0.047$&$2.00\pm0.88$&$0.94\pm0.47$&$14\pm78$& Shock? \\
        B5&$19:01:43.292$&$-36:59:12.28$&$0.746\pm0.040$&$0.858\pm0.052$&$0.71\pm0.17$&$0.304\pm0.036$&$5\pm9$  & Galaxy (5)\\
    \hline
  \end{tabular}
  \vspace{-1.5em}
  \footnotetext{\textbf{Notes.} Catalogue obtained from the X-band deep map at a resolution of $1''.84 \times 0''.78$. The positional uncertainties of the fits range between $0''.01$ and $0''.2$. Fits of the subcomponents were performed using the \texttt{estimates} parameter in \texttt{imfit}. Reported uncertainties, except coordinates and position angles, have two significant figures. P.S: Point Source, N.D: Non-Detection, M.D: Marginal Detection, S.T: Special Treatment.}
  \footnotetext{\textbf{References.} (1) \citet{Hsieh24}, (2) \citet{Dunham2015}, (3) \citet{Esplin22}, (4) \citet{Peterson11}, (5) \citet{Liu2014}}.
  \end{minipage}
 \end{table*}
%\end{landscape}

 \begin{table*}
  \caption{Coronet Cluster catalogue at 14.0 GHz or 2.1 cm.}
  \label{tab:survey_ku}
  \begin{minipage}{\textwidth}
  \begin{tabular}{cccccccc}
    \hline
    Source & R.A   & Decl.  & $I_{\rm pk}$ & $S_{\nu}$ & Major Axis FWHM & Minor Axis FWHM & P.A. \\
    Name & (J2000) & (J2000) & (mJy/beam) & (mJy) & (arcsec) & (arcsec) & (deg)  \\
    \hline 
        IRS 5N&$19:01:48.463$&$-36:57:15.33$&$0.145\pm0.028$&$0.222\pm0.070$&$1.56\pm0.92$&$0.46\pm0.30$&$11\pm9$  \\
        IRS 7A&$19:01:55.332$&$-36:57:22.72$&$4.61\pm0.29$&$8.61\pm0.64$&$1.33\pm0.18$&$0.74\pm0.22$&$150\pm13$\\
        B9&$19:01:55.302$&$-36:57:17.25$&$1.273\pm0.092$&$2.64\pm0.24$&$1.24\pm0.25$&$1.09\pm0.20$&$180\pm82$\\
        
        IRS 7E &$19:01:56.427$&$-36:57:28.36$&$1.386\pm0.096$&$3.72\pm0.31$& $2.46\pm0.25$&$0.962\pm0.082$&$14\pm4$ \\
        JVLA2&$19:01:58.560$&$-36:57:08.81$&$0.205\pm0.025$&$0.79\pm0.12$&$4.53\pm0.84$&$0.80\pm0.15$&$180\pm2$\\
        
        IRS 5 &$19:01:48.074$&$-36:57:22.14$& \multicolumn{2}{|c|}{S.T} & & &\\
        IRS 1&$19:01:50.693$&$-36:58:10.18$&$0.969\pm0.062$&$1.20\pm0.10$&$1.06\pm0.29$&$0.25\pm0.12$&$170\pm13$\\
        IRS 2&$19:01:41.586$&$-36:58:31.54$&$0.85\pm0.13$&$1.84\pm0.41$&$2.02\pm0.83$&$0.74\pm0.34$&$170\pm29$ \\
        IRS 6&$19:01:50.472$&$-36:56:38.38$& & $\leq0.095$& \multicolumn{3}{|c|}{N.D}\\
        TCrA&$19:01:58.792$&$-36:57:49.91$&$0.136\pm0.022$&$0.307\pm0.068$& $1.58\pm0.47$&$0.14\pm0.87$&$100\pm64$ \\
        RCrA&$19:01:53.691$&$-36:57:08.66$&$0.309\pm0.026$&$0.223\pm0.038$&\multicolumn{3}{|c|}{P.S}\\
        FPM13&$19:01:55.390$&$-36:57:13.16$&$0.187\pm0.030$&$1.24\pm0.23$&$5.5\pm1.2$&$1.44\pm0.26$&$180\pm4$\\
        FPM10&$19:01:54.987$&$-36:57:17.11$&$0.052\pm0.016$&$0.124\pm0.053$&$2.1\pm1.2$&$0.49\pm0.89$&$140\pm27$\\
        IRS 7AS &$19:01:55.282$&$-36:57:28.64$& &$\leq0.12$& \multicolumn{3}{|c|}{N.D} \\
        B5&$19:01:43.302$&$-36:59:12.82$&$1.18\pm0.11$&$1.89\pm0.28$&$\leq2.80$&$\leq 0.17$&\\
        
    \hline
  \end{tabular}
  \vspace{-1.5em}
  \footnotetext{\textbf{Notes.} Catalogue obtained from the Ku-band deep map at a resolution of $1''.84 \times 0''.78$. The positional uncertainties of the fits range between $0''.01$ and $0''.2$. The source JVLA1 lies outside the field of view of the Ku-band map. Reported uncertainties, except coordinates and position angles, have two significant figures. P.S: Point Source, N.D: Non-Detection, M.D: Marginal Detection, S.T: Special Treatment.}
  \end{minipage}
 \end{table*}

\subsection{Radio Spectral Index}
\label{sec:alpha_rad}

In both thermal and non-thermal emission mechanisms, the frequency dependence of the optical depth ($\tau$) determines the frequency dependence of the observed intensity, often characterized by the spectral index $\alpha$, where $I_\nu \propto \nu^\alpha$.
In both types of emission, $\tau$ decreases with increasing frequency, although the physical origin of the opacity and the characteristic turnover frequency at which $\tau \sim 1$ differ between the two mechanisms. In many cases, the turnover frequency associated with non-thermal emission occurs at lower frequencies than that of thermal free–free emission. 
Therefore, for centimetre-wavelength studies of YSOs, it is generally expected that non-thermal emission will be optically thin and have a negative spectral index, whereas thermal free-free emission will have spectral indices ranging from $\alpha = -0.1$ in the optically thin limit to $\alpha = 2.0$ in the optically-thick limit.

For each source detected at both frequencies, we computed the spectral index $\alpha$. The convolved \textit{Ku}-band deep map is noisier ($\approx23\ \mu\mathrm{Jy\ beam}^{-1}$ prior primary beam correction) and contains more artifacts. This map received special treatment and includes only 7 epochs in its concatenation, compared to the \textit{X}-band deep map, which was constructed using 31 epochs. To minimize the effect of differences in the data sets, we compared the sources in the two bands using the measured peak intensities in maps convolved to the same beam size, rather using flux densities. The $\alpha_{\mathrm{pk}}$ values were obtained using the following equation:

\begin{equation}
\alpha_{\mathrm{pk}} = \frac{\log(I_{\nu_2}/I_{\nu_1})}{\log(\nu_2/\nu_{1})},
\end{equation}

\noindent where $I_{\nu_1}$ is the peak intensity measured at  $\nu_1 = 9.$ GHz, and $I_{\nu_2}$ is the peak intensity measured at $\nu_2 = 14.0$ GHz. 
The resolution in the \textit{Ku}-band was degraded when convolved to match that of the \textit{X}-band; therefore, the spectral indices of sources IRS~7A, IRS~7B, and IRS~5 are reported without considering their multiplicity. For the sources JVLA1 and IRS~5, $\alpha_{\mathrm{pk}}$ was computed using the \textit{X}-band data divided into two sub-bands centred at 8.50 GHz and 9.50 GHz, respectively. The reason is that JVLA1 lies outside the field of view of the \textit{Ku}-band map, and IRS 5 required a special treatment in the \textit{Ku}-band deep map (see Sec. \ref{sec:challenges}). Their spectral index is therefore computed over a narrower frequency range compared to the other sources, which translates into  larger errors in $\alpha_{\mathrm{pk}}$. On the other hand, the sources IRS~7AS and IRS~6 show no detection in the \textit{Ku}-band; therefore, we use $3\sigma$ upper limits, where $\sigma$ is the local noise after primary-beam correction.

The $\alpha_{\mathrm{pk}}$ results are presented in Table~\ref{tab:alpha_pk}. Within the uncertainties, three emission regimes can be identified. Values of $\alpha_{\mathrm{pk}}$ in the range $-0.5$ to $0.5$ are consistent with optically thin free-free emission, although in the lowest range there could have contributions from non-thermal emission; values between $0.5$ and approaching $2$ correspond to partially optically thick emission; and values $\approx2.0$ indicate optically thick emission. Although some sources exhibit large uncertainties (e.g., IRS~5N, IRS~2, IRS~5, and TCrA), the YSOs can be broadly grouped into these three emission regimes.
Class~0 YSOs exhibit spectral index values ranging from $-0.05$ to $1.6$ with an average error of $0.2$.
The younger Class~I YSO JVLA2 and the Flat-spectrum source IRS~5 show lower spectral index values compared to the more evolved YSOs IRS~1 and IRS~2; they display a broad range between $0.1$ and $1.7$ with an average error of $0.3$.
The Class~II YSOs exhibit flatter spectral indices, spanning from about $0.1$ to $0.8$ with an average error of $0.3$. The Class~III YSO JVLA1 shows an  upper limit $\alpha_{\mathrm{pk}} < 0.5$.
On the other hand, the possible shocks FPM10 and IRS~7AS display clearly negative spectral indices, whereas $\alpha_{\rm pk}$ in FPM10 is consistent with zero. The interpretation of these is discussed in the following sections.

Although the noise levels of the original maps in the two bands are nearly identical, the noise in the \textit{Ku}-band map is larger after the convolution applied to homogenize their resolutions. This is properly taken into account in the spectral index calculations.

\begin{table}
	\centering
	\caption{Infrared properties and radio peak spectral indices derived from deep maps.}
	\label{tab:alpha_pk}
    \begin{minipage}{\columnwidth}
	\begin{tabular}{cccccc} 
		\hline
		Source  & Class & $T_{\mathrm{bol}}$ & $K_s-[8.0]$& $\alpha_{\rm pk}$ & Ref. \\
        Name  &  & (K) & (mag) & &\\
		\hline
        IRS 5N & 0 & 40 &  & $1.58\pm0.53$ & (1)\\
        IRS 7A & 0 & 54 &  & $-0.36\pm0.19$& (1)\\
        B9 & 0 & 54 &  & $-0.05\pm0.20$& (1)\\
        IRS 7B & 0 & 56 &  & $0.69\pm0.21$&(1)\\
        JVLA2  & I & 72 & $7.27\pm0.02$ & $0.22\pm0.35$&(1)\\
        IRS 5 & F & 208 & $6.02\pm0.05$ & $0.12\pm0.63$&(1)\\
        IRS 1 & I & 210 & $4.16\pm0.04$ & $1.23\pm0.19$&(2)\\
		IRS 2 & I & 270 & $4.66\pm0.03$ & $1.7\pm0.38$&(2)\\
        IRS 6  & II &  & $3.82\pm0.50$ & $\leq-0.030$&(3)\\
        TCrA & II &  &  $3.24\pm0.03$ & $0.10\pm0.45$ &(3)\\
        RCrA  & II &  & $2.86\pm0.26$ & $0.76\pm0.24$ &(3)\\
        JVLA1 & III &  & $0.18\pm0.03$ & $\leq 0.49$&(3)\\
        FPM13 & shock? &  &   & $-0.14\pm0.49$&\\
        FPM10 & shock? &  &   & $-1.16\pm0.80$&\\
        IRS 7AS & shock? &  &   & $\leq -0.80$&\\
        \hline
	\end{tabular}
    \vspace{-1.5em}
    \footnotetext{\textbf{Notes.} Peak spectral index for IRS 5 and JVLA1 was computed using the two sub-bands within the \textit{X}-band deep map.}
    \footnotetext{\textbf{References.} For YSO class and $T_{\mathrm{bol}}$: (1) \citet{Hsieh24}, (2) \citet{Dunham2015}, (3) \citet{Esplin22}. Photometry in $K_s$ band from \citet{Skrutskie2006} and photometry in the Spitzer 8.0 $\mu$m band from \citet{Esplin22}}.
    \end{minipage}
\end{table}

\subsection{Comparison of Radio and Infrared Properties}
\label{sec:radvsir}

We compared our radio photometry with near-IR photometry from the survey presented in \citet{Esplin22}. 
We compute the colour index $K_s-[8.0]$, where $K_s$ corresponds to the 2MASS band centred at 2.2~$\mu$m \citep{Skrutskie2006} and the 8.0~$\mu$m band comes from IRAC-Spitzer \citep{Fazio2004}. The radio spectral index was computed as described in Sec. \ref{sec:alpha_rad}. 
The near-IR vs. IRAC colour–colour diagrams can be used to identify sources with disks or circumstellar envelopes \citep[e.g.,][]{Allen2004, Teixeira12}. Colours such as $K_s - [4.5]$ or $K_s - [8.0]$ are typically associated with different classes: low values with Class III YSOs (disk-less), intermediate values with Class II YSOs (with disks), and high values with  the more embedded YSOs. Complementarily, the bolometric temperature is also useful to trace the evolutionary stage of the embedded YSOs, with lower $T_{\mathrm{bol}}$ values corresponding to younger YSOs and higher values to more evolved ones \citep[e.g.,][]{Myers1993, Dunham2015, Sandell2021}.
 
\begin{figure}
	\includegraphics[width=\columnwidth]{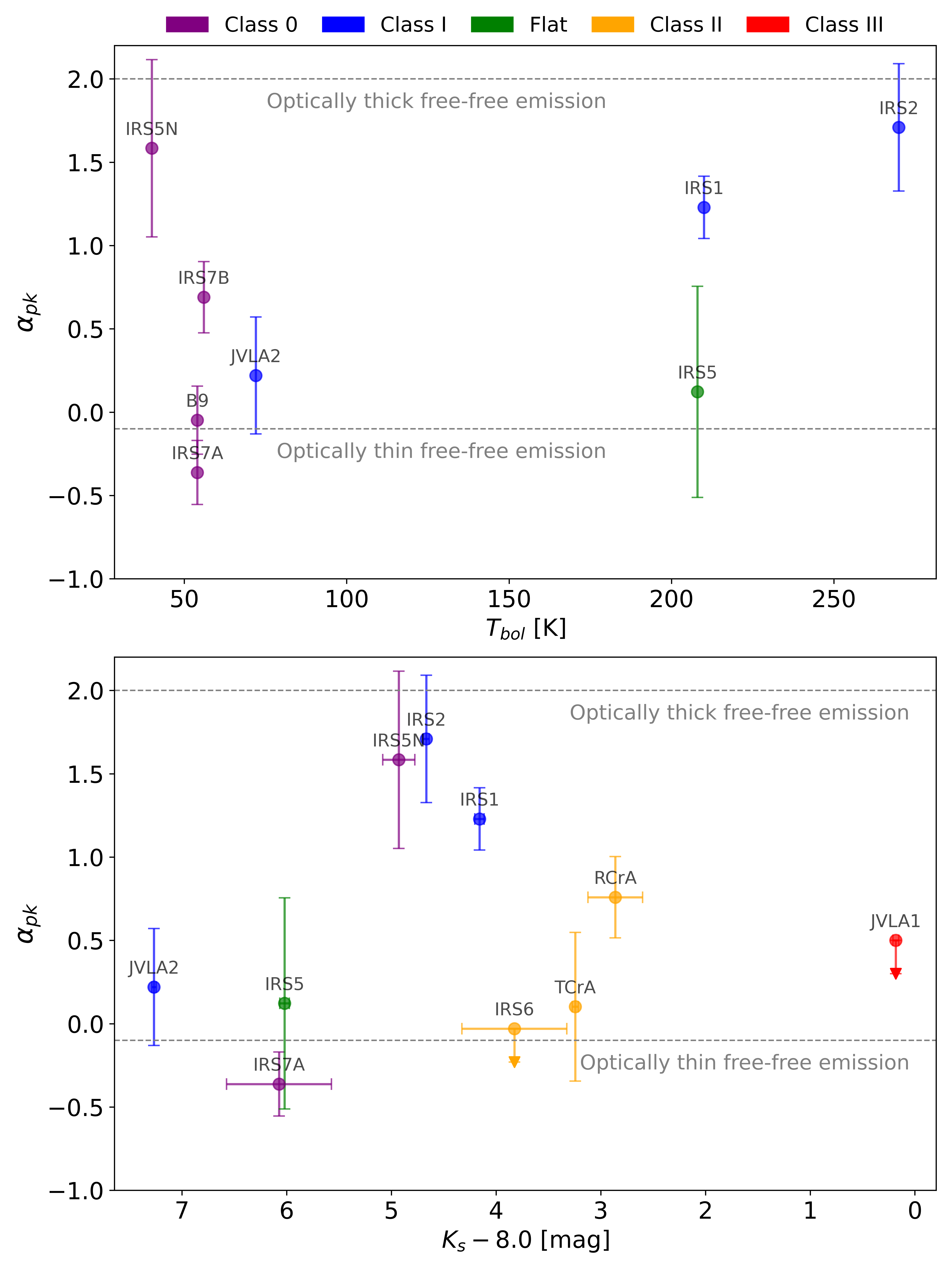}
    \caption{Peak spectral index ($\alpha_{\mathrm{pk}}$) as a function of the evolutionary stage (younger YSOs to the left). The upper panel orders the sources according to their bolometric temperature $T_{\mathrm{bol}}$ , while the lower panel orders them by the colour index $K_s-[8.0]$. The dashed lines indicate the optically thick and optically thin limits to free-free emission, respectively.}
    \label{fig:k-8}
\end{figure}

 In Fig.~\ref{fig:k-8}, we present the radio spectral index as a function of bolometric temperature $T_{\mathrm{bol}}$ (top panel) and as a function of the near-IR colour index (bottom panel). Regarding the radio spectral index, $\alpha_{\mathrm{pk}}$, Class~II YSOs show values consistent with optically thin free–free emission, while Class~0, I, and Flat-spectrum YSOs display values spanning both optically thin and thick regimes. For the Class~III YSO (JVLA1), the reported upper limit is consistent with the optically thin regime, but also suggests a possible non-thermal contribution to the emission.
 
\subsection{Radio Emission Variability}
\label{sec:radio_var}

This work provides a unique dataset to monitor the variability of YSO radio emission, spanning approximately one thousand days of observations. For this analysis, we use the epochs listed in Table~\ref{tab:obs}, except for epochs 14, 19, and 20. The angular resolution in the 14$^{\mathrm{th}}$ epoch was very low ($\theta_{\mathrm{syn}} = 27\farcs3 \times 5\farcs4$), and the data quality in the 19$^{\mathrm{th}}$ and 20$^{\mathrm{th}}$ epochs was poor, with artifacts present in the maps. As described in Sec. \ref{sec:observations}, the epochs were divided into five resolution groups, and variability was analysed within each group. Isolated sources that remain compact even at the highest resolution are analysed across all five groups; these include TCrA, RCrA, IRS~1, IRS~2, IRS~6, and JVLA1. For the remaining sources, we excluded from the analysis those epochs in which they were not isolated. The non-isolated components within the IRS~5, IRS~7A, and IRS~7B systems are discussed in the following sections. The light curves of each source are presented in the top panels of Fig.~\ref{fig:panel1}. 

We begin the variability analysis by defining a variability index, $VI$, based on the standard deviation of the peak intensities, $\sigma_{\mathrm{pk}}$, and the mean per-epoch error, $\bar{\epsilon}$, as

\begin{equation}
VI= \frac{\sigma_{pk}}{\bar \epsilon }.
\label{eq:VI}
\end{equation}

\begin{figure}
    \centering
    \includegraphics[width=1\columnwidth]{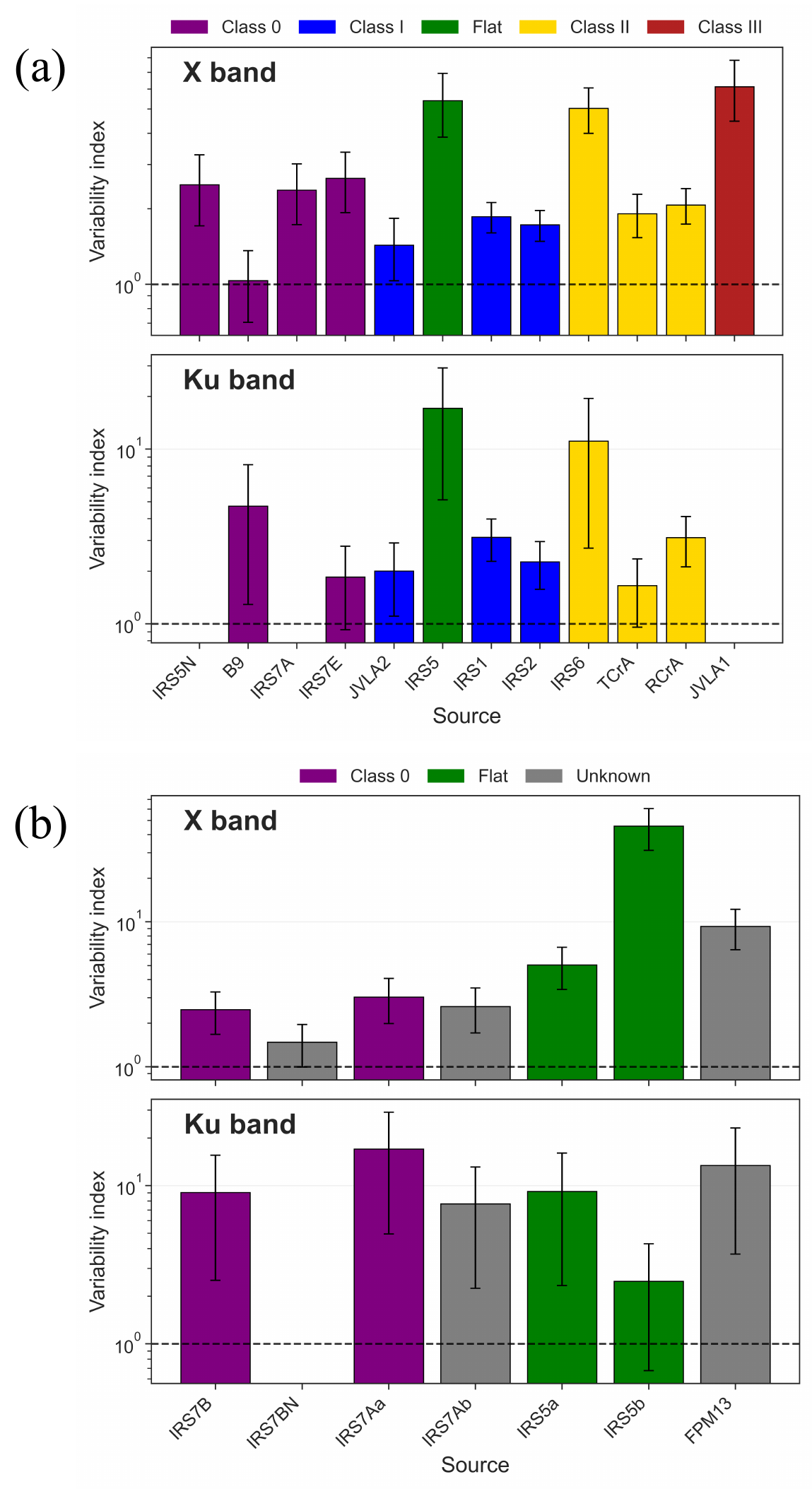}
    \caption{
    Variability index as a function of the evolutionary stage,  for the YSOs (a, without) and (b, with) consideration of their subcomponents. The dashed lines indicate the variability threshold $VI=1$. The upper panels correspond to variability in the \textit{X}-band, while the lower panels shows variability in the \textit{Ku}-band. }
    \label{fig:var_index_persource}
\end{figure}

The uncertainties in the variability index are estimated through standard error propagation. We therefore define a variability threshold such that the standard deviation of the peak intensities equals the mean per-epoch error of the sample, corresponding to $VI = 1$. Sources with variability indices $VI \gg 1$ are classified as highly variable, whereas sources with $VI \leq 1$ are considered non-variable within the uncertainties. The variability indices of the YSOs are presented in Fig.~\ref{fig:var_index_persource}a, and those of the subcomponents within IRS~5, IRS~7A, and IRS~7B are shown in Fig.~\ref{fig:var_index_persource}b. All sources exhibit $VI > 1$ within the uncertainties in both bands, implying variability in all YSOs.

\begin{figure*}
 \includegraphics[width=1\textwidth]{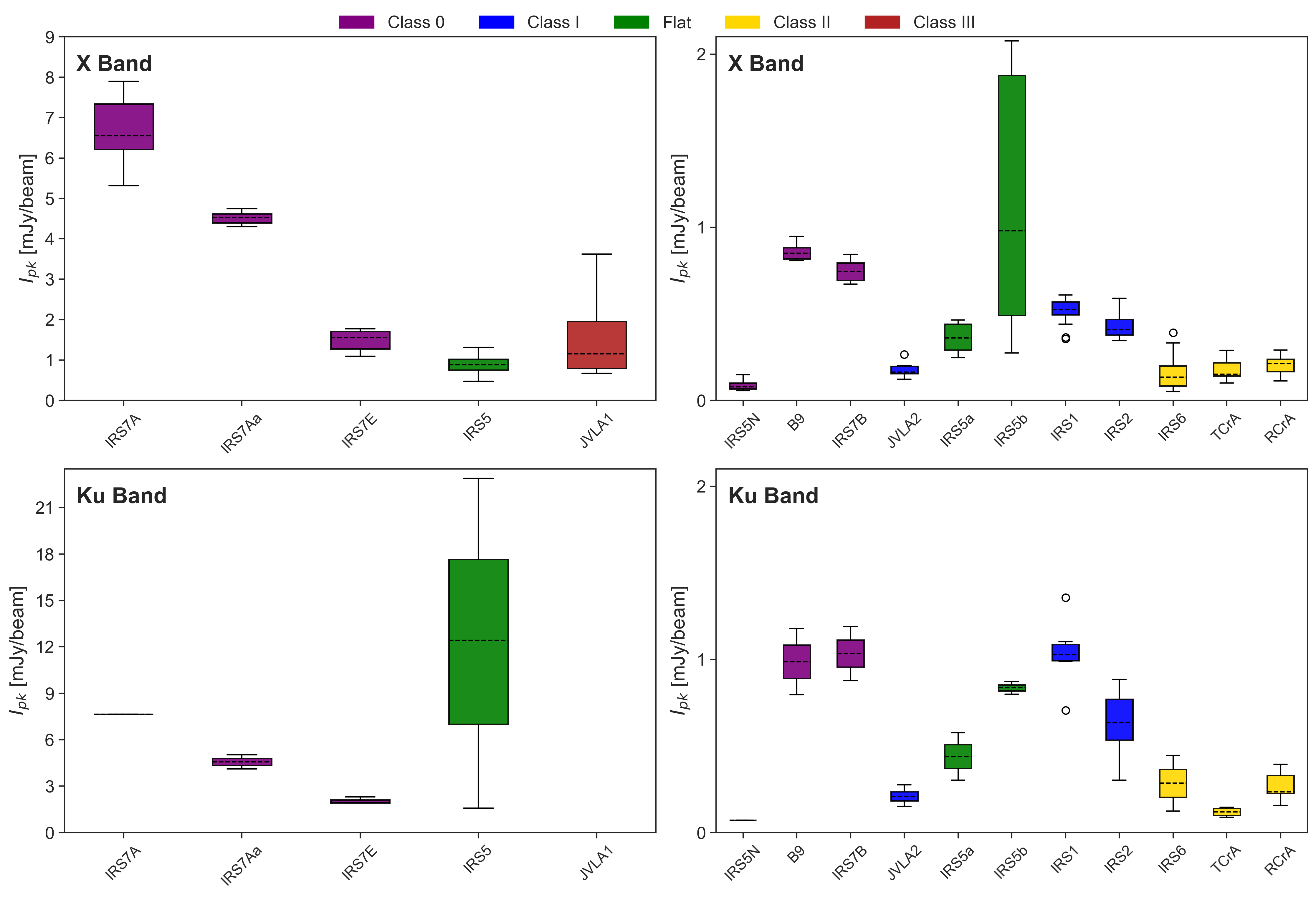}
 \caption{
 Peak intensity box plots in both bands, ordered as in Table \ref{tab:alpha_pk} from left to right. The upper panel corresponds to variability in the \textit{X}-band, while the lower panel shows variability in the \textit{Ku}-band. 
 The interquartile range (IQR) of the boxes extends from the first to the third quartile, with a horizontal line indicating the median. The whiskers extend to $\times 1.5$ the IQR from the box, while points outside the whiskers are considered outliers.}
    \label{fig:boxplots_peak}
\end{figure*}

\begin{figure*}
 \includegraphics[width=1\textwidth]{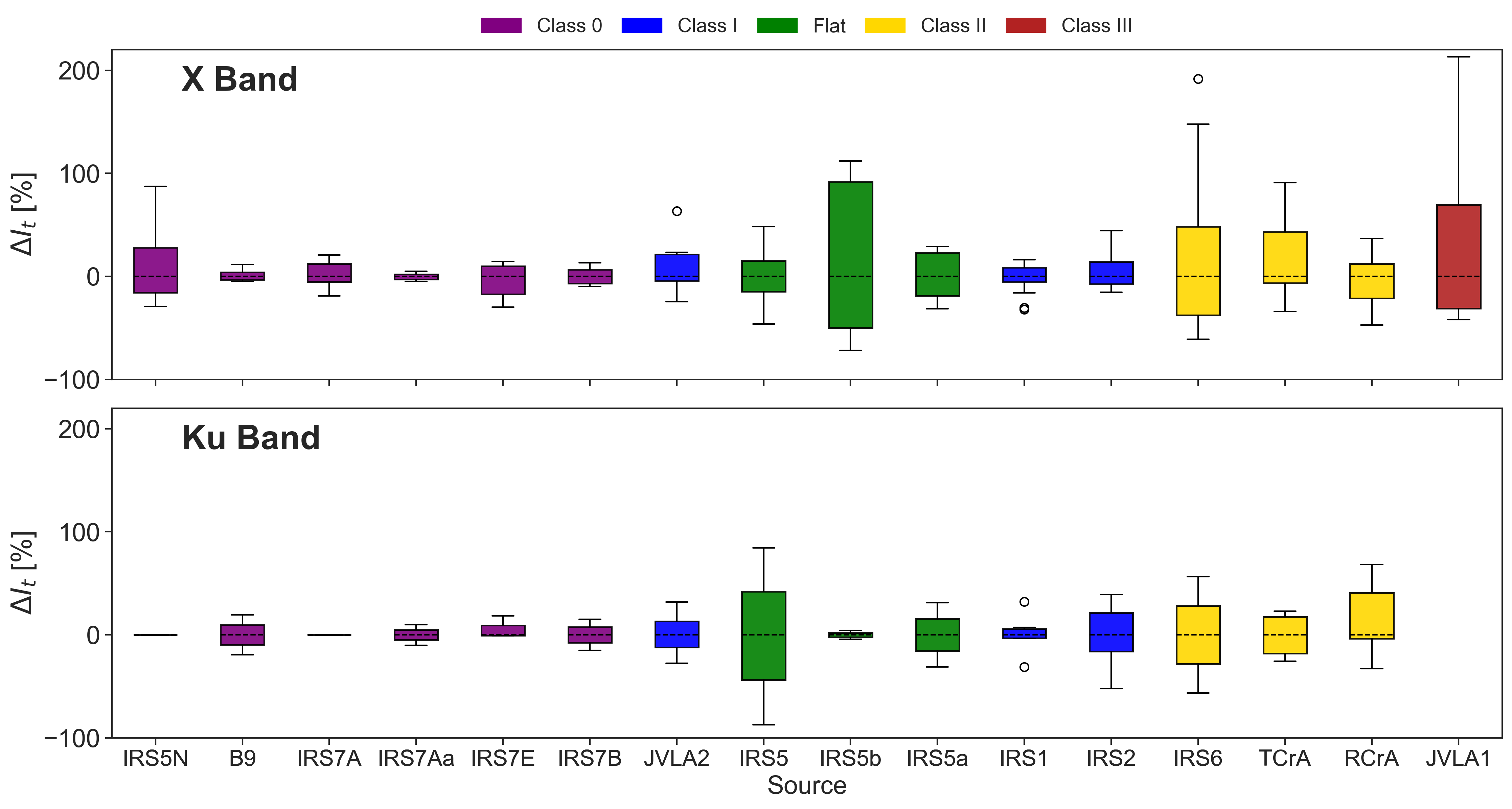}
 \caption{
 Percentage variability box plots.  
 Symbols and labels are defined as in Fig. \ref{fig:boxplots_peak}.
 ordered as in Table \ref{tab:alpha_pk} from left to right. The upper panel corresponds to variability in the \textit{X}-band, while the lower panel shows variability in the \textit{Ku}-band.}
    \label{fig:boxplots}
\end{figure*}

To examine variability in more detail, we generated box plots for each source to illustrate the distribution and skewness of the time-domain photometric data through their interquartile ranges (IQRs). We begin the analysis with the absolute variability by plotting the peak-intensity box plots (Fig.~\ref{fig:boxplots_peak}), followed by the fractional (percentage) variability (Fig.~\ref{fig:boxplots}). The sources in the plots are ordered as in Table~\ref{tab:alpha_pk}, from the youngest to the most evolved YSOs. 
The percentage variability of the YSOs was defined with respect to their median values and computed as:

\begin{equation}
\Delta I_t\ [\%]= \frac{I_t - \tilde{I}}{\tilde{I}}\times100,
\label{eq:var}
\end{equation}

\noindent 
where $I_t$ is the peak intensity at epoch $t$, and $\tilde{I}$ is the median peak intensity over the full dataset. The light curves of percentage variability for each source are presented in the bottom panels of Fig.~\ref{fig:panel1}; the standard deviation, computed from the set of detections, is also shown. These figures reveal variability of a few to several tens of percent across all evolutionary stages.

From Fig.~\ref{fig:boxplots_peak}, where we show the absolute variability, is noticeable that the Class 0 YSO IRS 7A is the brightest, and that Class 0 YSOs are among the brightest overall. The Flat-spectrum source IRS~5 exhibits a particularly wide range of values (large IQR), especially in the \textit{Ku}-band (Fig. \ref{fig:boxplots_peak} lower panel). As discussed in Sect.~\ref{sec:challenges}, the \textit{Ku}-band measurements suggest the presence of two flare events. The physical interpretation of these results is presented in Sect.~\ref{sec:yso_var}.

Fig.~\ref{fig:boxplots} shows that, in the \textit{X}-band, most YSOs vary within a similar fractional range, except for IRS~5b, IRS~6, and JVLA1, which exhibit a broader variability range. The interquartile range in the \textit{X}-band extends, on average, from $-15\%$ to $25\%$. In the \textit{Ku}-band, the more evolved YSOs (IRS~5, IRS~2, IRS~6, and RCrA) show a broader variability range compared to the younger sources. In this case, the interquartile range extends, on average, from $-10\%$ to $13\%$. Approximately half of the sources in the \textit{X}-band exhibit noticeable skewness — particularly IRS~5N, JVLA2, IRS6, TCrA and JVLA1 — whereas in the \textit{Ku}-band only RCrA shows significant asymmetry.

In summary, our survey reveals ubiquitous radio variability across all YSO evolutionary stages.

Following the variability analysis, we computed the Structure Function S($\Delta t$) as defined in \citet{Liu2014}. This function quantifies how much the flux density changes as a function of the time lag $\Delta t=t_i-t_j$ between measurements, as defined in: 
\begin{equation}
\mathrm{S (\Delta t)} = [I_{\nu i} - I_{\nu j}]^2 / \sigma^2_{bw},
\end{equation}
where $I_{\nu i}$ and $I_{\nu j}$ are the peak intensities observed at epochs $i$ and $j$, respectively, and $\sigma_{\mathrm{bw}}$ is the biweight standard deviation of the measured fluxes over the full observing period of approximately 1100 days (March 2012 to February 2015). The time lags ($\Delta t_{ij}$) were grouped into approximately logarithmic custom bins $[0.1, 1, 3, 10, 30, 100, 300, 1000]$. Since the \textit{X}-band data provide better temporal sampling, we restrict the analysis to this band. Some sources do not show a structure function because the number of available epochs was insufficient to compute sufficient time-lag pairs.

\begin{figure}
    \centering 
	\includegraphics[width=0.75\columnwidth]{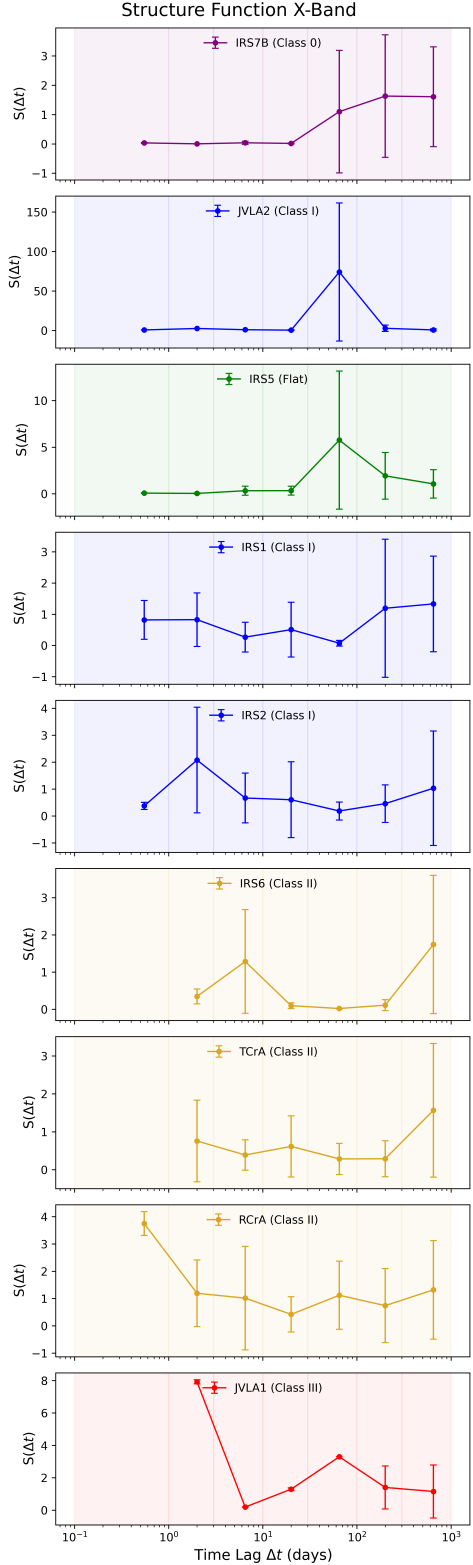}
    \caption{Structure functions for a subset of YSOs over a time span of approximately 1100 d. The sources are ordered by evolutionary stage, from younger to more evolved. Vertical lines indicate the time-lag bins of [0.1, 1, 3, 10, 30, 100, 300, 1000] d. The error bars correspond to the standard deviation of the measurements within each bin.}
    \label{fig:sf}
\end{figure}

Representative structure functions for a subset of YSOs are shown in Fig.~\ref{fig:sf}, ordered by evolutionary stage from top to bottom. In most cases, the errors are large and prevent robust identification of characteristic variability timescales. Although some bins suggest enhanced variability at particular $\Delta t$ values, these features remain within the error bars.
An exception is JVLA1 (Class~III), which shows a clear enhancement at short time lags ($\Delta t \sim$ few days) that stands above the associated error. This behaviour is consistent with flare-like variability on short timescales. Apart from this case, and possibly RCrA also on the shortest timescales, we find no statistically significant trend between the structure function amplitude and evolutionary class. Overall, within the current uncertainties, we find no compelling evidence for preferred variability timescales in the sample, suggesting that the variability observed in most of these YSOs is mostly stochastic in nature.

\subsection{Spectral Index Variability}

Most of the YSOs in our sample exhibit variability in their radio emission, 
with different levels of absolute and percentage variability between bands,
which could translate into spectral index variability. Although the analysis is complicated by the non-simultaneity of the observations in both bands, the analysis is possible on timescales of $\approx$~90~days, which is acceptable for slow thermal processes. We therefore adopted the following approach to examine possible variations in $\alpha_{\mathrm{pk}}$. First, for each source, a \textit{Ku}-band detection was identified in the light curve. We then searched for \textit{X}-band detections within a 90-day temporal window, corresponding to the maximum time gap between epochs within the same \textit{Ku}-band resolution group. If multiple detections were found, their mean value was computed. The uncertainty $\delta\alpha$ was estimated using standard error propagation. Non-detections were excluded from the analysis.

\begin{figure*}
    \centering
    % Fila 1
    \begin{subfigure}{0.5\textwidth}
        \centering
        \includegraphics[width=\linewidth]{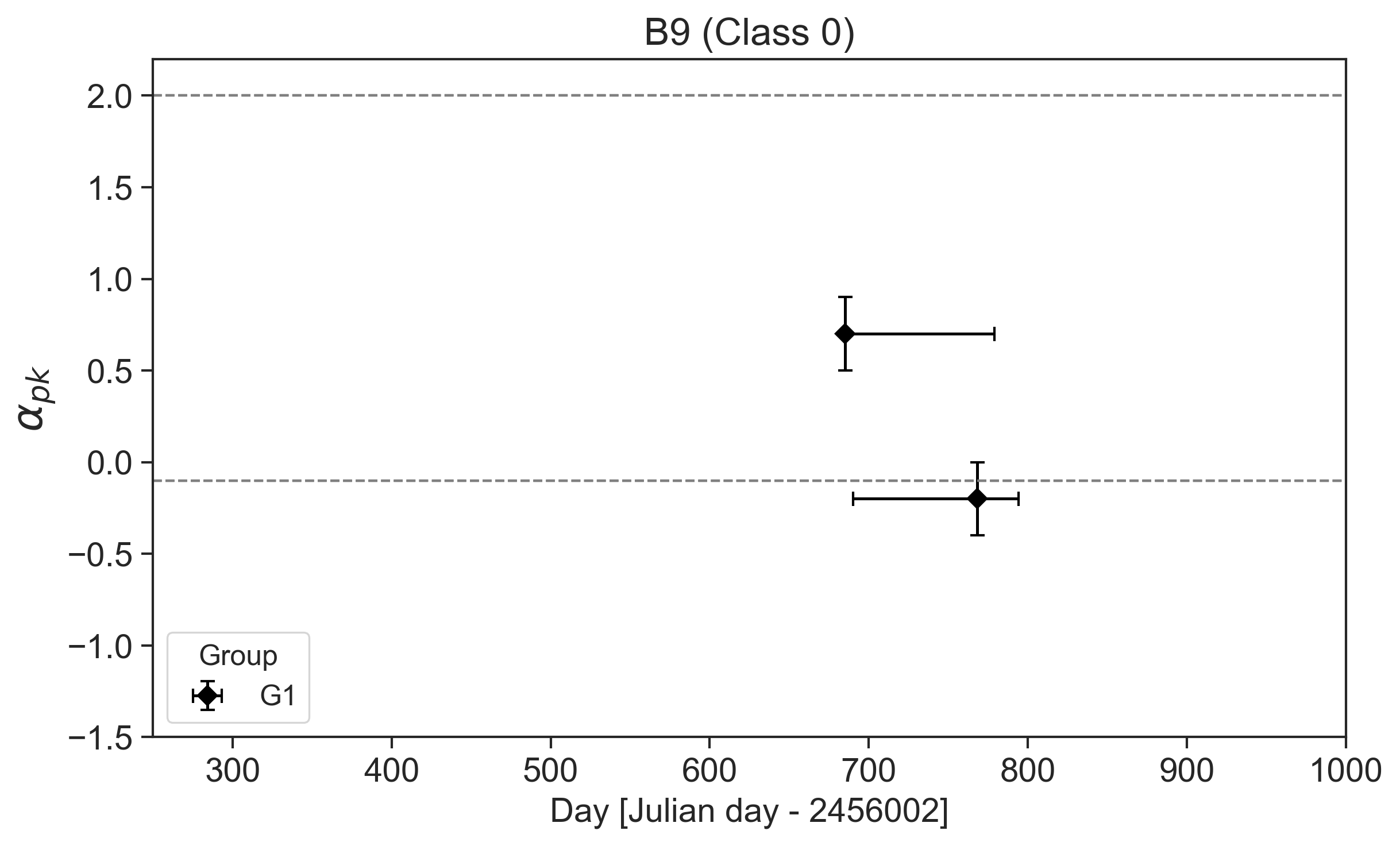}
    \end{subfigure}%
    \begin{subfigure}{0.5\textwidth}
        \centering
        \includegraphics[width=\linewidth]{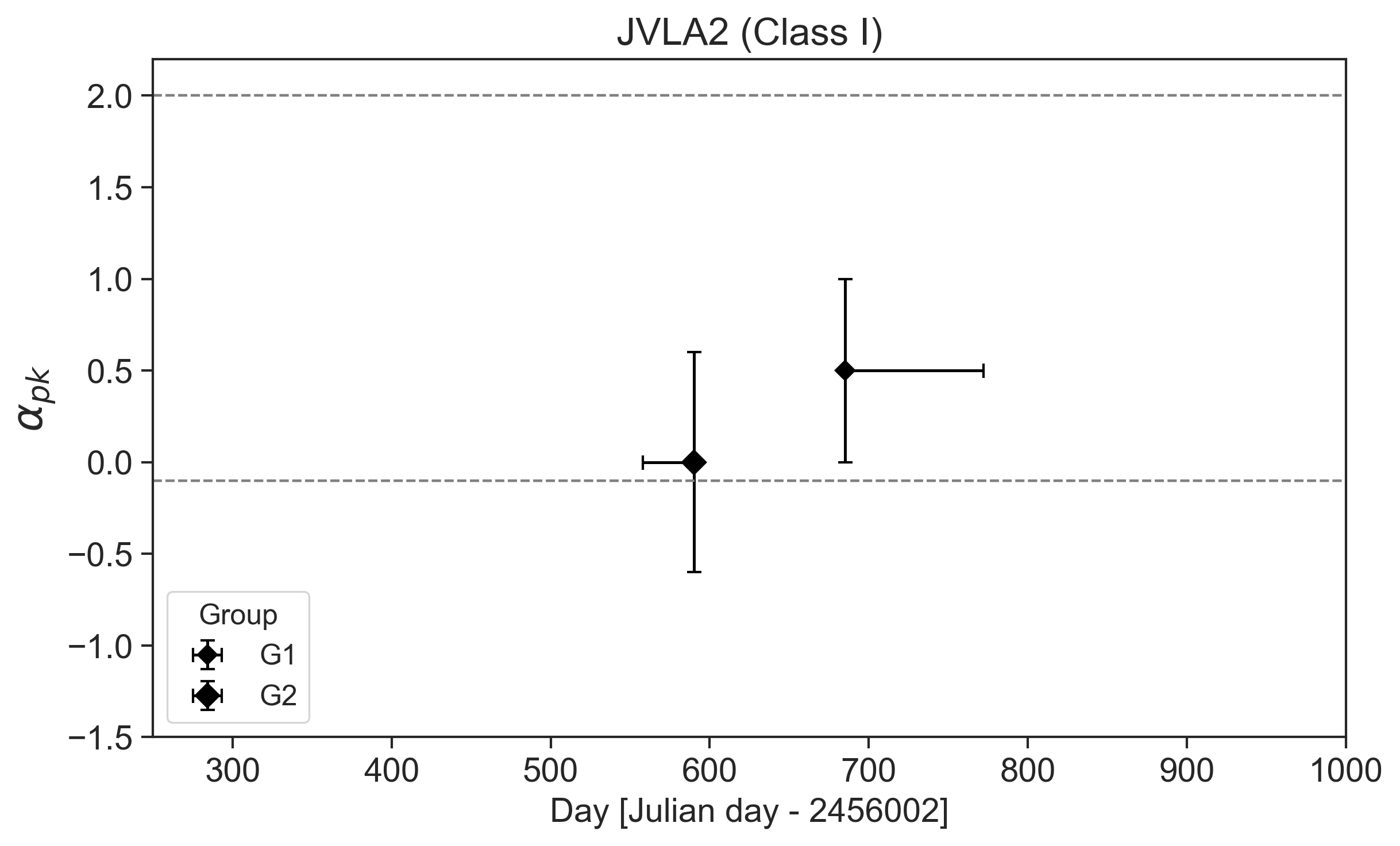}
    \end{subfigure} 

    % Fila 2
    \begin{subfigure}{0.5\textwidth}
        \centering
        \includegraphics[width=\linewidth]{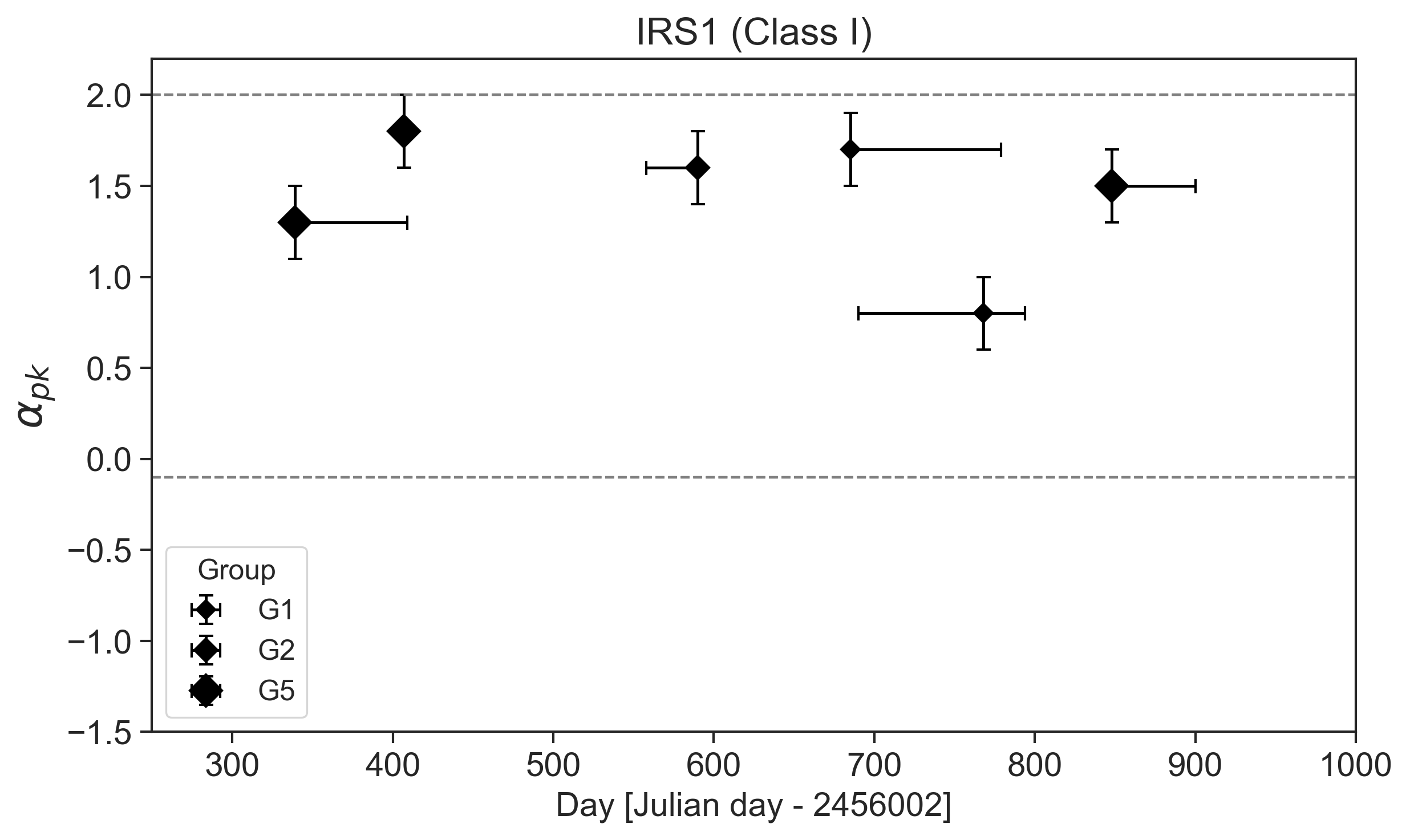}
    \end{subfigure}%
    \begin{subfigure}{0.5\textwidth}
        \centering
        \includegraphics[width=\linewidth]{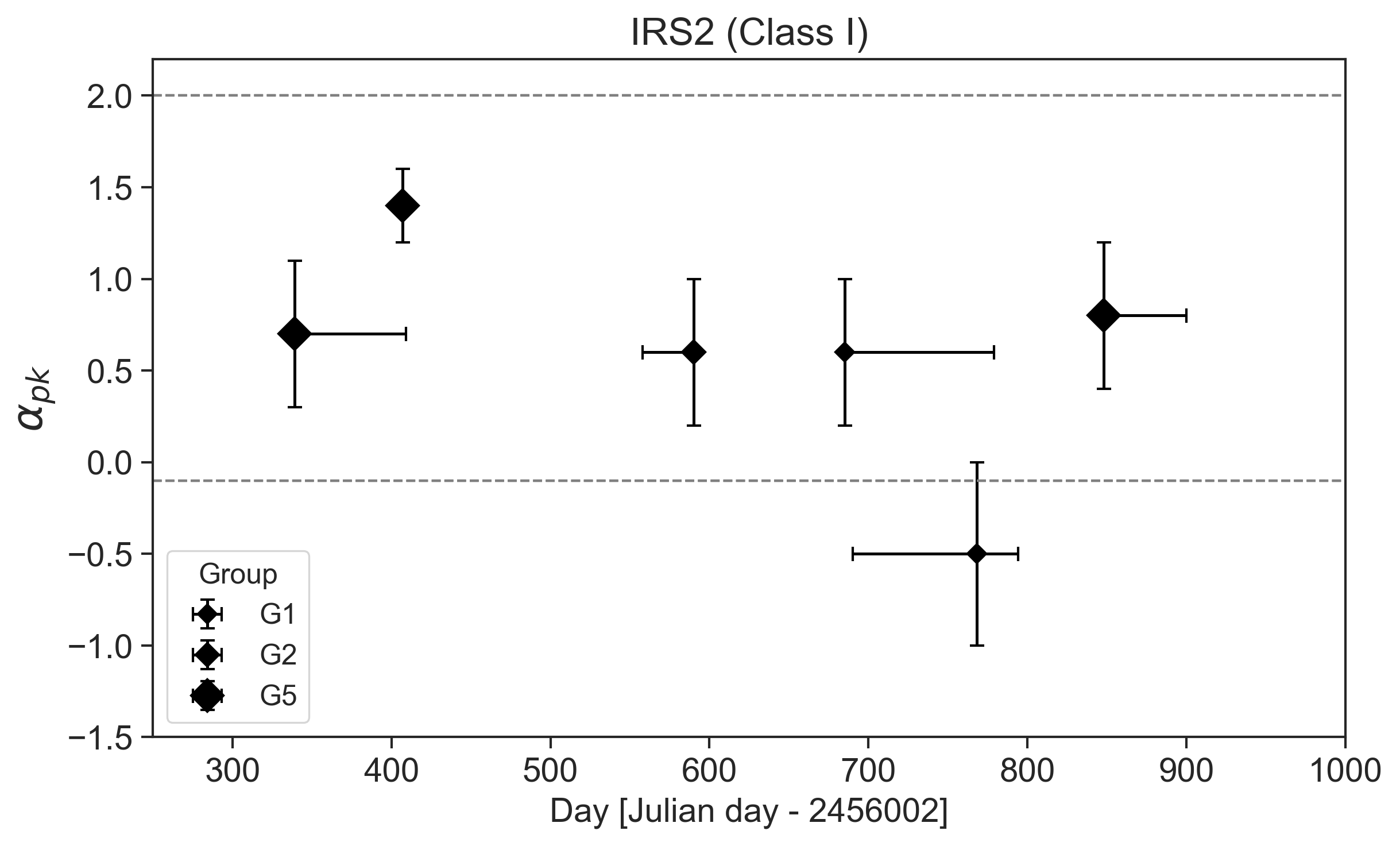}
    \end{subfigure} 

    % Fila 3
    \begin{subfigure}{0.5\textwidth}
        \centering
        \includegraphics[width=\linewidth]{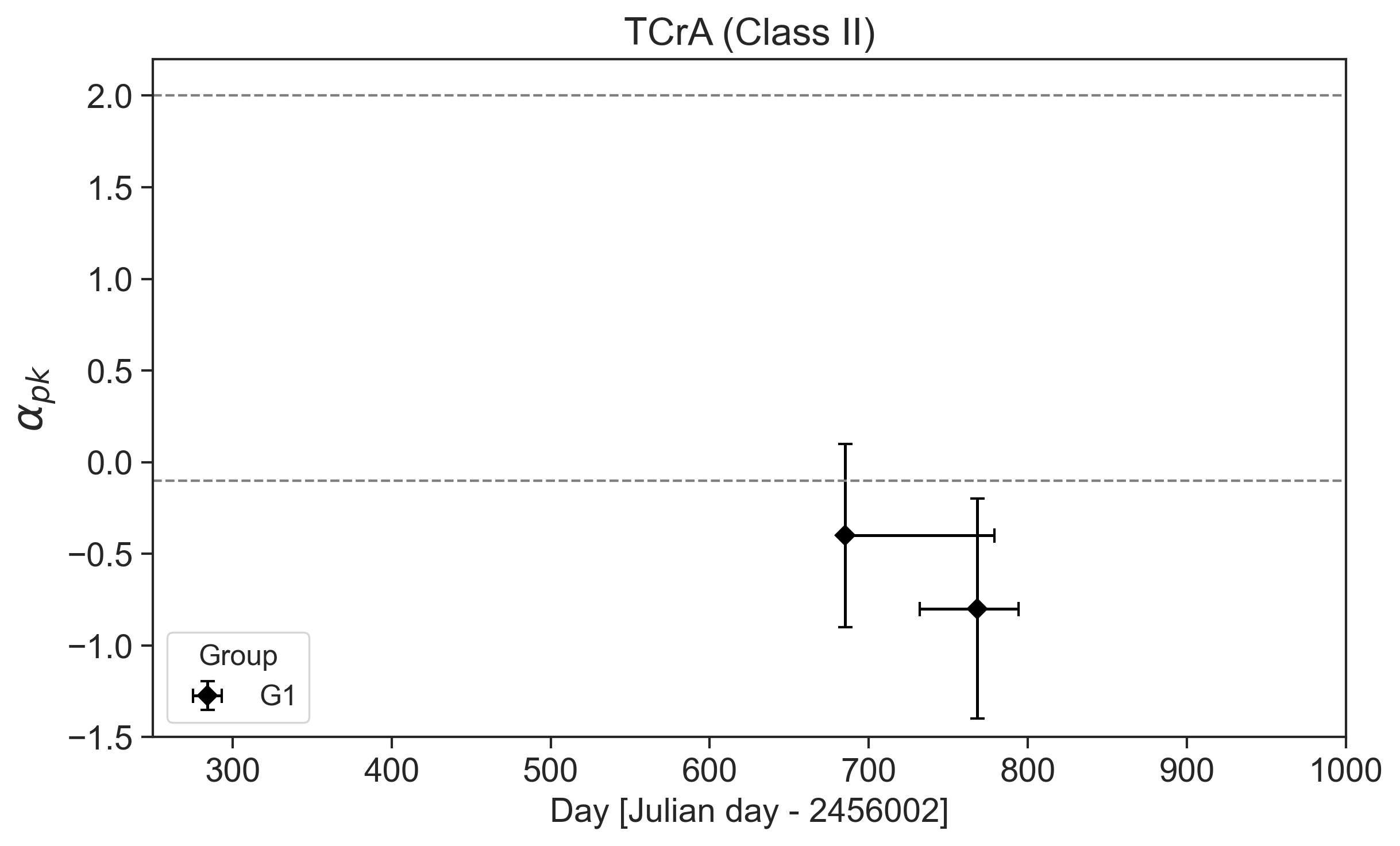}
    \end{subfigure}%
    \begin{subfigure}{0.5\textwidth}
        \centering
        \includegraphics[width=\linewidth]{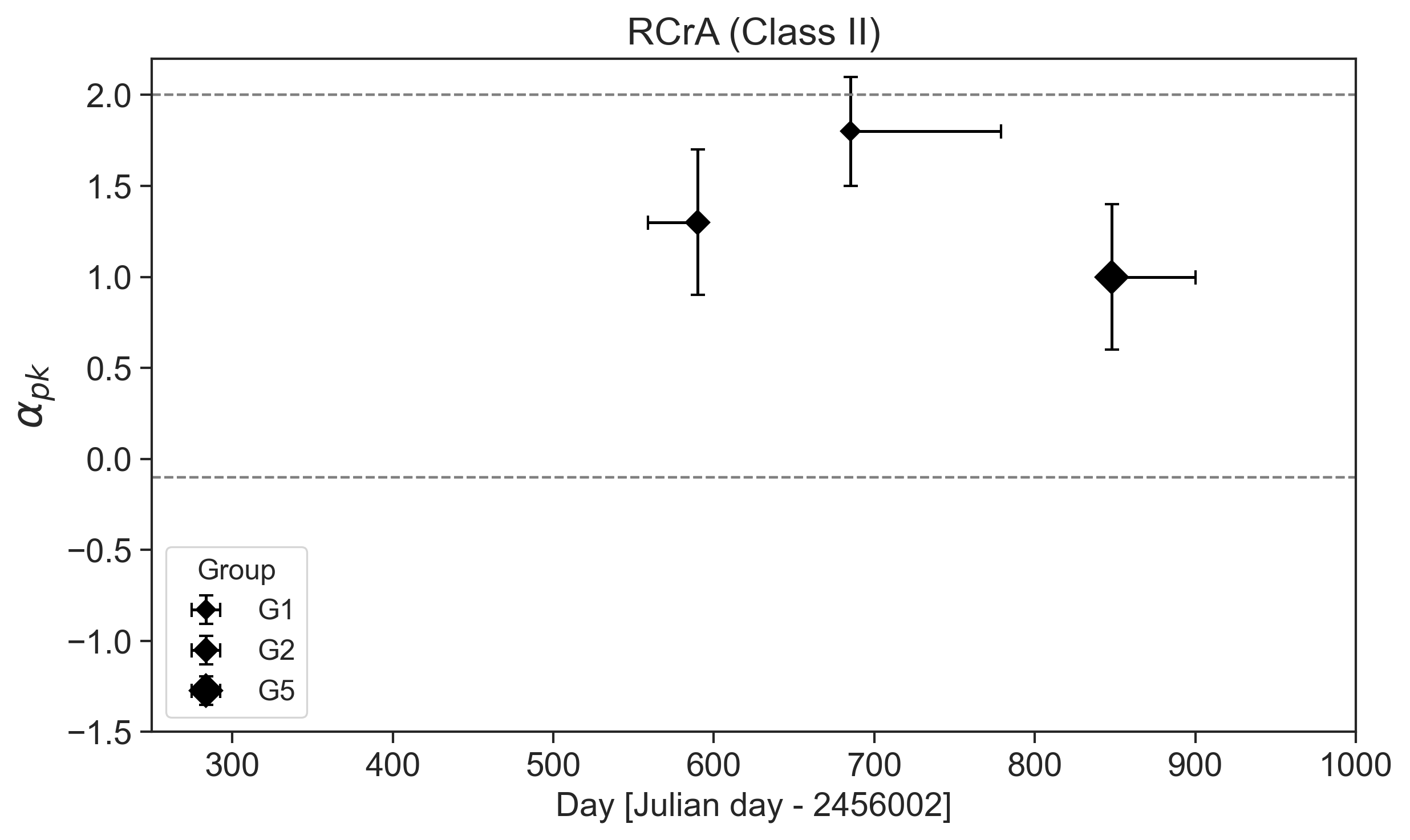}
    \end{subfigure} 

    \caption{Time variability of the peak spectral index. Symbol sizes indicate the resolution group. The horizontal dashed lines represent the optically thick (upper) and optically thin (lower) free–free emission limits. The x-axis location of the diamond symbols denote the epoch of the \textit{Ku}-band detections, while the horizontal error bars indicate the time intervals during which \textit{X}-band detections were obtained and $\alpha_{\mathrm{pk}}$ was computed. Vertical error bars correspond to the uncertainty in $\alpha_{\mathrm{pk}}$. Sources are ordered by evolutionary stage from top-left to bottom-right.}
    \label{fig:alpha_var}
\end{figure*}

We were able to perform the spectral index variability analysis for six sources (Fig.~\ref{fig:alpha_var}) with enough available measurements. The sources JVLA2, TCrA, and RCrA show no significant variation of $\alpha_{\mathrm{pk}}$ within the errors. The younger YSOs B9 and JVLA2 remain in the optically thin free–free emission regime. 
 
Among the more evolved Class~II YSOs, TCrA is consistent with optically thin free–free emission, with a possible non-thermal contribution. In contrast, RCrA shows partially optically thick spectral indices in all epochs.
Similarly, the Class~I source IRS~1 exhibits partially optically thick free–free emission at all times, whereas the Class I IRS~2 also shows intermediate spectral indices, although closer to the optically-thin limit, and decreasing sharply around one epoch.

\section{Discussion}
\label{sec:discussion}

Free-free emission in young YSOs is linked to the amount of material being ejected. Therefore, by studying the radio emission, we gain indirect access to the accretion processes of protostars. This connection arises from the strong coupling between accretion and the mechanisms responsible for wind and jet launching \citep[e.g.,][]{Ray2007}. 

Observational evidence supports this accretion–ejection connection. Previous studies of embedded protostars, summarised in \citet{Anglada2018}, have shown a correlation between radio and bolometric luminosities, suggesting a physical link between accretion and ejection processes. More recently, studies of Class~II YSOs—where accretion rates can be directly measured through infrared recombination lines—have revealed a sub-linear relationship between the accretion rate $\dot{M}_{\mathrm{acc}}$ and the ionised jet mass-loss rate $\dot{M}_{\mathrm{ion}}$ \citep[e.g.,][]{Rota24, Rota25, Garuffi2025}. This relation has now been extended to Class~I YSOs \citep[][]{Ghosh2026}. 

\begin{figure*}
	\includegraphics[width=0.8\textwidth]{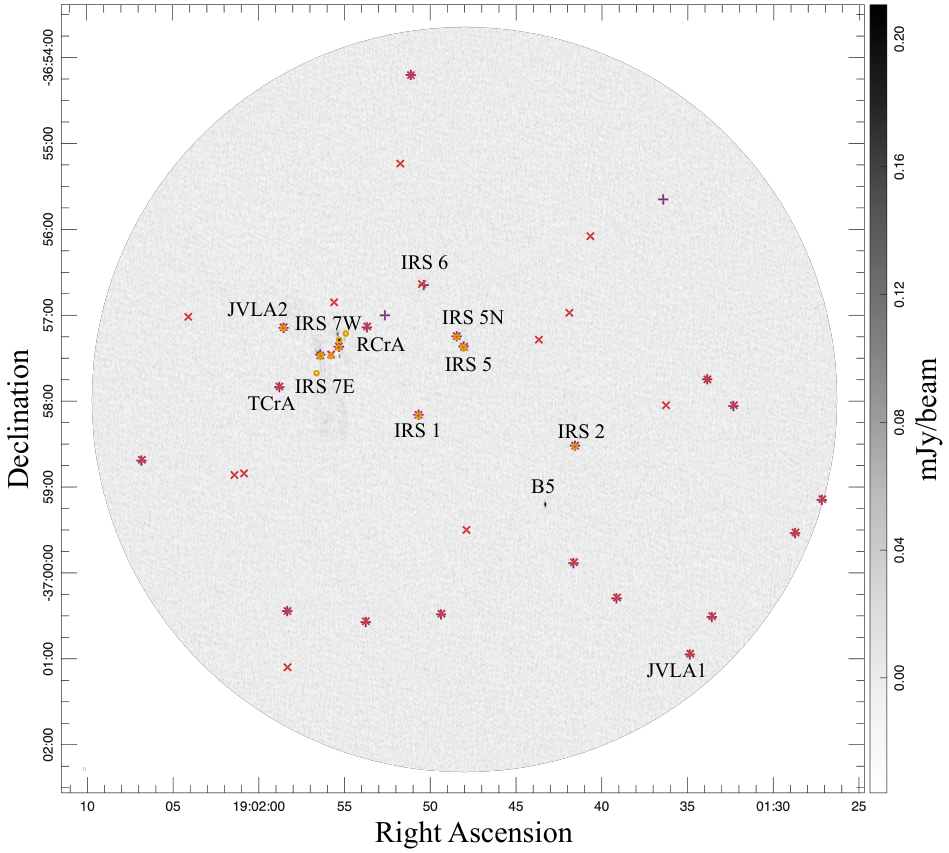}
    \caption{YSOs in the Coronet cluster within our field of view, previously detected by \citet{Peterson11} ($\times$ symbols), \citet{Esplin22} (crosses), and \citet{Hsieh24} (open circles).}
    \label{fig:coronet_sources}
\end{figure*}

The Coronet is a well-known young stellar cluster that contains a very compact group of protostars in early evolutionary stages, which have been previously studied at different wavelengths, including near-infrared \citep[e.g.,][]{Peterson11, Esplin22}, mid-infrared \citep[e.g.,][]{Sandell2021}, X-ray \citep[e.g.,][]{Forbrich07}, submillimetre \citep[e.g.,][]{Hsieh24, Maureira2025}, and radio \citep[e.g.,][]{Choi2008, Liu2014}. From these studies, we identified 42 YSOs within our field of view (see Fig. \ref{fig:coronet_sources}). In our radio observations we detected approximately 60\% (5/8) of the Class~0, 60\% of the Class~I (3/5), 50\% (2/4) of the Flat-spectrum, 20\% (3/14) of the Class~II, and 10\% (1/11) of the Class~III sources. We report 20 sources in our catalogue, half of which are Class~0 and Class~I YSOs, including flat-spectrum objects.  
In contrast, our radio observations are less sensitive to Flat-spectrum, Class~II, and Class~III YSOs. These more evolved objects can range from very faint to very bright owing to their intrinsic variability \citep[e.g.,][]{Forbrich2006, Liu2014}. Our results demonstrate the capability of deep radio continuum surveys to detect a significant fraction of the earliest YSOs.

\subsection{Characterizing radio emission in all evolutionary stages}

Photometry in the near-IR and the bolometric temperature ($T_{\mathrm{bol}}$) derived from submillimetre data allowed us to assign a continuous classification to the YSOs. We can characterise the radio emission of our sample through the spectral index (see Fig.~\ref{fig:k-8}), which provides information about the emission mechanism and, therefore, about the physical processes responsible for the observed emission. Although our statistics are limited, our small sample still displays notable trends.

The younger YSOs (Class~0 and I) fall within the range of free–free emission ($-0.1<\alpha<2$), although non-thermal contamination cannot be ruled out for IRS~7A. These objects are in the most active accretion phase, during which episodes of intense accretion and ejection — and consequently, dense winds — are expected. The ionised gas in sources IRS~1, IRS~2, IRS~5N, and RCrA  appears to be denser and therefore exhibits partially optically thick emission. 
In contrast, JVLA2 lies within the optically thin emission regime.

The Class~II YSOs (TCrA and IRS~6) and the single flat-spectrum source IRS~5 show spectral indices consistent with the optically thin regime. Their emission may originate from faint thermal jets or photoevaporating disks \citep[e.g.,][]{Galvan2014, Pasucci2014,Rodriguez2014}; however, non-thermal contributions cannot be discarded and are known to be present for IRS 5 \citep{Feigelson1998}.  

For the Class~III YSO JVLA1, we report an upper limit ($\alpha_{\mathrm{pk}} \leq 0.5$) consistent with the optically thin regime, although also compatible with non-thermal emission. The short-timescale variability in this object also suggests non-thermal emission. In Class~III sources, significant thermal radio emission is not expected; instead, their emission is dominated by non-thermal (gyro)synchrotron radiation produced by magnetospheric activity \citep[e.g.,][]{Dzib2013, Ortiz-Leon2015}. 

Furthermore, we investigated spectral index variability in a small subset of YSOs (see Fig.~\ref{fig:alpha_var}) to explore potential intrinsic changes in their emission mechanisms. We find that although some sources exhibit variations exceeding their uncertainties, they remain within the same emission regime. The only exception is the Class~I YSO IRS~2, which appears to transition from partially optically thick to optically thin emission.

\subsection{YSOs variability}
\label{sec:yso_var}

Radio emission variability was investigated through several statistical diagnostics applied to our YSO sample. First, the variability index $VI$ indicates that all sources exhibit variability (i.e., $VI > 1$). From the analysis of absolute variability, we find that some YSOs display significantly larger interquartile ranges (IQRs) than others. In the \textit{X}-band, IRS~5b, JVLA1, and IRS~7A are the most variable sources, followed by IRS~7E and IRS~6. IRS~5b is known to exhibit gyrosynchrotron emission \citep[e.g.,][]{Feigelson1998, Choi2008}, which may account for its enhanced variability. The light curve of JVLA1 shows bursts followed by extended periods of non-detections, suggesting episodic activity. In the cases of IRS~7A, IRS~7E, and IRS~6 — all multiple systems — interactions between their components may contribute to the observed variability. In the \textit{Ku}-band, IQR values are generally smaller than in the \textit{X}-band, with the notable exception of IRS~5.

Although the absolute peak intensities span different ranges among sources, the fractional variability analysis reveals a more uniform behaviour: YSOs across all evolutionary stages exhibit typical percentage variability of a few to several tens of percent. In general, variability amplitudes are slightly higher in the \textit{X}-band. IRS~5, IRS~6 and JVLA1 again stand out as the most variable sources, in terms of their IQRs. Aside from these two cases, we find ubiquitous fractional variability across evolutionary stages.

Finally, the structure functions computed for a subset of YSOs show no clear evidence of preferred variability timescales, suggesting that the radio variability is predominantly stochastic. JVLA1 exhibits enhanced variability at short time lags, which, together with the results above, supports the interpretation that its radio emission is linked to magnetospheric activity typical of Class~III YSOs.  The nature of the radio emission variability in the younger YSOs could be related to the ionized ejection, linked to fluctuations in the accretion. From these, R CrA also shows evidence of enhanced variability at short timescales. This Class II YSO is a triple system where the components of the close binary are relatively massive \citep[2 to 3 $M_\odot$,][]{Rigliaco2019}. The spectral indices of this object are consistent with partially optically thick free-free emission.

\subsection{Nature of individual YSOs in the Coronet}
\label{sec:natureofsubcomps}

In this work, we investigate several noteworthy sources within the Coronet cluster. Our high angular resolution (around 120 au) allowed us to identify young binary systems associated with the origin of ionised and molecular outflow components in IRS~7A and IRS~7B \citep{Liu2014,Sabatini2024}. We also report sources that may correspond to shocks, including FPM13, FPM10, IRS~7AS, IRS~7Ab and IRS~7BN. The classification of these objects as shock candidates is based on several observational properties. First, they lack clear infrared and submillimetre counterparts, which indicates that they are not embedded protostars. Second, they reside  within extended emission associated with the prominent radio jets. Finally, their spectral indices are compatible with the non-thermal emission expected from shocks (see Sec. \ref{sec:shocks} and Table \ref{tab:alpha_pk}).

\begin{figure}
\centering
\includegraphics[width=0.9\columnwidth]{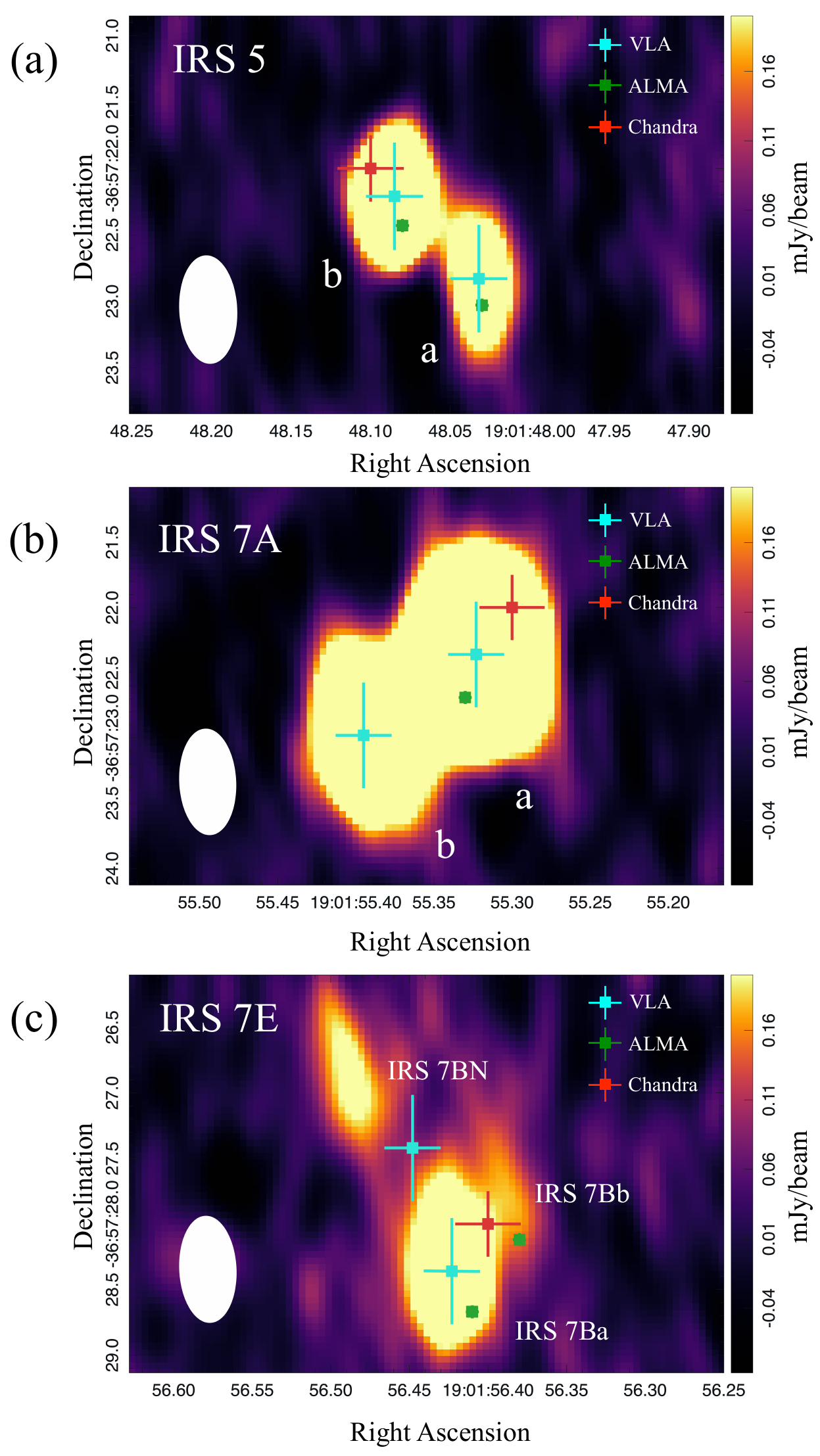}
\caption{Zoom-in views of the subcomponents in IRS~5 (a), IRS~7A (b), and IRS~7B (c) at 9.0 GHz (3.3 cm), imaged with uniform weighting. The synthesized beam is $\theta_{\rm syn}=0\farcs81 \times 0\farcs43$, and the rms noise is $\sim27\ \mu$Jy beam$^{-1}$ before primary-beam correction. Symbols mark the coordinates reported in this work with the VLA, by \citet{Hsieh24} with ALMA, and by \citet{Forbrich2006} with Chandra. Over plotted lines indicate the angular resolution of each instrument: for the VLA, the synthesized beam; for ALMA, $0\farcs1$; and for Chandra, $0\farcs5$.}
\label{fig:subcomponents}
\end{figure}

These remarkable YSOs are described in detail in the following subsections. In Fig.~\ref{fig:subcomponents}, we show the subcomponents within IRS~5, IRS~7A, IRS~7B, and their coordinates as reported in this work, together with those identified with ALMA by \citet{Hsieh24} and with \textit{Chandra} by \citet{Forbrich2006}. We also indicate the angular resolution of each observation. For the VLA, we adopt the synthesized beam of the uniform-weighted \textit{X}-band deep map; for ALMA, we use the $0\farcs1$ (15~au) beam reported by \citet{Hsieh24}; and for \textit{Chandra}, the nominal resolution of $0\farcs5$ (75~au).

\subsubsection{IRS 5}

IRS~5 is a particularly interesting binary system. It is the first YSO known to have non-thermal gyrosynchrotron emission \citep[e.g.,][]{Feigelson1998} and the first firm radio detection using very long baseline interferometry \citep[][]{Deller2013}. Its components (see Fig.~\ref{fig:subcomponents}a), IRS~5b (north-east) and IRS~5a (south-west), are separated by about $0''.9$ (135~au). \citet{Hsieh24} classified both objects as Flat-spectrum YSOs. \citet{Forbrich2006} reported X-ray emission from IRS~5, whose position lies closer—within the uncertainties—to our VLA detection of IRS~5b. Moreover, \citet{Choi2008} detected circularly polarised emission at 3.5~cm from IRS~5b, confirming gyrosynchrotron radiation, whereas IRS~5a shows no detectable polarization. The radio positions of both components coincide with their respective near-IR counterparts reported by \citet{Nisini2005}.

We also present their light curves and box plots, which show that IRS~5b exhibits stronger variability than IRS~5a (see Fig.~\ref{fig:panel3} and Fig. \ref{fig:boxplots}). We suggest that interactions between the two components may produce bright radio flares, accounting for the known high variability reported for IRS~5 as a whole; however, the variability may also be dominated solely by the activity of IRS~5b. The average radio spectral index computed for IRS~5 ($\alpha_\mathrm{pk}=0.12\pm0.63$) indicates that optically thin free–free emission could be contaminated by a non-thermal contribution, consistent with the gyrosynchrotron emission detected from IRS~5b. 

\subsubsection{IRS 7A}

IRS~7A is one of the brightest sources in the Coronet cluster and a binary centimetric system (see Fig.~\ref{fig:subcomponents}b). \citet{Hsieh24} classified the component IRS~7Aa as a Class~0 YSO, and their reported millimetre position is consistent, within the uncertainties, with our VLA detection. We detect a new centimetre source to the southeast, which we designate IRS~7Ab. This object has not been previously classified in the literature and was not detected in recent 1.3 mm and 3 mm observations \citep[][]{Hsieh24, Maureira2025}. The lack of a millimetre compact source suggests that the centimetre emission may be non-thermal in origin, possibly arising from a shock.

Additionally, IRS~7A and the other components within the region surrounding IRS~7W (B9, FPM10, FPM13, IRS~7AS) may be driving a roughly north–south–oriented radio jet (see Fig.~\ref{fig:jet}). The source IRS~7Ab could trace a shock produced by the interaction of these components within the jet. In the study at 3~mm (see Fig. 1 in \citet{Maureira2025}) the emission from IRS~7A appears extended from south-east to north-west, IRS~7Ab may be related to the emission at the southeastern end.

We report a marginally negative radio spectral index for IRS~7A as a whole, $\alpha_{\mathrm{pk}} =-0.36 \pm 0.19$, which could arise from non-thermal processes in IRS~7Aa. \citet{Forbrich07} reported X-ray emission close to IRS~7Aa, which may originate from a protostellar magnetosphere. The detection of circularly polarised radio emission in IRS~7Aa further supports the presence of an active magnetosphere \citep[][]{Choi2008}. 

The light curves show that IRS~7Aa is brighter and varies within approximately $\pm15\%$ of the median value, while IRS~7Ab exhibits variability within $\pm20\%$ of the median value (see Fig.~\ref{fig:panel3}). Moreover, \citet{Choi2008} reported an increasing flux density, a trend that we confirm with our observations (see Fig.~\ref{fig:panel1}).

\subsubsection{IRS 7B}

IRS~7E is resolved as a multiple system at centimetre wavelengths (see Fig.~\ref{fig:subcomponents}c). It is composed of IRS~7Ba, IRS~7Bb and a northern source that we designate IRS~7BN. These components are embedded within an extended radio structure oriented northeast–southwest. IRS~7B is a known young and compact submillimetre binary system. Recent work by \citet{Maureira2025} reports millimetre emission from the circumstellar disks around IRS~7Ba and IRS~7Bb. The two subcomponents, IRS~7Ba (southwest) and IRS~7Bb (northeast), are classified as Class~0 YSOs \citep[][]{Hsieh24}. In contrast, IRS~7BN, which has no previous classification, appears to be associated exclusively with the extended radio emission.

X-ray emission has been reported by \citet{Forbrich2006}; the X-ray spectrum shows a hard component, and a soft component likely produced by shock-heated plasma associated with a jet or outflow \citep[e.g.,][]{Hamaguchi2005, ForbrichyPreibisch2007}. Although the multiple system is only marginally resolved in our highest-resolution map, it is unclear whether the X-ray emission arises from IRS~7Ba or IRS~7Bb.

In this work, we find flux variations in IRS~7E and its components of up to approximately $\pm20\%$ around the median value (see Fig.~\ref{fig:panel3}), although the source IRS~7BN remains closest to the non-variable regime ($VI=1$) in the \textit{X}-band. 
IRS~7E exemplifies the coexistence of magnetic processes and outflow activity during the earliest stages of star formation.

\begin{figure}
\centering	\includegraphics[width=0.8\columnwidth]{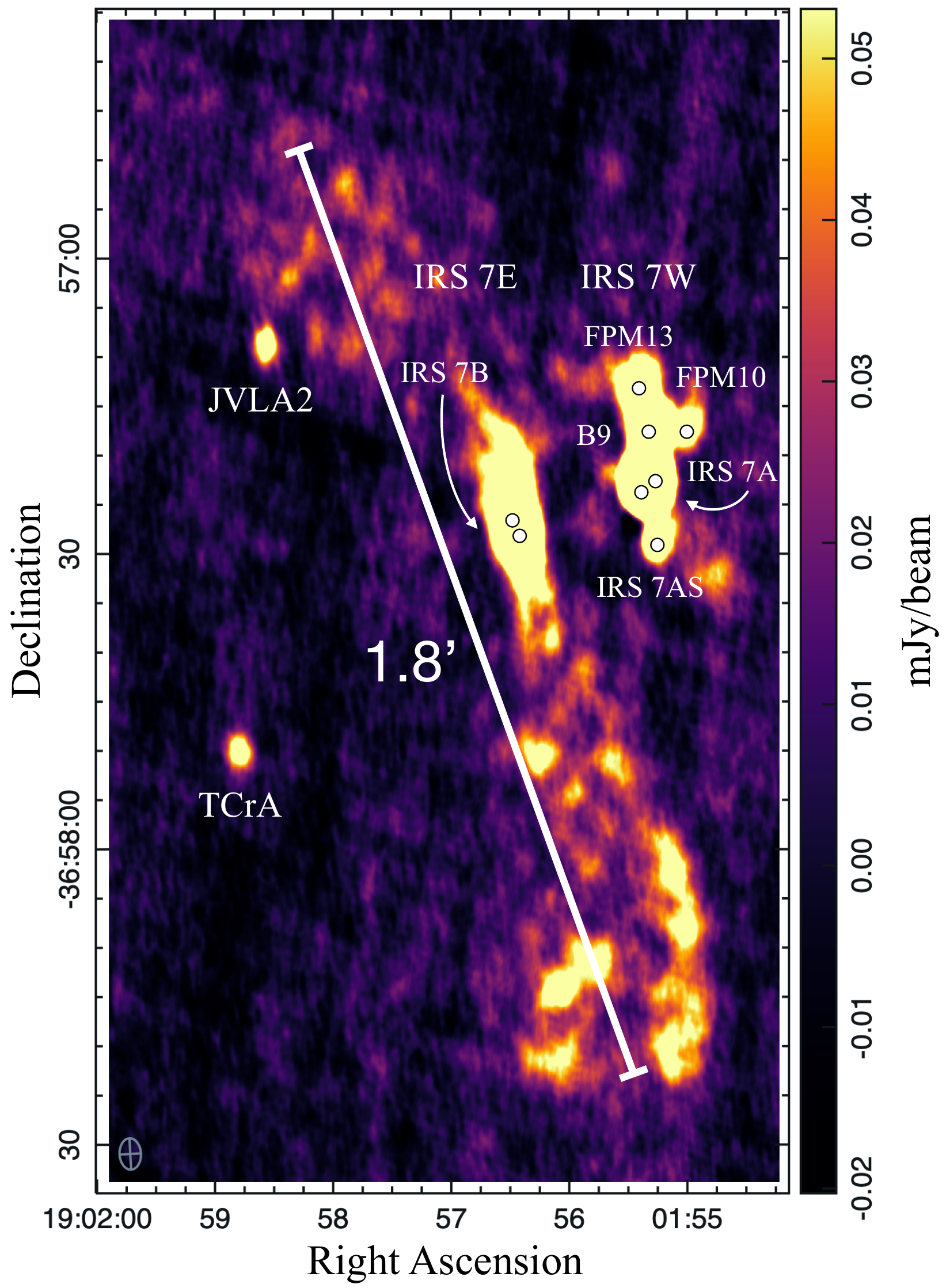}
    \caption{Zoom-in of the map generated with natural weighting, prior to the primary beam correction, showing the extended emission toward the source IRS7E. The synthesized beam size is $3''.19 \times 2''.18$ with an rms of $\sim\ 7\mu$Jy beam$^{-1}$. }
    \label{fig:jet}
\end{figure}

As described in Sec. \ref{sec:observations}, the deep maps were generated using different weightings to study the emission at various resolutions. In the \textit{X}-band deep map with natural weighting, we can appreciate the extended emission associated with the source IRS~7B (see Fig. \ref{fig:jet}). This structure has an extent of about 1.8 arcmin (16 200~au at 150~pc). 
 
This is a very atypical radio jet because of its large extension; radio jets are typically compact, with sizes of about $\sim10^2$~au \citep[e.g.,][]{Moscadelli2016,Anglada2018}. Extended jets have been observed in a few massive YSOs, which also exhibit non-thermal emission and shocks with the surrounding environment \citep[e.g.,][]{AdrianaRK2016, Carrasco21}. We speculate that the observed morphology could result from the interaction of several aligned jets, which could enhance the large-scale emission. 

Moreover, it is important to emphasize that IRS~7Ba and IRS~7Bb are Class~0, and therefore are in their most active accretion phase. Also, the two millimetre discs in IRS 7B are extremely well aligned with each other \citep{Maureira2025}, and perpendicular to the radio jet emission.

\subsubsection{FPM10, FPM13 and IRS 7AS}
\label{sec:shocks}

\begin{figure}
	\includegraphics[width=\columnwidth]{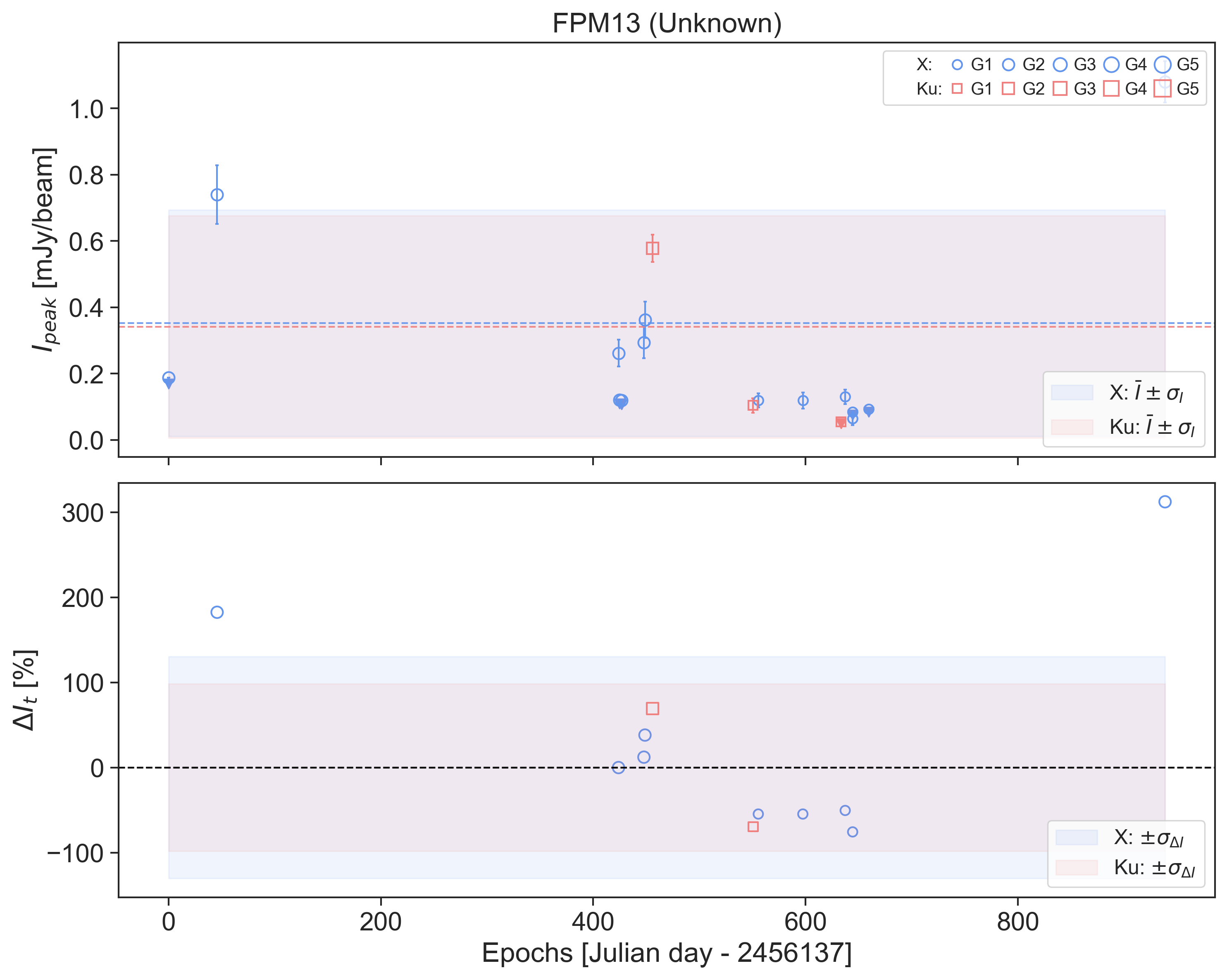}
    \caption{FPM13 light curve: The top panel show the peak intensity at each epoch, while the bottom panel show the percentage variability of the detections. The symbol sizes indicate the five resolution groups; circles correspond to measurements in the \textit{X}-band and squares to those in the \textit{Ku}-band. Dashed lines, in top panel, indicate the mean intensity in each band. In the top panel, the shaded regions span $\pm$ one standard deviation of the peak-intensity distribution, while in the bottom panel it span $\pm$ one standard deviation of the percentage-variability distribution.}
    \label{fig:fpm13}
\end{figure}

We report three extended sources within the region surrounding IRS~7W (see Fig.~\ref{fig:coraus_zoom}), two of them exhibiting negative spectral indices that may be associated with non-thermal emission. FPM10 traces the ionized component of a molecular outflow previously reported by \citet{Groppi2004}, likely driven by the protostar B9 \citep{Choi2008}. Its spectral index, $\alpha_{\mathrm{pk}} = -0.14 \pm 0.49$, lies within the optically thin regime and is consistent with predominantly thermal emission possibly contaminated by a non-thermal component.
The source FPM13 has no identified molecular outflow counterpart. 
We find significant variability, with peak intensity variations of up to $\pm100\%$ around the median value (see Fig.~\ref{fig:fpm13}). The spectral index, $\alpha_{\mathrm{pk}} = -1.16 \pm 0.80$, is consistent with non-thermal emission, and we suggest that FPM13 may correspond to a non-thermal shock \citep[e.g.,][]{Rodriguez-Kamenetzky2022}; however, a gyrosynchrotron origin cannot be ruled out.
Finally, the newly detected source IRS~7AS is detected only in the deep \textit{X}-band map. It is likely a non-thermal shock launched by one of the components of IRS~7A, as its spectral index, $\alpha_{\mathrm{pk}} < -0.8$, is consistent with non-thermal emission.

\section{Conclusions}

We presented 9.0 and 14.0 GHz VLA radio continuum observations, obtained between March 2012 and February 2015, toward a sample of 20 YSOs in the Coronet Cluster. We analysed 39 epochs across all available VLA configurations and homogenized the analysis by defining five resolution groups. From this data set, we produced deep \textit{X} and \textit{Ku}-band maps at multiple angular resolutions, ranging from uniform weighting ($\theta_{\rm syn}=0\farcs81 \times 0\farcs43$) to natural weighting ($\theta_{\rm syn}=3\farcs19 \times 2\farcs18$), i.e., down to a physical resolution of 65 au. The achieved noise levels are 7–9 $\mu$Jy beam$^{-1}$. Our main results are summarized below.

We report two new centimetre detections near IRS~7A—IRS~7AS (south) and IRS~7Ab (east)—and a new source northeast of IRS~7B (IRS~7BN). We also resolve IRS~5 into two subcomponents, consistent with its previously reported multiplicity. As expected, the relatively high frequencies observed in this project made our radio survey more sensitive to free-free than to non-thermal emission. Therefore, our survey is more complete to the detection of Class~0 and I YSOs than to the more evolved sources.

Radio spectral indices were computed to characterize the emission mechanisms. In general, Class 0, I, II and Flat-spectrum YSOs fall within the expected range for free–free emission, although non-thermal contamination cannot be ruled out for some sources. For the Class III source, we report an upper limit ($\alpha_{\mathrm{pk}} \leq 0.5$) suggestive of non-thermal emission.

We found that variability is ubiquitous across all evolutionary stages in both bands. 
The variability structure functions, S$(\Delta t)$, show no significant correlation with evolutionary class. We find no clear evidence for preferred variability timescales, suggesting that the observed variability in this subset of YSOs is stochastic in nature. The main exception is seen in the most evolved YSOs in the sample, which exhibit enhanced variability at short time lags.

We report extended \textit{X}-band emission toward IRS~7B with an extent of $\sim$1.8~arcmin (16 200~au). Its morphology may result from the interaction of multiple aligned jets that enhance the large-scale emission. The sources near the emission peak likely interact with each other and with nearby members of IRS~7A. Moreover, some subcomponents within IRS~7B and IRS~7A exhibit negative spectral indices, suggestive of non-thermal processes.

The Coronet Cluster offers a unique laboratory for investigating multiple stages of star formation. In this work we have characterized the radio continuum properties and variability of several notable YSOs, revealing a complex interplay between thermal and non-thermal processes, multiplicity, and stochastic variability. We additionally report sources whose radio emission may trace shocked gas associated with outflows or interactions among nearby YSOs.

\section*{Acknowledgements}

%The Acknowledgements section is not numbered. Here you can thank helpful
%colleagues, acknowledge funding agencies, telescopes and facilities used etc.
This research made use of data obtained with the Karl G. Jansky Very Large Array (VLA), operated by the National Radio Astronomy Observatory (NRAO). The NRAO is a facility of the National Science Foundation operated under cooperative agreement by Associated Universities, Inc.
R.G.M, C.C.-G., and J.R.A acknowledge support from DGAPA-PAPIIT projects IN105225, and IG101224.
The authors made use of the following software:
\texttt{CASA} \citep{Casa22}, \texttt{astropy} \citep{astropy} and \texttt{CARTA} \citep{CARTA}.

%%%%%%%%%%%%%%%%%%%%%%%%%%%%%%%%%%%%%%%%%%%%%%%%%%
\section*{Data Availability}

Images in FITS format will be made publicly available through Zenodo at DOI: https://doi.org/10.5281/zenodo.19078941.



\bibliographystyle{mnras}
\bibliography{example} % if your bibtex file is called example.bib

\begin{thebibliography}{}
\makeatletter
\relax
\def\mn@urlcharsother{\let\do\@makeother \do\$\do\&\do\#\do\^\do\_\do\%\do\~}
\def\mn@doi{\begingroup\mn@urlcharsother \@ifnextchar [ {\mn@doi@}
  {\mn@doi@[]}}
\def\mn@doi@[#1]#2{\def\@tempa{#1}\ifx\@tempa\@empty \href
  {http://dx.doi.org/#2} {doi:#2}\else \href {http://dx.doi.org/#2} {#1}\fi
  \endgroup}
\def\mn@eprint#1#2{\mn@eprint@#1:#2::\@nil}
\def\mn@eprint@arXiv#1{\href {http://arxiv.org/abs/#1} {{\tt arXiv:#1}}}
\def\mn@eprint@dblp#1{\href {http://dblp.uni-trier.de/rec/bibtex/#1.xml}
  {dblp:#1}}
\def\mn@eprint@#1:#2:#3:#4\@nil{\def\@tempa {#1}\def\@tempb {#2}\def\@tempc
  {#3}\ifx \@tempc \@empty \let \@tempc \@tempb \let \@tempb \@tempa \fi \ifx
  \@tempb \@empty \def\@tempb {arXiv}\fi \@ifundefined
  {mn@eprint@\@tempb}{\@tempb:\@tempc}{\expandafter \expandafter \csname
  mn@eprint@\@tempb\endcsname \expandafter{\@tempc}}}

\bibitem[\protect\citeauthoryear{{Allen} et~al.,}{{Allen}
  et~al.}{2004}]{Allen2004}
{Allen} L.~E.,  et~al., 2004, \mn@doi [\apjs] {10.1086/422715}, \href
  {https://ui.adsabs.harvard.edu/abs/2004ApJS..154..363A} {154, 363}

\bibitem[\protect\citeauthoryear{{Andre}, {Ward-Thompson}  \&
  {Barsony}}{{Andre} et~al.}{1993}]{Andre1993}
{Andre} P.,  {Ward-Thompson} D.,   {Barsony} M.,  1993, \mn@doi [\apj]
  {10.1086/172425}, \href
  {https://ui.adsabs.harvard.edu/abs/1993ApJ...406..122A} {406, 122}

\bibitem[\protect\citeauthoryear{{Anglada}, {Villuendas}, {Estalella},
  {Beltr{\'a}n}, {Rodr{\'\i}guez}, {Torrelles}  \& {Curiel}}{{Anglada}
  et~al.}{1998}]{Anglada1998}
{Anglada} G.,  {Villuendas} E.,  {Estalella} R.,  {Beltr{\'a}n} M.~T.,
  {Rodr{\'\i}guez} L.~F.,  {Torrelles} J.~M.,   {Curiel} S.,  1998, \mn@doi
  [\aj] {10.1086/300637}, \href
  {https://ui.adsabs.harvard.edu/abs/1998AJ....116.2953A} {116, 2953}

\bibitem[\protect\citeauthoryear{{Anglada}, {Rodr{\'\i}guez}  \&
  {Carrasco-Gonz{\'a}lez}}{{Anglada} et~al.}{2018}]{Anglada2018}
{Anglada} {Rodr{\'\i}guez}  {Carrasco-Gonz{\'a}lez} 2018, \mn@doi [\aapr]
  {10.1007/s00159-018-0107-z}, \href
  {https://ui.adsabs.harvard.edu/abs/2018A&ARv..26....3A} {26, 3}

\bibitem[\protect\citeauthoryear{{Astropy Collaboration} et~al.,}{{Astropy
  Collaboration} et~al.}{2018}]{astropy}
{Astropy Collaboration} et~al., 2018, \mn@doi [\aj] {10.3847/1538-3881/aabc4f},
  \href {https://ui.adsabs.harvard.edu/abs/2018AJ....156..123A} {156, 123}

\bibitem[\protect\citeauthoryear{{CASA Team} et~al.,}{{CASA Team}
  et~al.}{2022}]{Casa22}
{CASA Team} et~al., 2022, \mn@doi [\pasp] {10.1088/1538-3873/ac9642}, \href
  {https://ui.adsabs.harvard.edu/abs/2022PASP..134k4501C} {134, 114501}

\bibitem[\protect\citeauthoryear{{Carrasco-Gonz{\'a}lez}, {Sanna},
  {Rodr{\'\i}guez-Kamenetzky}, {Moscadelli}, {Hoare}, {Torrelles},
  {Galv{\'a}n-Madrid}  \& {Izquierdo}}{{Carrasco-Gonz{\'a}lez}
  et~al.}{2021}]{Carrasco21}
{Carrasco-Gonz{\'a}lez} C.,  {Sanna} A.,  {Rodr{\'\i}guez-Kamenetzky} A.,
  {Moscadelli} L.,  {Hoare} M.,  {Torrelles} J.~M.,  {Galv{\'a}n-Madrid} R.,
  {Izquierdo} A.~F.,  2021, \mn@doi [\apjl] {10.3847/2041-8213/abf735}, \href
  {https://ui.adsabs.harvard.edu/abs/2021ApJ...914L...1C} {914, L1}

\bibitem[\protect\citeauthoryear{{Cazzoletti} et~al.,}{{Cazzoletti}
  et~al.}{2019}]{Cazzoletti2019}
{Cazzoletti} P.,  et~al., 2019, \mn@doi [\aap] {10.1051/0004-6361/201935273},
  \href {https://ui.adsabs.harvard.edu/abs/2019A&A...626A..11C} {626, A11}

\bibitem[\protect\citeauthoryear{{Choi}, {Hamaguchi}, {Lee}  \&
  {Tatematsu}}{{Choi} et~al.}{2008}]{Choi2008}
{Choi} M.,  {Hamaguchi} K.,  {Lee} J.-E.,   {Tatematsu} K.,  2008, \mn@doi
  [\apj] {10.1086/591540}, \href
  {https://ui.adsabs.harvard.edu/abs/2008ApJ...687..406C} {687, 406}

\bibitem[\protect\citeauthoryear{Comrie et~al.,}{Comrie et~al.}{2026}]{CARTA}
Comrie A.,  et~al., 2026, CARTA: The Cube Analysis and Rendering Tool for
  Astronomy, \mn@doi{10.5281/zenodo.18477253}, \url
  {https://doi.org/10.5281/zenodo.18477253}

\bibitem[\protect\citeauthoryear{{Deller}, {Forbrich}  \& {Loinard}}{{Deller}
  et~al.}{2013}]{Deller2013}
{Deller} A.~T.,  {Forbrich} J.,   {Loinard} L.,  2013, \mn@doi [\aap]
  {10.1051/0004-6361/201321085}, \href
  {https://ui.adsabs.harvard.edu/abs/2013A&A...552A..51D} {552, A51}

\bibitem[\protect\citeauthoryear{{D{\'\i}az-M{\'a}rquez}
  et~al.,}{{D{\'\i}az-M{\'a}rquez} et~al.}{2024}]{Diaz-Marquez2024}
{D{\'\i}az-M{\'a}rquez} E.,  et~al., 2024, \mn@doi [\aap]
  {10.1051/0004-6361/202348085}, \href
  {https://ui.adsabs.harvard.edu/abs/2024A&A...682A.180D} {682, A180}

\bibitem[\protect\citeauthoryear{{Dunham} et~al.,}{{Dunham}
  et~al.}{2015}]{Dunham2015}
{Dunham} M.~M.,  et~al., 2015, \mn@doi [\apjs] {10.1088/0067-0049/220/1/11},
  \href {https://ui.adsabs.harvard.edu/abs/2015ApJS..220...11D} {220, 11}

\bibitem[\protect\citeauthoryear{{Dzib} et~al.,}{{Dzib}
  et~al.}{2013}]{Dzib2013}
{Dzib} S.~A.,  et~al., 2013, \mn@doi [\apj] {10.1088/0004-637X/775/1/63}, \href
  {https://ui.adsabs.harvard.edu/abs/2013ApJ...775...63D} {775, 63}

\bibitem[\protect\citeauthoryear{{Esplin} \& {Luhman}}{{Esplin} \&
  {Luhman}}{2022}]{Esplin22}
{Esplin} T.,  {Luhman} K.,  2022, \mn@doi [\aj] {10.3847/1538-3881/ac3e64},
  \href {https://ui.adsabs.harvard.edu/abs/2022AJ....163...64E} {163, 64}

\bibitem[\protect\citeauthoryear{{Evans} Neal~J. et~al.,}{{Evans}
  et~al.}{2009}]{Evans2009}
{Evans} Neal~J. I.,  et~al., 2009, \mn@doi [\apjs]
  {10.1088/0067-0049/181/2/321}, \href
  {https://ui.adsabs.harvard.edu/abs/2009ApJS..181..321E} {181, 321}

\bibitem[\protect\citeauthoryear{{Fazio} et~al.,}{{Fazio}
  et~al.}{2004}]{Fazio2004}
{Fazio} G.~G.,  et~al., 2004, \mn@doi [\apjs] {10.1086/422843}, \href
  {https://ui.adsabs.harvard.edu/abs/2004ApJS..154...10F} {154, 10}

\bibitem[\protect\citeauthoryear{{Feigelson}, {Carkner}  \&
  {Wilking}}{{Feigelson} et~al.}{1998}]{Feigelson1998}
{Feigelson} E.~D.,  {Carkner} L.,   {Wilking} B.~A.,  1998, \mn@doi [\apjl]
  {10.1086/311190}, \href
  {https://ui.adsabs.harvard.edu/abs/1998ApJ...494L.215F} {494, L215}

\bibitem[\protect\citeauthoryear{{Forbrich} \& {Preibisch}}{{Forbrich} \&
  {Preibisch}}{2007}]{ForbrichyPreibisch2007}
{Forbrich} J.,  {Preibisch} T.,  2007, \mn@doi [\aap]
  {10.1051/0004-6361:20066342}, \href
  {https://ui.adsabs.harvard.edu/abs/2007A&A...475..959F} {475, 959}

\bibitem[\protect\citeauthoryear{{Forbrich}, {Preibisch}  \&
  {Menten}}{{Forbrich} et~al.}{2006}]{Forbrich2006}
{Forbrich} J.,  {Preibisch} T.,   {Menten} K.~M.,  2006, \mn@doi [\aap]
  {10.1051/0004-6361:20052871}, \href
  {https://ui.adsabs.harvard.edu/abs/2006A&A...446..155F} {446, 155}

\bibitem[\protect\citeauthoryear{Forbrich et~al.,}{Forbrich
  et~al.}{2007a}]{Forbrich07}
Forbrich J.,  et~al., 2007a, \mn@doi [{Astronomy \& Astrophysics - A\&A}]
  {10.1051/0004-6361:20066158}, 464, 1003

\bibitem[\protect\citeauthoryear{{Forbrich}, {Massi}, {Ros}, {Brunthaler}  \&
  {Menten}}{{Forbrich} et~al.}{2007b}]{Forbrich2007}
{Forbrich} J.,  {Massi} M.,  {Ros} E.,  {Brunthaler} A.,   {Menten} K.~M.,
  2007b, \mn@doi [\aap] {10.1051/0004-6361:20077113}, \href
  {https://ui.adsabs.harvard.edu/abs/2007A&A...469..985F} {469, 985}

\bibitem[\protect\citeauthoryear{{Frank} et~al.,}{{Frank}
  et~al.}{2014}]{Frank2014}
{Frank} A.,  et~al., 2014, in {Beuther} H.,  {Klessen} R.~S.,  {Dullemond}
  C.~P.,   {Henning} T.,  eds, Protostars and Planets VI. pp 451--474
  (\mn@eprint {arXiv} {1402.3553}),
  \mn@doi{10.2458/azu_uapress_9780816531240-ch020}

\bibitem[\protect\citeauthoryear{{Galli}, {Bouy}, {Olivares}, {Miret-Roig},
  {Sarro}, {Barrado}, {Berihuete}  \& {Brandner}}{{Galli}
  et~al.}{2020}]{Galli2020}
{Galli} P.~A.~B.,  {Bouy} H.,  {Olivares} J.,  {Miret-Roig} N.,  {Sarro} L.~M.,
   {Barrado} D.,  {Berihuete} A.,   {Brandner} W.,  2020, \mn@doi [\aap]
  {10.1051/0004-6361/201936708}, \href
  {https://ui.adsabs.harvard.edu/abs/2020A&A...634A..98G} {634, A98}

\bibitem[\protect\citeauthoryear{{Galv{\'a}n-Madrid}
  et~al.,}{{Galv{\'a}n-Madrid} et~al.}{2014}]{Galvan2014}
{Galv{\'a}n-Madrid} R.,  et~al., 2014, \mn@doi [\aap]
  {10.1051/0004-6361/201424630}, \href
  {https://ui.adsabs.harvard.edu/abs/2014A&A...570L...9G} {570, L9}

\bibitem[\protect\citeauthoryear{{Garufi} et~al.,}{{Garufi}
  et~al.}{2025}]{Garuffi2025}
{Garufi} A.,  et~al., 2025, \mn@doi [\aap] {10.1051/0004-6361/202452496}, \href
  {https://ui.adsabs.harvard.edu/abs/2025A&A...694A.290G} {694, A290}

\bibitem[\protect\citeauthoryear{{Ghosh} et~al.,}{{Ghosh}
  et~al.}{2026}]{Ghosh2026}
{Ghosh} A.,  et~al., 2026, \mn@doi [\mnras] {10.1093/mnras/stag154}, \href
  {https://ui.adsabs.harvard.edu/abs/2026MNRAS.tmp..156G} {}

\bibitem[\protect\citeauthoryear{{Groppi}, {Kulesa}, {Walker}  \&
  {Martin}}{{Groppi} et~al.}{2004}]{Groppi2004}
{Groppi} C.~E.,  {Kulesa} C.,  {Walker} C.,   {Martin} C.~L.,  2004, \mn@doi
  [\apj] {10.1086/422168}, \href
  {https://ui.adsabs.harvard.edu/abs/2004ApJ...612..946G} {612, 946}

\bibitem[\protect\citeauthoryear{{G{\"u}del}}{{G{\"u}del}}{2002}]{Gudel2002}
{G{\"u}del} M.,  2002, \mn@doi [\araa]
  {10.1146/annurev.astro.40.060401.093806}, \href
  {https://ui.adsabs.harvard.edu/abs/2002ARA&A..40..217G} {40, 217}

\bibitem[\protect\citeauthoryear{{Hamaguchi}, {Corcoran}, {Petre}, {White},
  {Stelzer}, {Nedachi}, {Kobayashi}  \& {Tokunaga}}{{Hamaguchi}
  et~al.}{2005}]{Hamaguchi2005}
{Hamaguchi} K.,  {Corcoran} M.~F.,  {Petre} R.,  {White} N.~E.,  {Stelzer} B.,
  {Nedachi} K.,  {Kobayashi} N.,   {Tokunaga} A.~T.,  2005, \mn@doi [\apj]
  {10.1086/428434}, \href
  {https://ui.adsabs.harvard.edu/abs/2005ApJ...623..291H} {623, 291}

\bibitem[\protect\citeauthoryear{{Hartmann}, {Herczeg}  \& {Calvet}}{{Hartmann}
  et~al.}{2016}]{Hartmann2016}
{Hartmann} L.,  {Herczeg} G.,   {Calvet} N.,  2016, \mn@doi [\araa]
  {10.1146/annurev-astro-081915-023347}, \href
  {https://ui.adsabs.harvard.edu/abs/2016ARA&A..54..135H} {54, 135}

\bibitem[\protect\citeauthoryear{{Hern{\'a}ndez} et~al.,}{{Hern{\'a}ndez}
  et~al.}{2007}]{Hernandez2007}
{Hern{\'a}ndez} J.,  et~al., 2007, \mn@doi [\apj] {10.1086/522882}, \href
  {https://ui.adsabs.harvard.edu/abs/2007ApJ...671.1784H} {671, 1784}

\bibitem[\protect\citeauthoryear{{Hsieh}, {Arce}, {Maureira}, {Pineda},
  {Segura-Cox}, {Mardones}, {Dunham}  \& {Arun}}{{Hsieh}
  et~al.}{2024}]{Hsieh24}
{Hsieh} C.-H.,  {Arce} H.~G.,  {Maureira} M.~J.,  {Pineda} J.~E.,  {Segura-Cox}
  D.,  {Mardones} D.,  {Dunham} M.~M.,   {Arun} A.,  2024, \mn@doi [\apj]
  {10.3847/1538-4357/ad6152}, \href
  {https://ui.adsabs.harvard.edu/abs/2024ApJ...973..138H} {973, 138}

\bibitem[\protect\citeauthoryear{{Lada}}{{Lada}}{1987}]{Lada1987}
{Lada} C.~J.,  1987, in {Peimbert} M.,  {Jugaku} J.,  eds,  IAU Symposium Vol.
  115, Star Forming Regions. p.~1

\bibitem[\protect\citeauthoryear{{Liu} et~al.,}{{Liu} et~al.}{2014}]{Liu2014}
{Liu} H.~B.,  et~al., 2014, \mn@doi [\apj] {10.1088/0004-637X/780/2/155}, \href
  {https://ui.adsabs.harvard.edu/abs/2014ApJ...780..155L} {780, 155}

\bibitem[\protect\citeauthoryear{{Mac{\'\i}as} et~al.,}{{Mac{\'\i}as}
  et~al.}{2016}]{Macias2016}
{Mac{\'\i}as} E.,  et~al., 2016, \mn@doi [\apj] {10.3847/0004-637X/829/1/1},
  \href {https://ui.adsabs.harvard.edu/abs/2016ApJ...829....1M} {829, 1}

\bibitem[\protect\citeauthoryear{{Manara}, {Ansdell}, {Rosotti}, {Hughes},
  {Armitage}, {Lodato}  \& {Williams}}{{Manara} et~al.}{2023}]{Manara2023}
{Manara} C.~F.,  {Ansdell} M.,  {Rosotti} G.~P.,  {Hughes} A.~M.,  {Armitage}
  P.~J.,  {Lodato} G.,   {Williams} J.~P.,  2023, in {Inutsuka} S.,  {Aikawa}
  Y.,  {Muto} T.,  {Tomida} K.,   {Tamura} M.,  eds,  Astronomical Society of
  the Pacific Conference Series Vol. 534, Protostars and Planets VII. p.~539
  (\mn@eprint {arXiv} {2203.09930}), \mn@doi{10.48550/arXiv.2203.09930}

\bibitem[\protect\citeauthoryear{{Maureira} et~al.,}{{Maureira}
  et~al.}{2026}]{Maureira2025}
{Maureira} M.~J.,  et~al., 2026, \mn@doi [\aap] {10.1051/0004-6361/202556063},
  \href {https://ui.adsabs.harvard.edu/abs/2026A&A...705A..96M} {705, A96}

\bibitem[\protect\citeauthoryear{{Miettinen}, {Kontinen}, {Harju}  \&
  {Higdon}}{{Miettinen} et~al.}{2008}]{Miettinen2008}
{Miettinen} O.,  {Kontinen} S.,  {Harju} J.,   {Higdon} J.~L.,  2008, \mn@doi
  [\aap] {10.1051/0004-6361:200809348}, \href
  {https://ui.adsabs.harvard.edu/abs/2008A&A...486..799M} {486, 799}

\bibitem[\protect\citeauthoryear{{Moscadelli} et~al.,}{{Moscadelli}
  et~al.}{2016}]{Moscadelli2016}
{Moscadelli} L.,  et~al., 2016, \mn@doi [\aap] {10.1051/0004-6361/201526238},
  \href {https://ui.adsabs.harvard.edu/abs/2016A&A...585A..71M} {585, A71}

\bibitem[\protect\citeauthoryear{{Myers} \& {Ladd}}{{Myers} \&
  {Ladd}}{1993}]{Myers1993}
{Myers} P.~C.,  {Ladd} E.~F.,  1993, \mn@doi [\apjl] {10.1086/186956}, \href
  {https://ui.adsabs.harvard.edu/abs/1993ApJ...413L..47M} {413, L47}

\bibitem[\protect\citeauthoryear{{Neuh{\"a}user} \& {Forbrich}}{{Neuh{\"a}user}
  \& {Forbrich}}{2008}]{NeuhauserForbrich2008}
{Neuh{\"a}user} R.,  {Forbrich} J.,  2008, in {Reipurth} B.,  ed., , Vol.~5,
  Handbook of Star Forming Regions, Volume II.
p.~735, \mn@doi{10.48550/arXiv.0808.3374}

\bibitem[\protect\citeauthoryear{{Nisini}, {Antoniucci}, {Giannini}  \&
  {Lorenzetti}}{{Nisini} et~al.}{2005}]{Nisini2005}
{Nisini} B.,  {Antoniucci} S.,  {Giannini} T.,   {Lorenzetti} D.,  2005,
  \mn@doi [\aap] {10.1051/0004-6361:20041409}, \href
  {https://ui.adsabs.harvard.edu/abs/2005A&A...429..543N} {429, 543}

\bibitem[\protect\citeauthoryear{{Ortiz-Le{\'o}n} et~al.,}{{Ortiz-Le{\'o}n}
  et~al.}{2015}]{Ortiz-Leon2015}
{Ortiz-Le{\'o}n} G.~N.,  et~al., 2015, \mn@doi [\apj]
  {10.1088/0004-637X/805/1/9}, \href
  {https://ui.adsabs.harvard.edu/abs/2015ApJ...805....9O} {805, 9}

\bibitem[\protect\citeauthoryear{{Pascucci}, {Gorti}  \&
  {Hollenbach}}{{Pascucci} et~al.}{2012}]{Pascucci2012}
{Pascucci} I.,  {Gorti} U.,   {Hollenbach} D.,  2012, \mn@doi [\apjl]
  {10.1088/2041-8205/751/2/L42}, \href
  {https://ui.adsabs.harvard.edu/abs/2012ApJ...751L..42P} {751, L42}

\bibitem[\protect\citeauthoryear{{Pascucci}, {Ricci}, {Gorti}, {Hollenbach},
  {Hendler}, {Brooks}  \& {Contreras}}{{Pascucci} et~al.}{2014}]{Pasucci2014}
{Pascucci} I.,  {Ricci} L.,  {Gorti} U.,  {Hollenbach} D.,  {Hendler} N.~P.,
  {Brooks} K.~J.,   {Contreras} Y.,  2014, \mn@doi [\apj]
  {10.1088/0004-637X/795/1/1}, \href
  {https://ui.adsabs.harvard.edu/abs/2014ApJ...795....1P} {795, 1}

\bibitem[\protect\citeauthoryear{{Pattle} et~al.,}{{Pattle}
  et~al.}{2025}]{Pattle2025}
{Pattle} K.,  et~al., 2025, \mn@doi [\mnras] {10.1093/mnras/staf009}, \href
  {https://ui.adsabs.harvard.edu/abs/2025MNRAS.537.2127P} {537, 2127}

\bibitem[\protect\citeauthoryear{{Peterson} et~al.,}{{Peterson}
  et~al.}{2011}]{Peterson11}
{Peterson} D.~E.,  et~al., 2011, \mn@doi [\apjs] {10.1088/0067-0049/194/2/43},
  \href {https://ui.adsabs.harvard.edu/abs/2011ApJS..194...43P} {194, 43}

\bibitem[\protect\citeauthoryear{{Phillips}, {Lonsdale}  \&
  {Feigelson}}{{Phillips} et~al.}{1993}]{Phillips1993}
{Phillips} R.~B.,  {Lonsdale} C.~J.,   {Feigelson} E.~D.,  1993, \mn@doi
  [\apjl] {10.1086/186717}, \href
  {https://ui.adsabs.harvard.edu/abs/1993ApJ...403L..43P} {403, L43}

\bibitem[\protect\citeauthoryear{{Ray}, {Muxlow}, {Axon}, {Brown}, {Corcoran},
  {Dyson}  \& {Mundt}}{{Ray} et~al.}{1997}]{Ray1997}
{Ray} T.~P.,  {Muxlow} T.~W.~B.,  {Axon} D.~J.,  {Brown} A.,  {Corcoran} D.,
  {Dyson} J.,   {Mundt} R.,  1997, \mn@doi [\nat] {10.1038/385415a0}, \href
  {https://ui.adsabs.harvard.edu/abs/1997Natur.385..415R} {385, 415}

\bibitem[\protect\citeauthoryear{{Ray}, {Dougados}, {Bacciotti},
  {Eisl{\"o}ffel}  \& {Chrysostomou}}{{Ray} et~al.}{2007}]{Ray2007}
{Ray} T.,  {Dougados} C.,  {Bacciotti} F.,  {Eisl{\"o}ffel} J.,
  {Chrysostomou} A.,  2007, in {Reipurth} B.,  {Jewitt} D.,   {Keil} K.,  eds,
  Protostars and Planets V. p.~231 (\mn@eprint {arXiv} {astro-ph/0605597}),
  \mn@doi{10.48550/arXiv.astro-ph/0605597}

\bibitem[\protect\citeauthoryear{{Rigliaco} et~al.,}{{Rigliaco}
  et~al.}{2019}]{Rigliaco2019}
{Rigliaco} E.,  et~al., 2019, \mn@doi [\aap] {10.1051/0004-6361/201936707},
  \href {https://ui.adsabs.harvard.edu/abs/2019A&A...632A..18R} {632, A18}

\bibitem[\protect\citeauthoryear{{Rodr{\'\i}guez-Kamenetzky},
  {Carrasco-Gonz{\'a}lez}, {Araudo}, {Torrelles}, {Anglada}, {Mart{\'\i}},
  {Rodr{\'\i}guez}  \& {Valotto}}{{Rodr{\'\i}guez-Kamenetzky}
  et~al.}{2016}]{AdrianaRK2016}
{Rodr{\'\i}guez-Kamenetzky} A.,  {Carrasco-Gonz{\'a}lez} C.,  {Araudo} A.,
  {Torrelles} J.~M.,  {Anglada} G.,  {Mart{\'\i}} J.,  {Rodr{\'\i}guez} L.~F.,
   {Valotto} C.,  2016, \mn@doi [\apj] {10.3847/0004-637X/818/1/27}, \href
  {https://ui.adsabs.harvard.edu/abs/2016ApJ...818...27R} {818, 27}

\bibitem[\protect\citeauthoryear{{Rodr{\'\i}guez-Kamenetzky}
  et~al.,}{{Rodr{\'\i}guez-Kamenetzky} et~al.}{2022}]{Rodriguez-Kamenetzky2022}
{Rodr{\'\i}guez-Kamenetzky} A.~R.,  et~al., 2022, \mn@doi [\apjl]
  {10.3847/2041-8213/ac6fd1}, \href
  {https://ui.adsabs.harvard.edu/abs/2022ApJ...931L..26R} {931, L26}

\bibitem[\protect\citeauthoryear{{Rodr{\'\i}guez}, {Zapata}, {Dzib},
  {Ortiz-Le{\'o}n}, {Loinard}, {Mac{\'\i}as}  \& {Anglada}}{{Rodr{\'\i}guez}
  et~al.}{2014}]{Rodriguez2014}
{Rodr{\'\i}guez} L.~F.,  {Zapata} L.~A.,  {Dzib} S.~A.,  {Ortiz-Le{\'o}n}
  G.~N.,  {Loinard} L.,  {Mac{\'\i}as} E.,   {Anglada} G.,  2014, \mn@doi
  [\apjl] {10.1088/2041-8205/793/1/L21}, \href
  {https://ui.adsabs.harvard.edu/abs/2014ApJ...793L..21R} {793, L21}

\bibitem[\protect\citeauthoryear{{Rota}, {Meijerhof}, {van der Marel},
  {Francis}, {van der Tak}  \& {Sellek}}{{Rota} et~al.}{2024}]{Rota24}
{Rota} A.~A.,  {Meijerhof} J.~D.,  {van der Marel} N.,  {Francis} L.,  {van der
  Tak} F.~F.~S.,   {Sellek} A.~D.,  2024, \mn@doi [\aap]
  {10.1051/0004-6361/202348387}, \href
  {https://ui.adsabs.harvard.edu/abs/2024A&A...684A.134R} {684, A134}

\bibitem[\protect\citeauthoryear{{Rota} et~al.,}{{Rota} et~al.}{2025}]{Rota25}
{Rota} A.~A.,  et~al., 2025, \mn@doi [\aap] {10.1051/0004-6361/202554259},
  \href {https://ui.adsabs.harvard.edu/abs/2025A&A...700A..32R} {700, A32}

\bibitem[\protect\citeauthoryear{{Sabatini} et~al.,}{{Sabatini}
  et~al.}{2024}]{Sabatini2024}
{Sabatini} G.,  et~al., 2024, \mn@doi [\aap] {10.1051/0004-6361/202449616},
  \href {https://ui.adsabs.harvard.edu/abs/2024A&A...684L..12S} {684, L12}

\bibitem[\protect\citeauthoryear{{Sandell}, {Reipurth}, {Vacca}  \&
  {Bajaj}}{{Sandell} et~al.}{2021}]{Sandell2021}
{Sandell} G.,  {Reipurth} B.,  {Vacca} W.~D.,   {Bajaj} N.~S.,  2021, \mn@doi
  [\apj] {10.3847/1538-4357/ac133d}, \href
  {https://ui.adsabs.harvard.edu/abs/2021ApJ...920....7S} {920, 7}

\bibitem[\protect\citeauthoryear{{Sicilia-Aguilar}, {Henning}, {Linz},
  {Andr{\'e}}, {Stutz}, {Eiroa}  \& {White}}{{Sicilia-Aguilar}
  et~al.}{2013}]{SiciliaAguilar2013}
{Sicilia-Aguilar} A.,  {Henning} T.,  {Linz} H.,  {Andr{\'e}} P.,  {Stutz} A.,
  {Eiroa} C.,   {White} G.~J.,  2013, \mn@doi [\aap]
  {10.1051/0004-6361/201220170}, \href
  {https://ui.adsabs.harvard.edu/abs/2013A&A...551A..34S} {551, A34}

\bibitem[\protect\citeauthoryear{{Skrutskie} et~al.,}{{Skrutskie}
  et~al.}{2006}]{Skrutskie2006}
{Skrutskie} M.~F.,  et~al., 2006, \mn@doi [\aj] {10.1086/498708}, \href
  {https://ui.adsabs.harvard.edu/abs/2006AJ....131.1163S} {131, 1163}

\bibitem[\protect\citeauthoryear{{Teixeira}, {Lada}, {Marengo}  \&
  {Lada}}{{Teixeira} et~al.}{2012}]{Teixeira12}
{Teixeira} P.~S.,  {Lada} C.~J.,  {Marengo} M.,   {Lada} E.~A.,  2012, \mn@doi
  [\aap] {10.1051/0004-6361/201015326}, \href
  {https://ui.adsabs.harvard.edu/abs/2012A&A...540A..83T} {540, A83}

\bibitem[\protect\citeauthoryear{{Tobin} \& {Sheehan}}{{Tobin} \&
  {Sheehan}}{2024}]{Tobin2024}
{Tobin} J.~J.,  {Sheehan} P.~D.,  2024, \mn@doi [\araa]
  {10.1146/annurev-astro-052920-103752}, \href
  {https://ui.adsabs.harvard.edu/abs/2024ARA&A..62..203T} {62, 203}

\makeatother
\end{thebibliography}

%\begin{thebibliography}{99}
%\end{thebibliography}

%%%%%%%%%%%%%%%%%%%%%%%%%%%%%%%%%%%%%%%%%%%%%%%%%%

%%%%%%%%%%%%%%%%% APPENDICES %%%%%%%%%%%%%%%%%%%%%

\appendix
\section{Figures and Tables}

\begin{figure*}
    \centering
    % Fila 1
    \begin{subfigure}{0.5\textwidth}
        \centering
        \includegraphics[width=\linewidth]{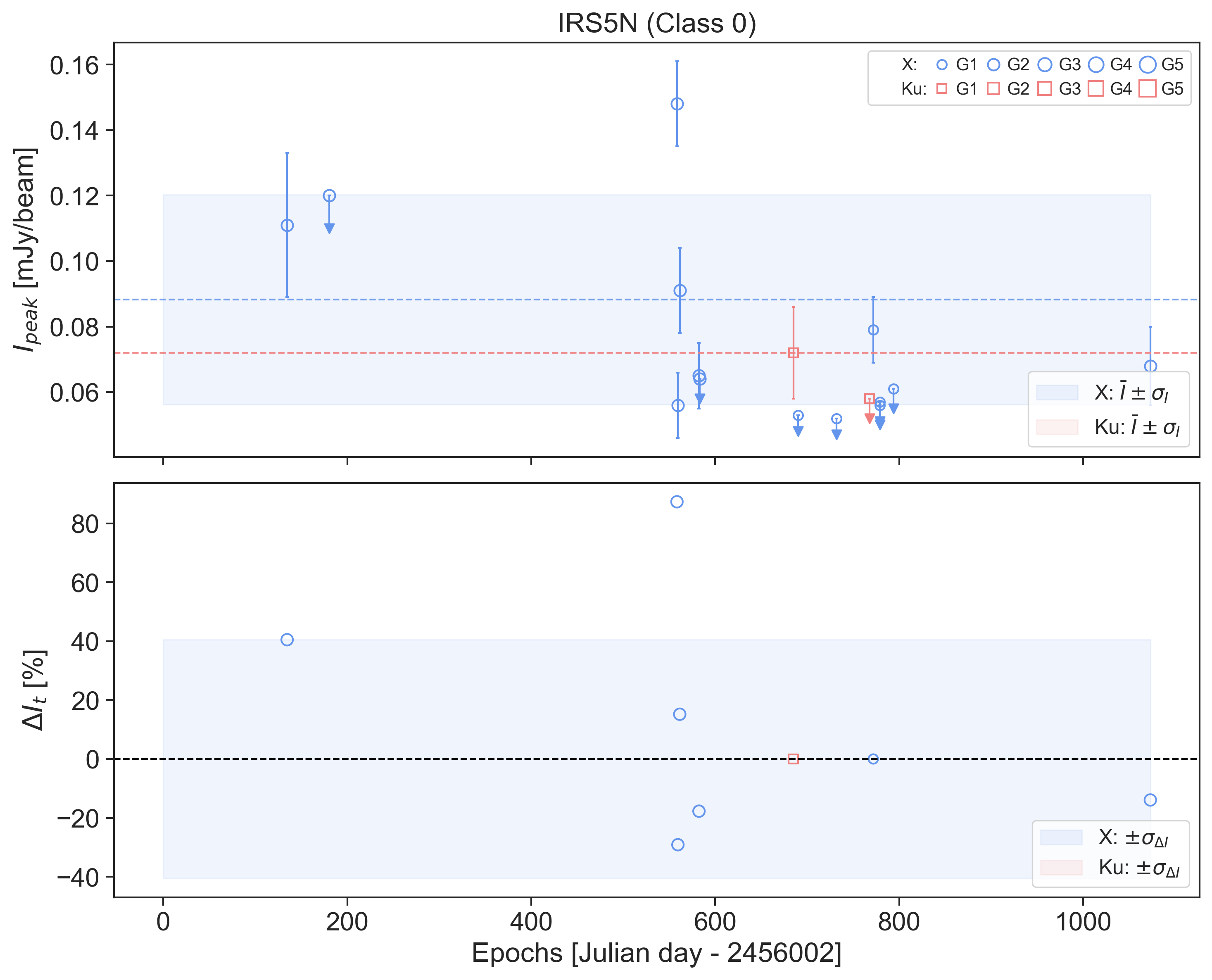}
    \end{subfigure}%
    \begin{subfigure}{0.5\textwidth}
        \centering
        \includegraphics[width=\linewidth]{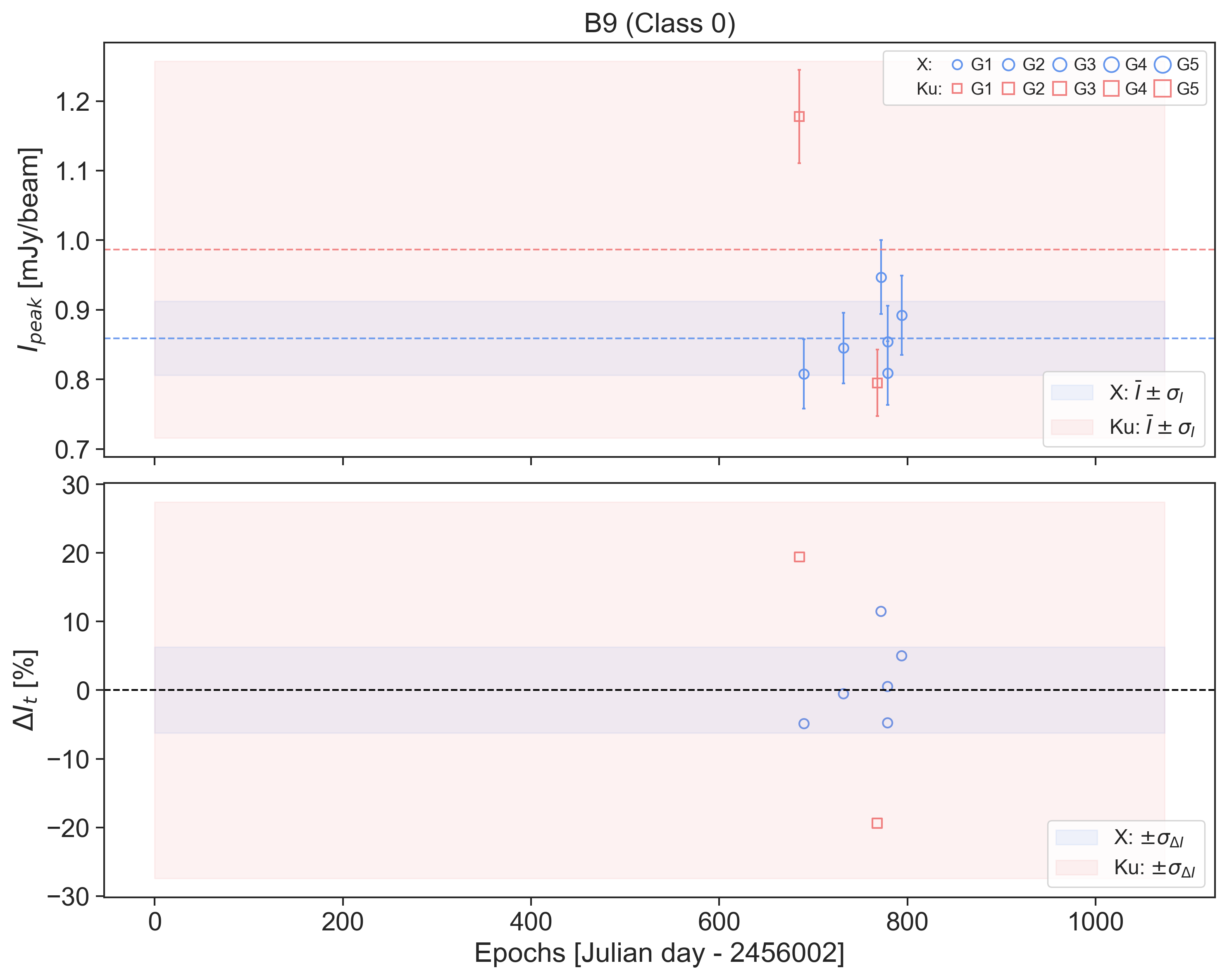}
    \end{subfigure} 
    
    % Fila 2
    \begin{subfigure}{0.5\textwidth}
        \centering
        \includegraphics[width=\linewidth]{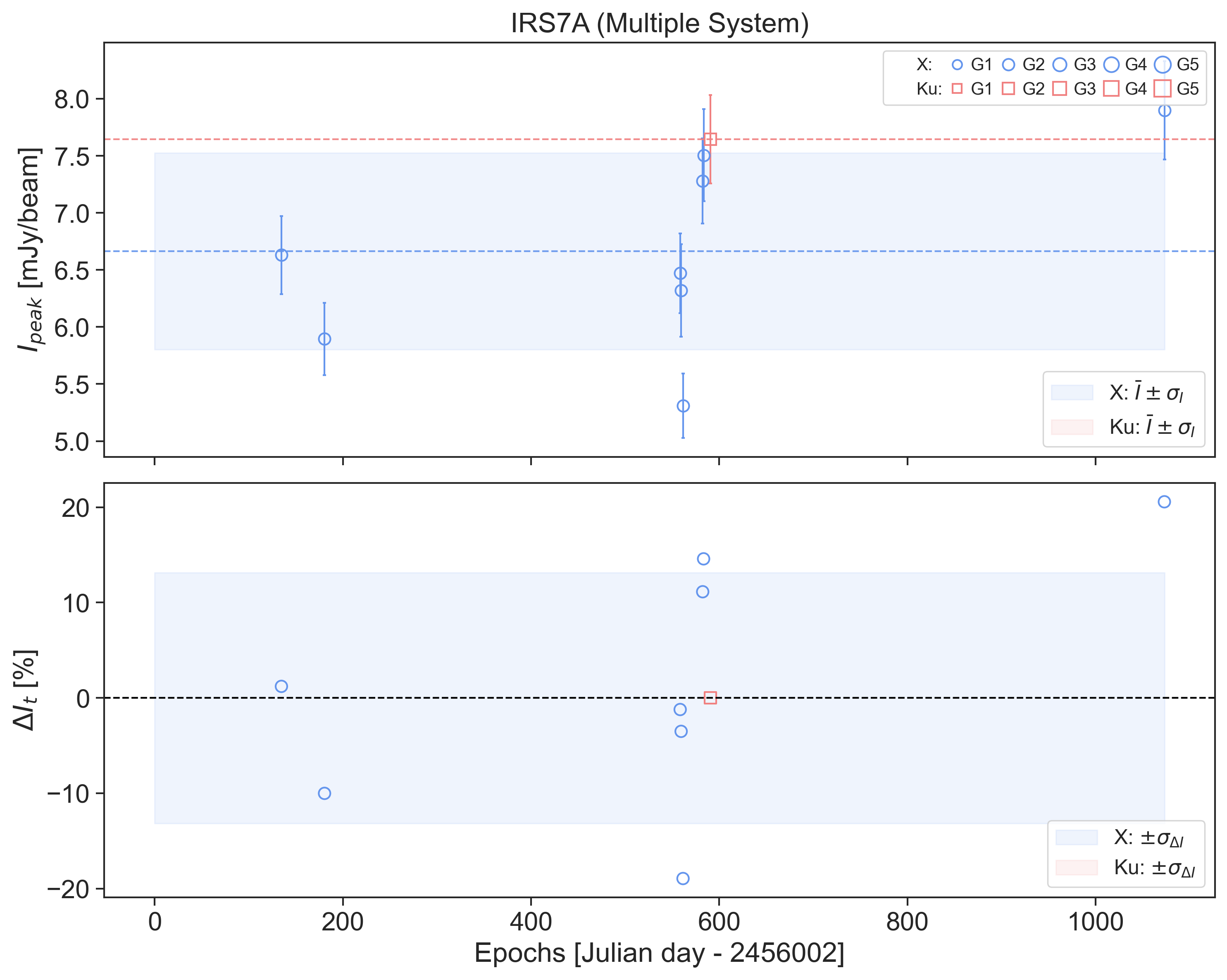}
    \end{subfigure}%
    \begin{subfigure}{0.5\textwidth}
        \centering
        \includegraphics[width=\linewidth]{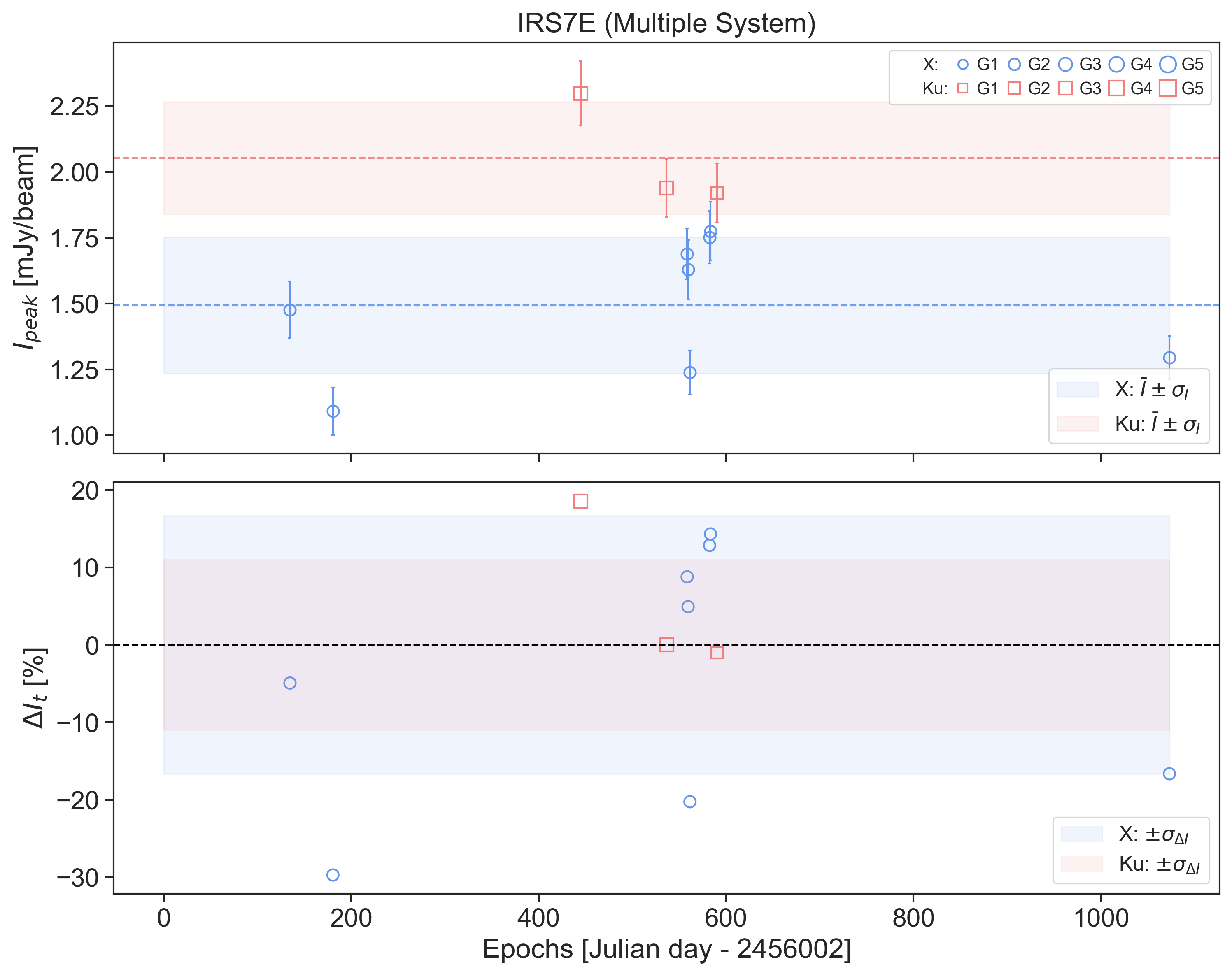}
    \end{subfigure}%
    
    % Fila 3
    \begin{subfigure}{0.5\textwidth}
        \centering
        \includegraphics[width=\linewidth]{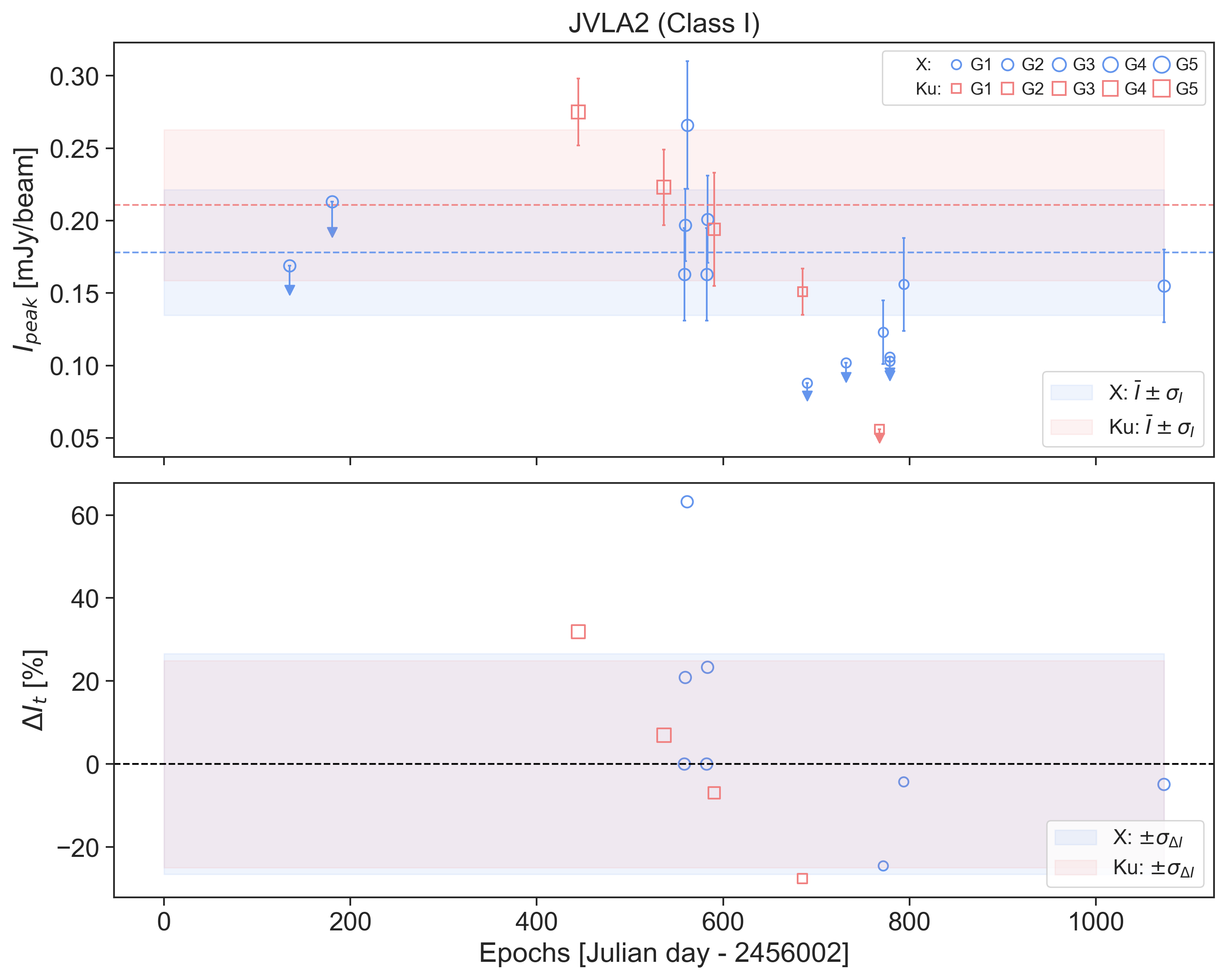}
    \end{subfigure}%
    \begin{subfigure}{0.5\textwidth}
        \centering
        \includegraphics[width=\linewidth]{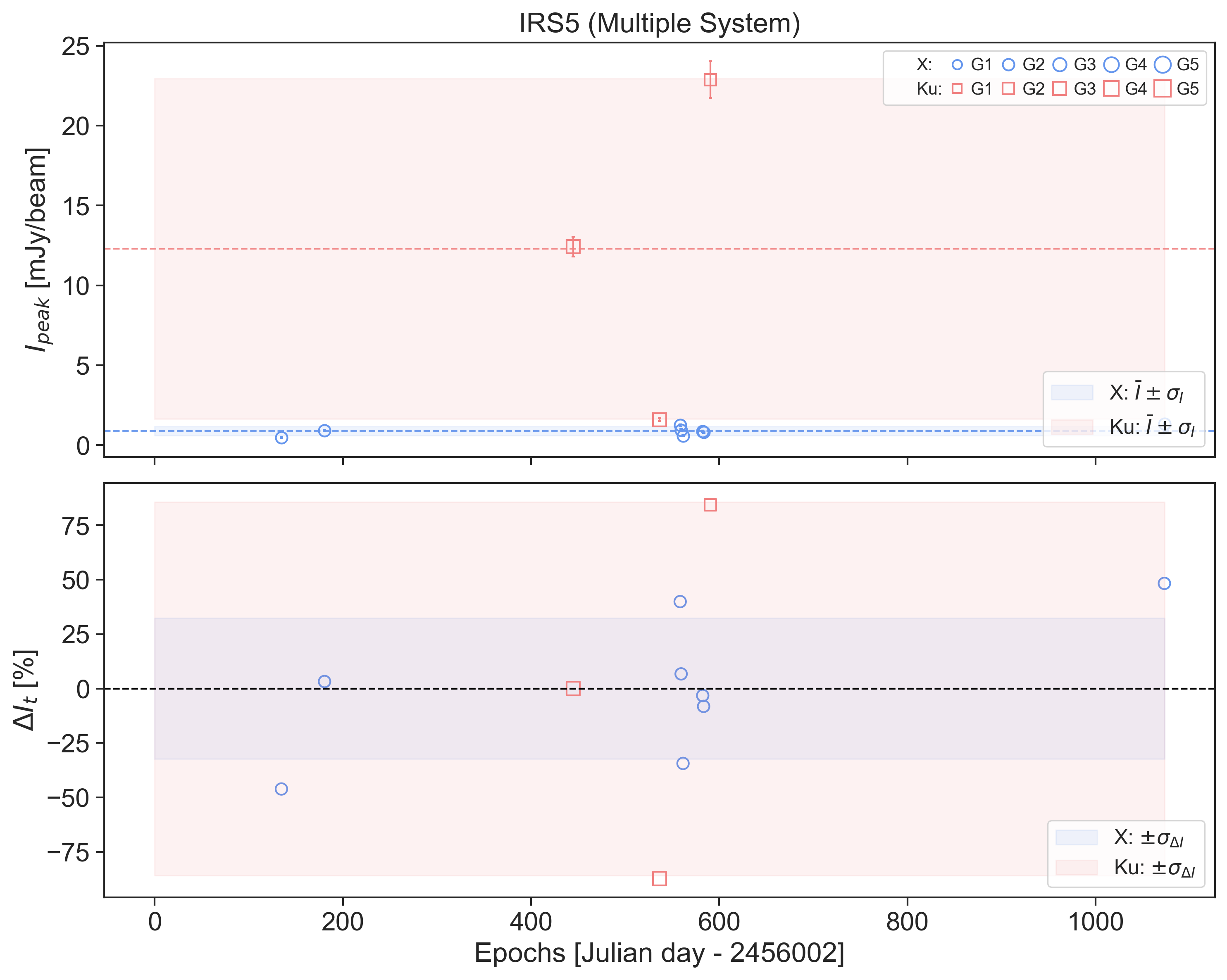}
    \end{subfigure}%
    
    \caption{
    %\textcolor{red}{This should be a single A1 figure that continues over several pages. And I suggest it to have only 1 column and 4 rows (2 YSOs) per page.}
    Light curves of the sources ordered by evolutionary stage. The top panels show the peak intensity at each epoch, while the bottom panels show the percentage variability of the detections. The symbol sizes indicate the five resolution groups; circles correspond to measurements in the \textit{X}-band and squares to those in the \textit{Ku}-band. Dashed lines, in top panels, indicate the mean intensity in each band. In the top panels, the shaded regions span $\pm$ one standard deviation of the peak-intensity distribution, while in the bottom panels they span $\pm$ one standard deviation of the percentage-variability distribution.}
    \label{fig:panel1}
\end{figure*}

\begin{figure*}\ContinuedFloat
    \centering
    % Fila 1
    \begin{subfigure}{0.5\textwidth}
        \centering
        \includegraphics[width=\linewidth]{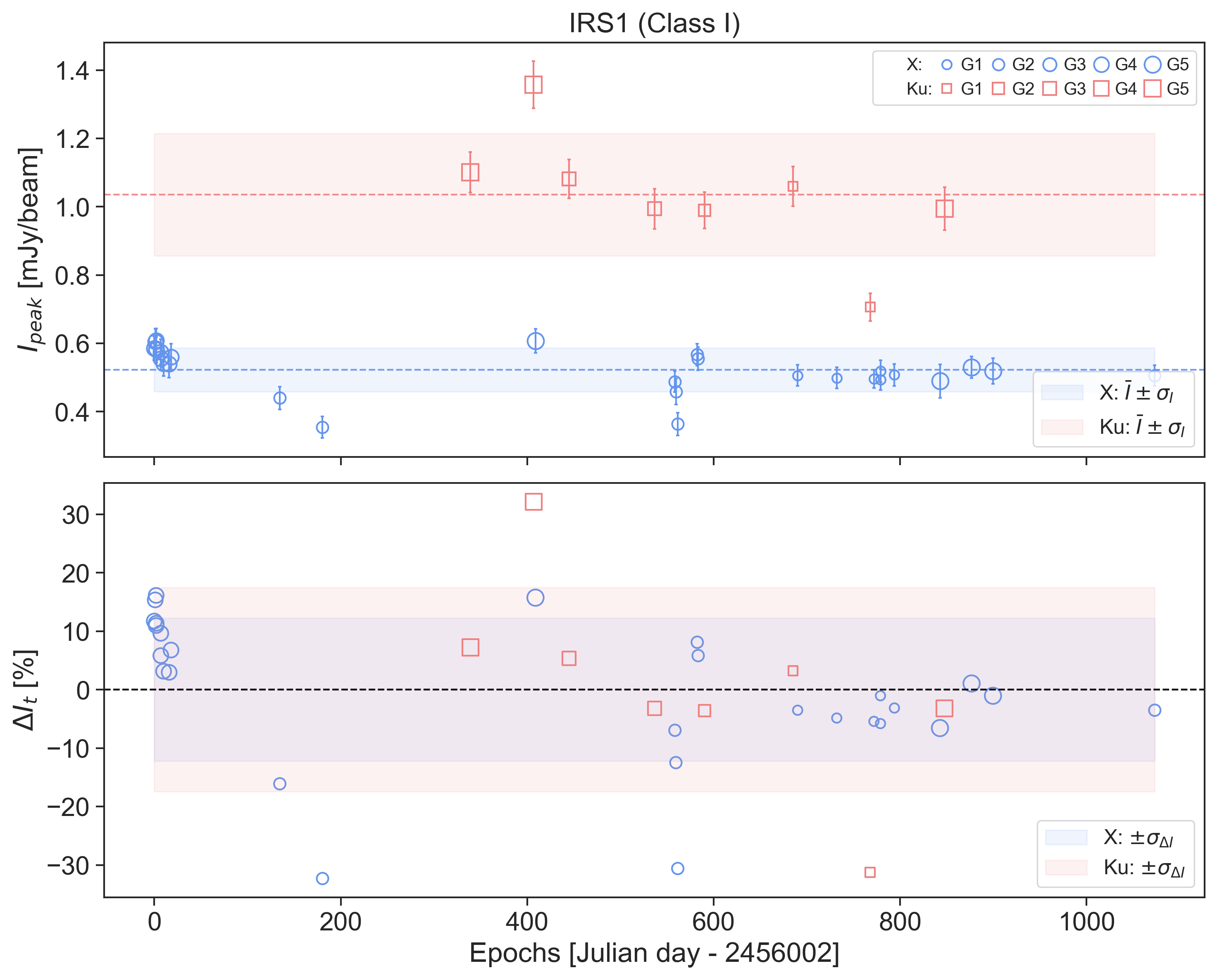}
    \end{subfigure}%
    \begin{subfigure}{0.5\textwidth}
        \centering
        \includegraphics[width=\linewidth]{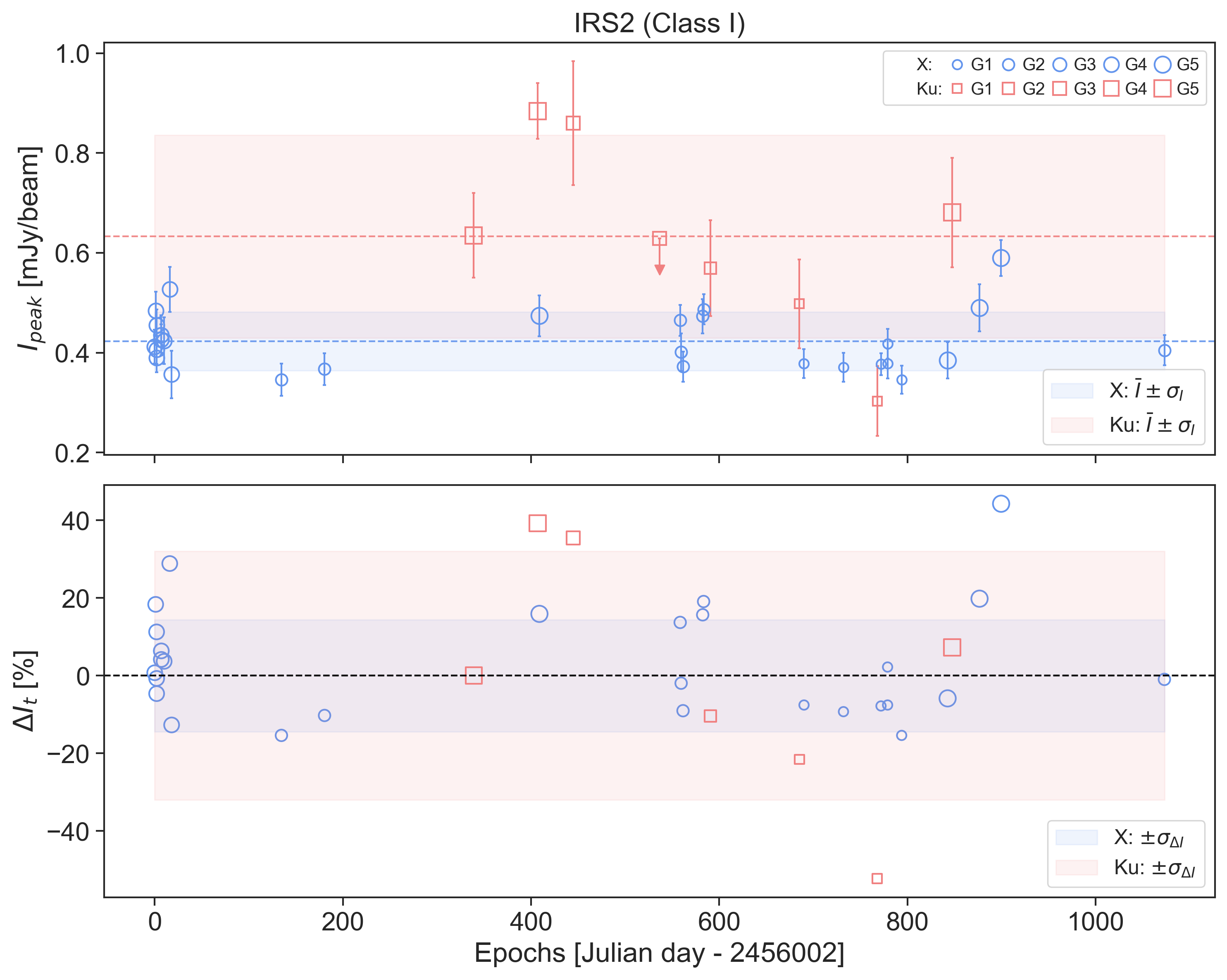}
    \end{subfigure} 
    
    % Fila 2
    \begin{subfigure}{0.5\textwidth}
        \centering
        \includegraphics[width=\linewidth]{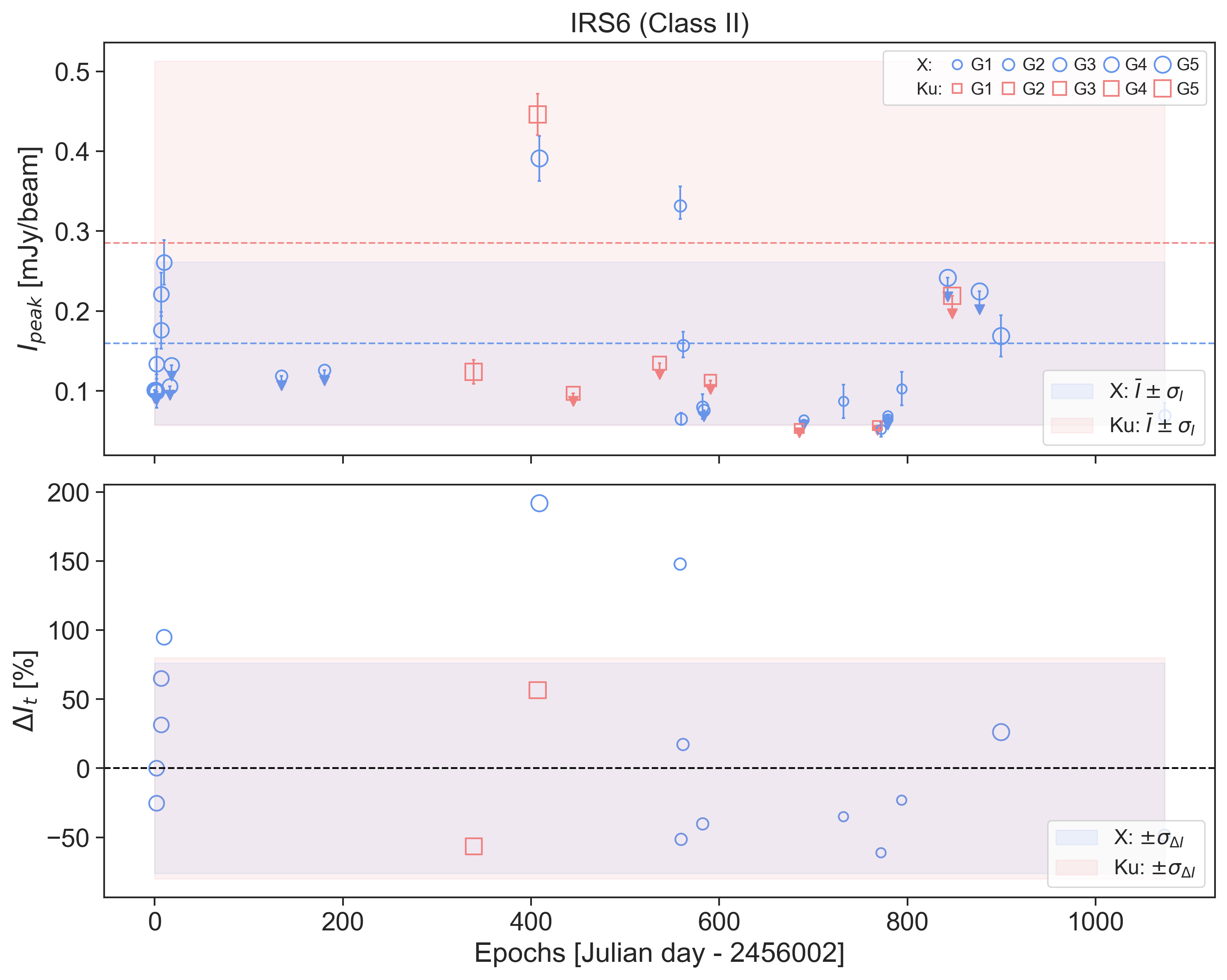}
    \end{subfigure}%
    \begin{subfigure}{0.5\textwidth}
        \centering
        \includegraphics[width=\linewidth]{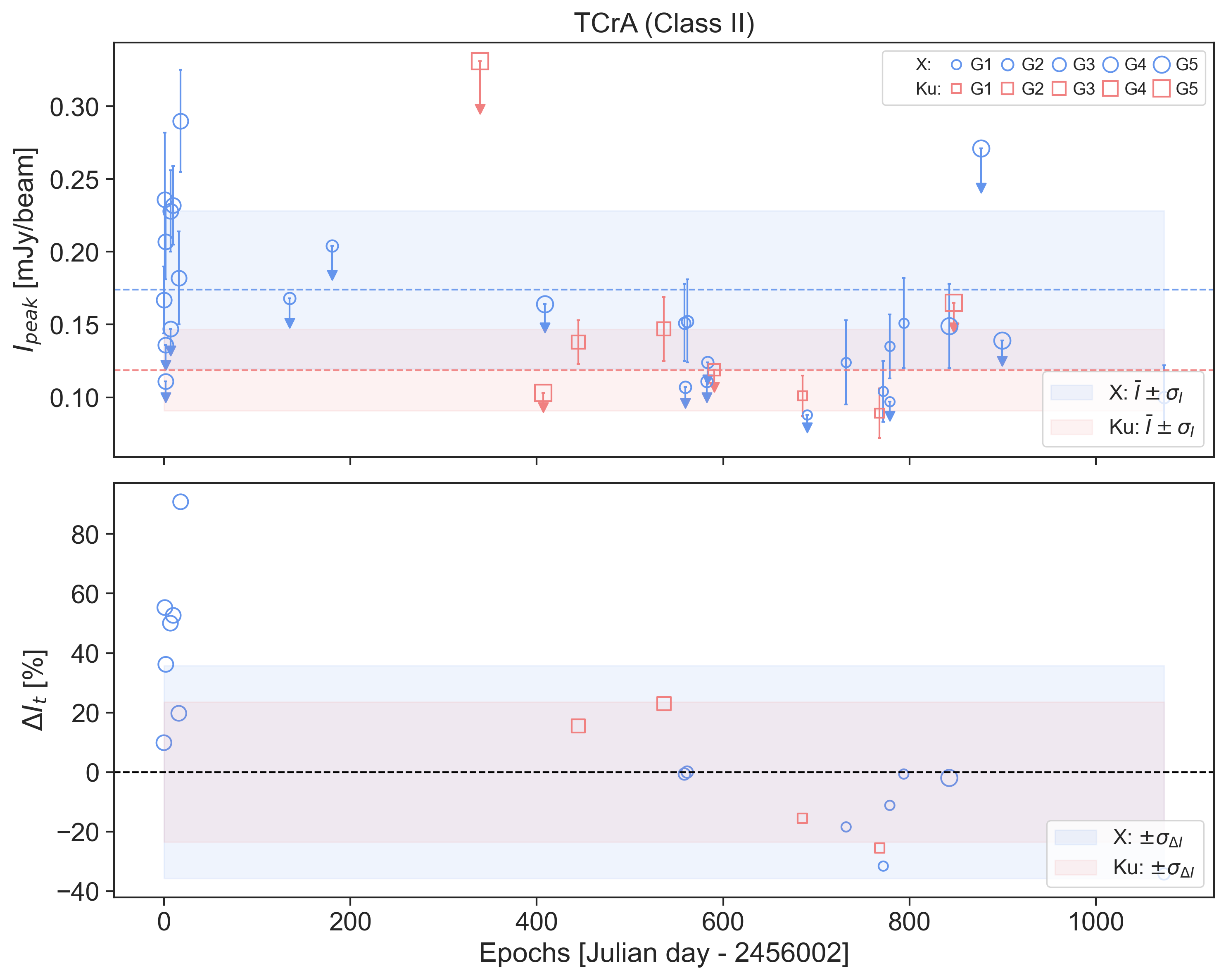}
    \end{subfigure}%
    
    % Fila 3
    \begin{subfigure}{0.5\textwidth}
        \centering
        \includegraphics[width=\linewidth]{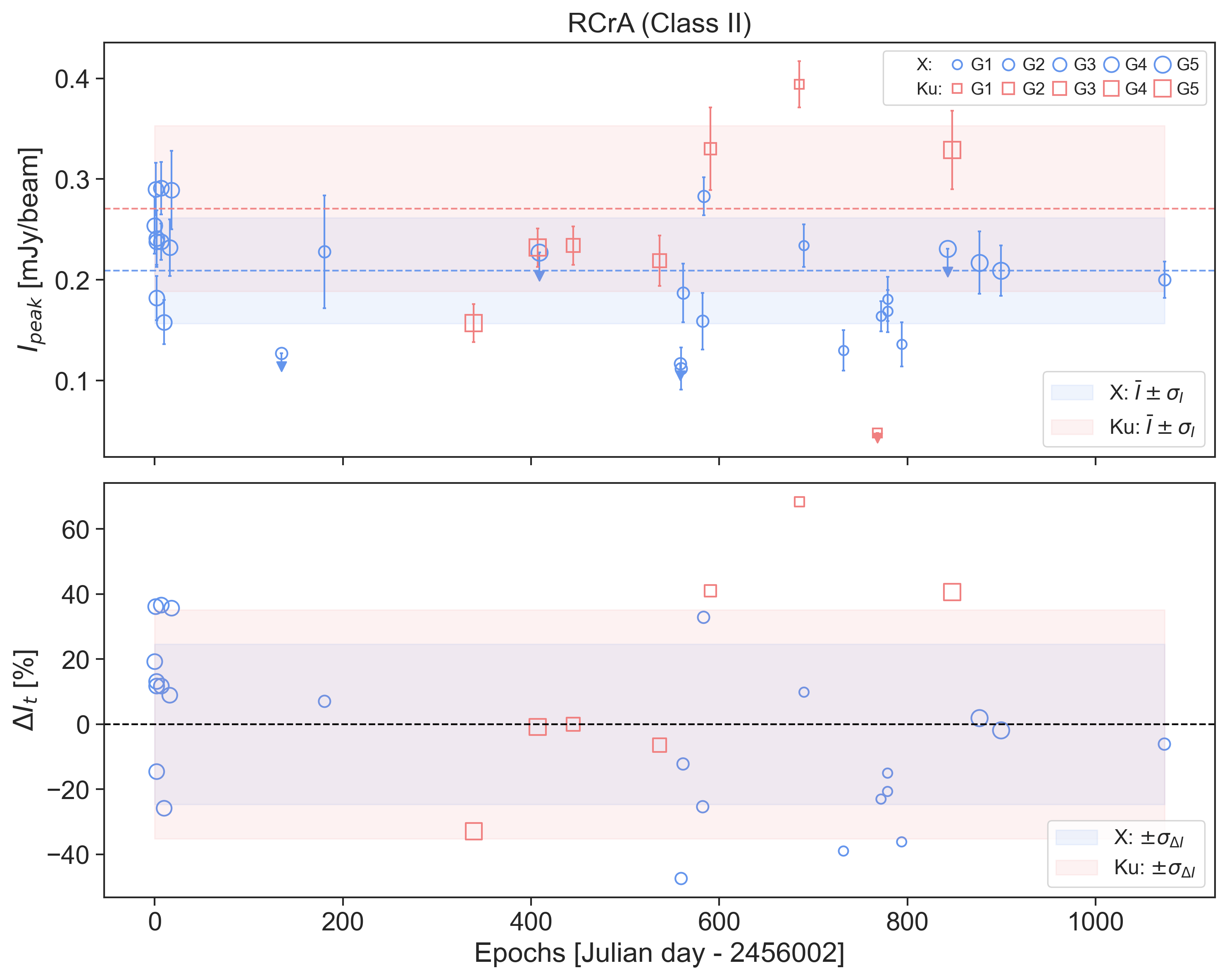}
    \end{subfigure}%
    \begin{subfigure}{0.5\textwidth}
        \centering
        \includegraphics[width=\linewidth]{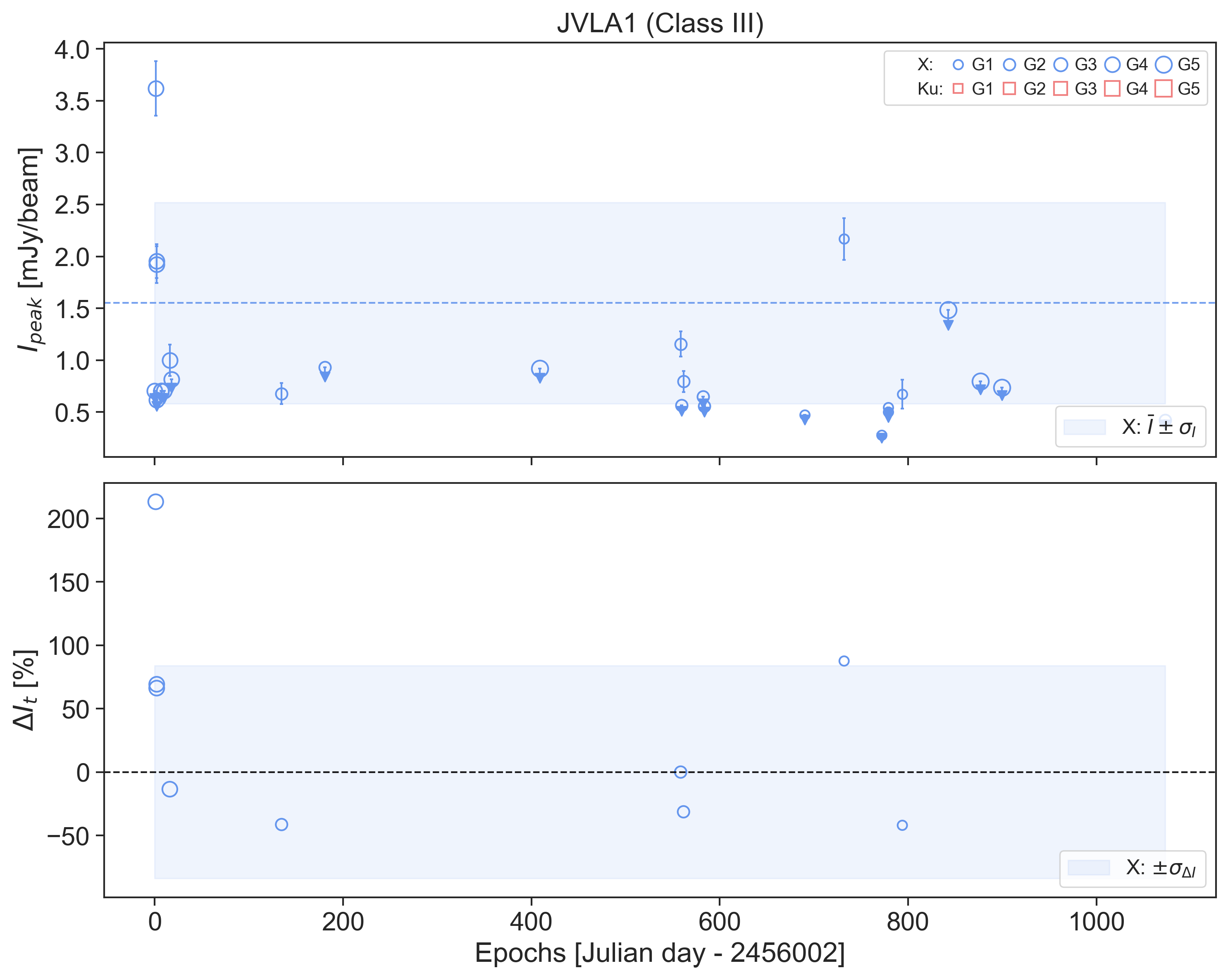}
    \end{subfigure}%
    
    \caption{Light curves of the sources ordered by evolutionary stage (continued).}
    \label{fig:panel2}
\end{figure*}

\begin{figure*}\ContinuedFloat
    \centering
    % Fila 1
    \begin{subfigure}{0.5\textwidth}
        \centering
        \includegraphics[width=\linewidth]{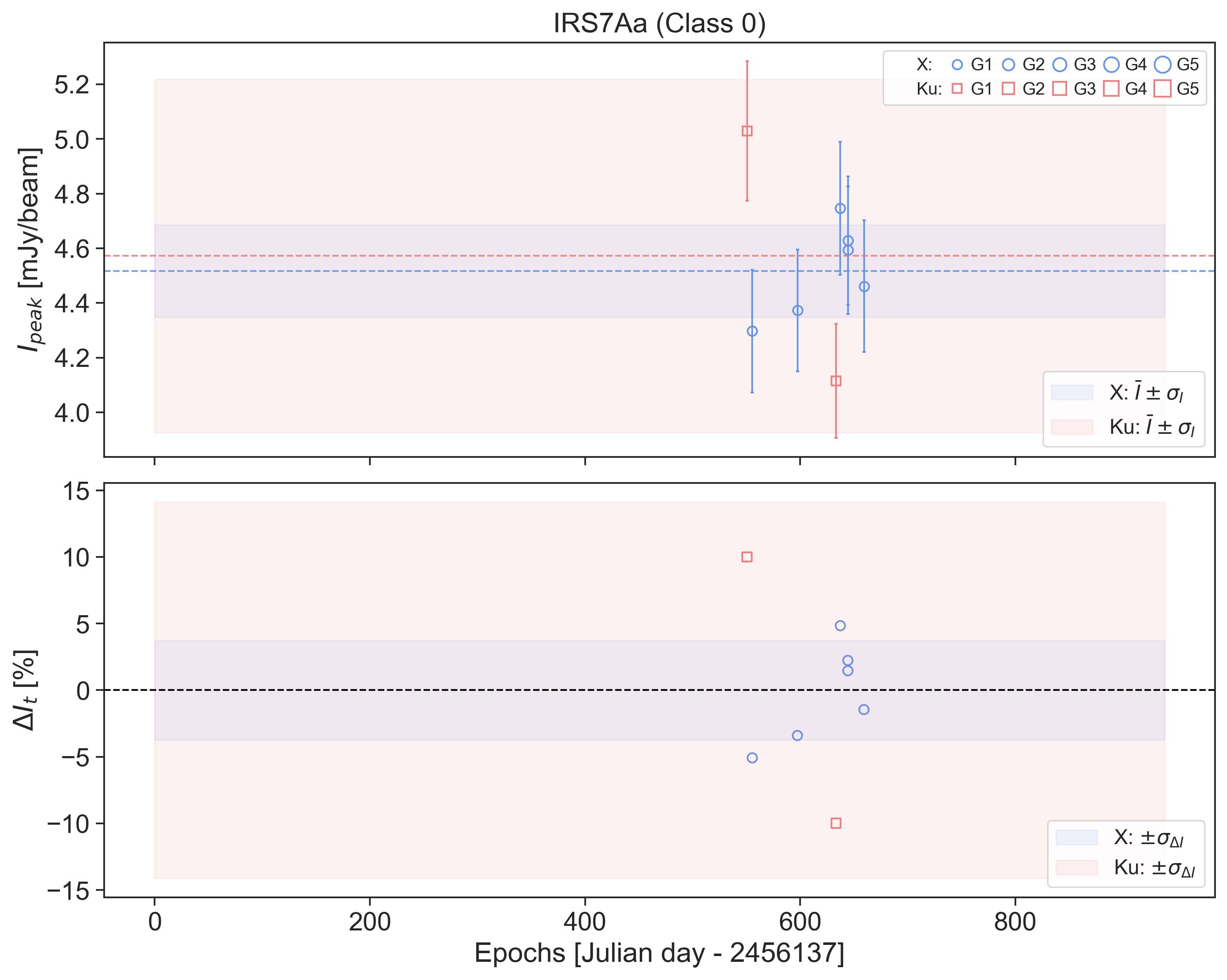}
    \end{subfigure}%
    \begin{subfigure}{0.5\textwidth}
        \centering
        \includegraphics[width=\linewidth]{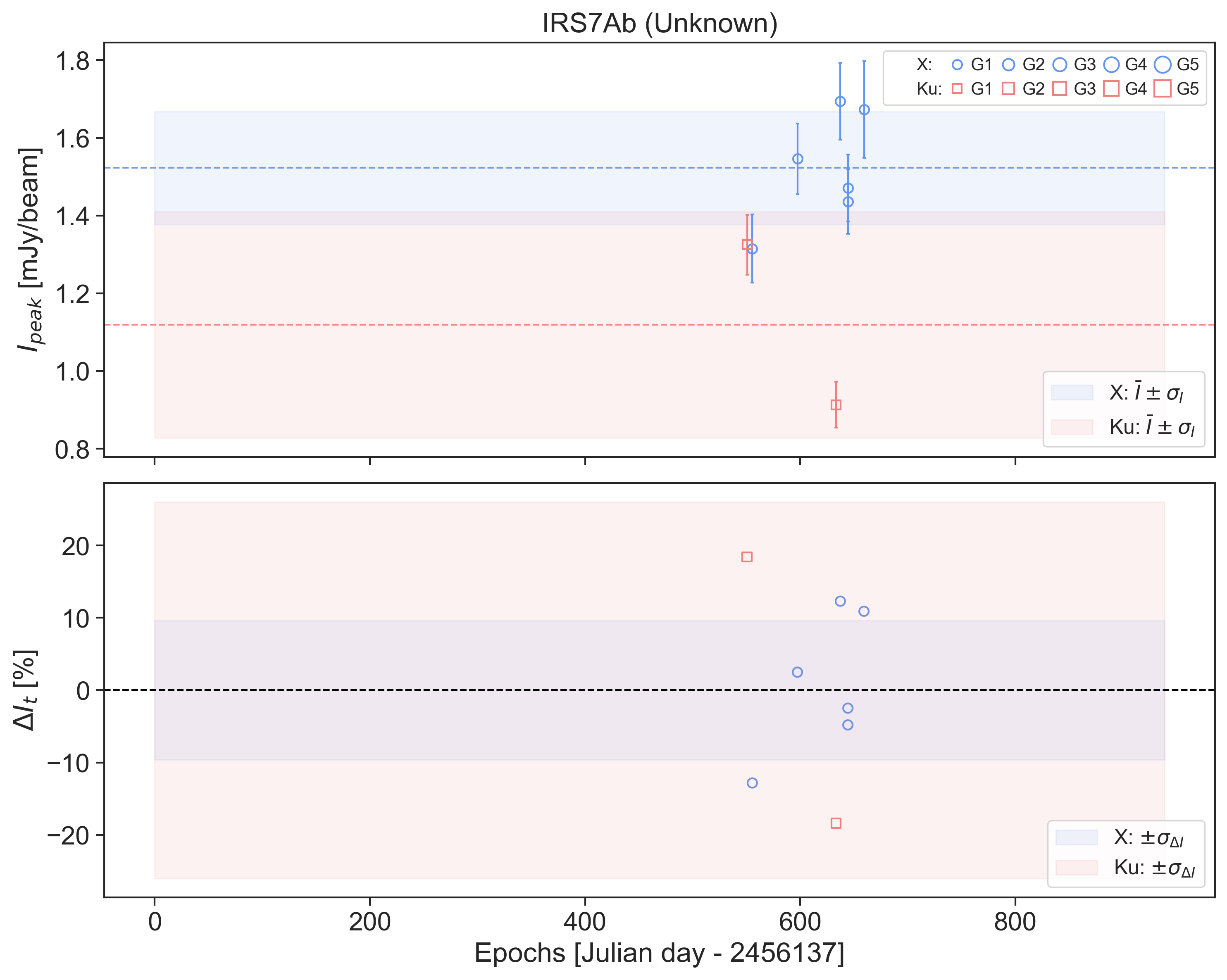}
    \end{subfigure} 
    
    % Fila 2
    \begin{subfigure}{0.5\textwidth}
        \centering
        \includegraphics[width=\linewidth]{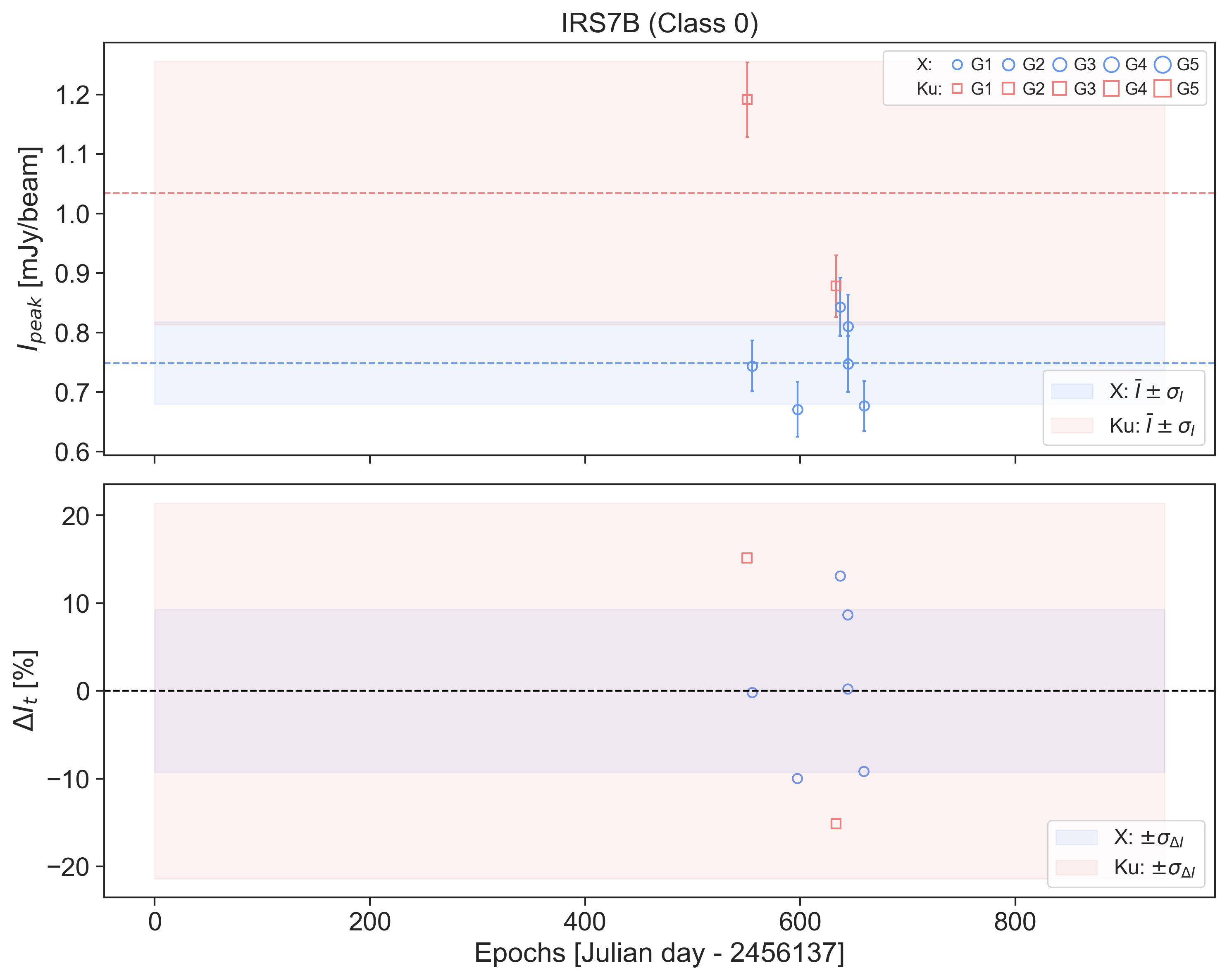}
    \end{subfigure}%
    \begin{subfigure}{0.5\textwidth}
        \centering
        \includegraphics[width=\linewidth]{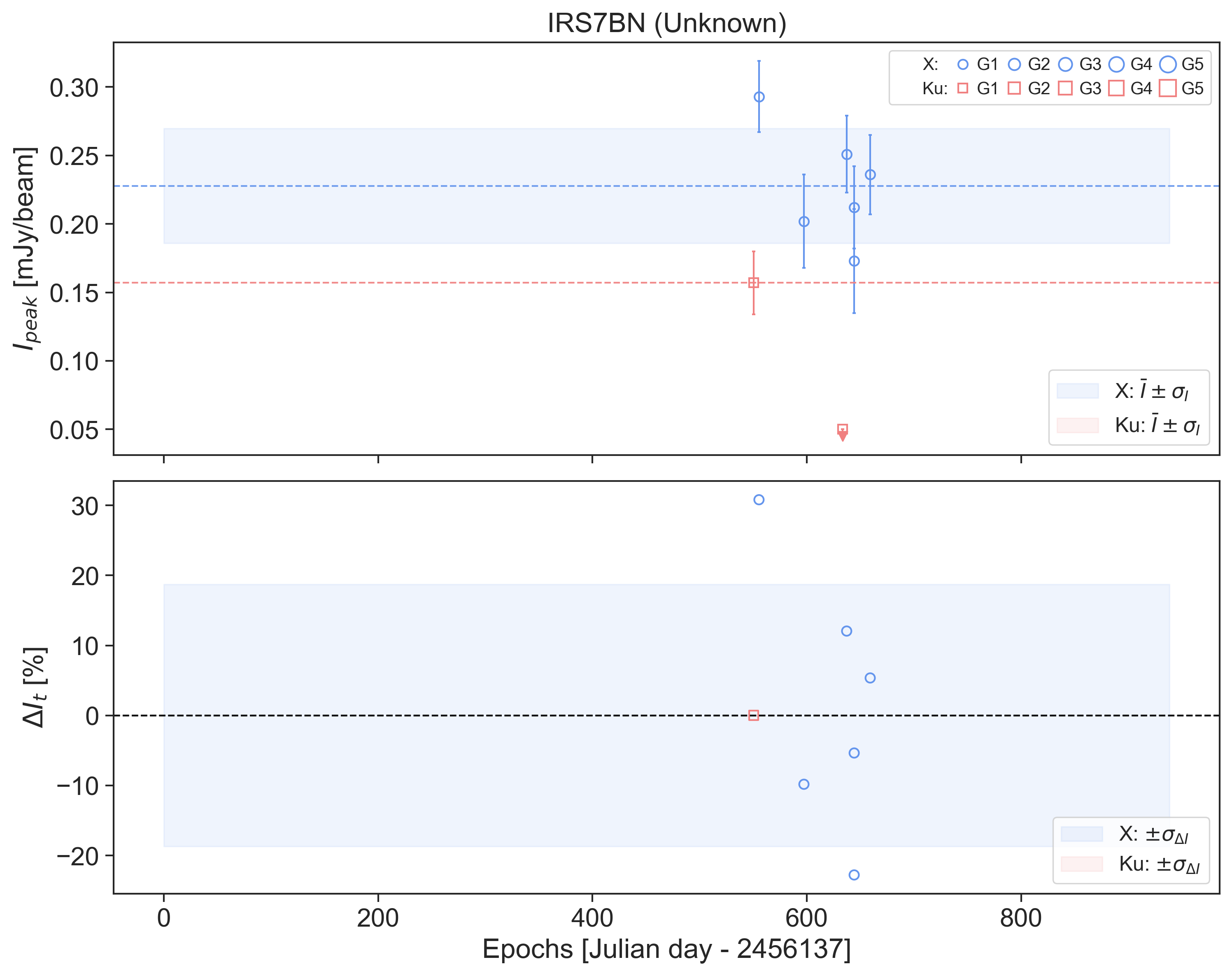}
    \end{subfigure}%
    
    % Fila 3
    \begin{subfigure}{0.5\textwidth}
        \centering
        \includegraphics[width=\linewidth]{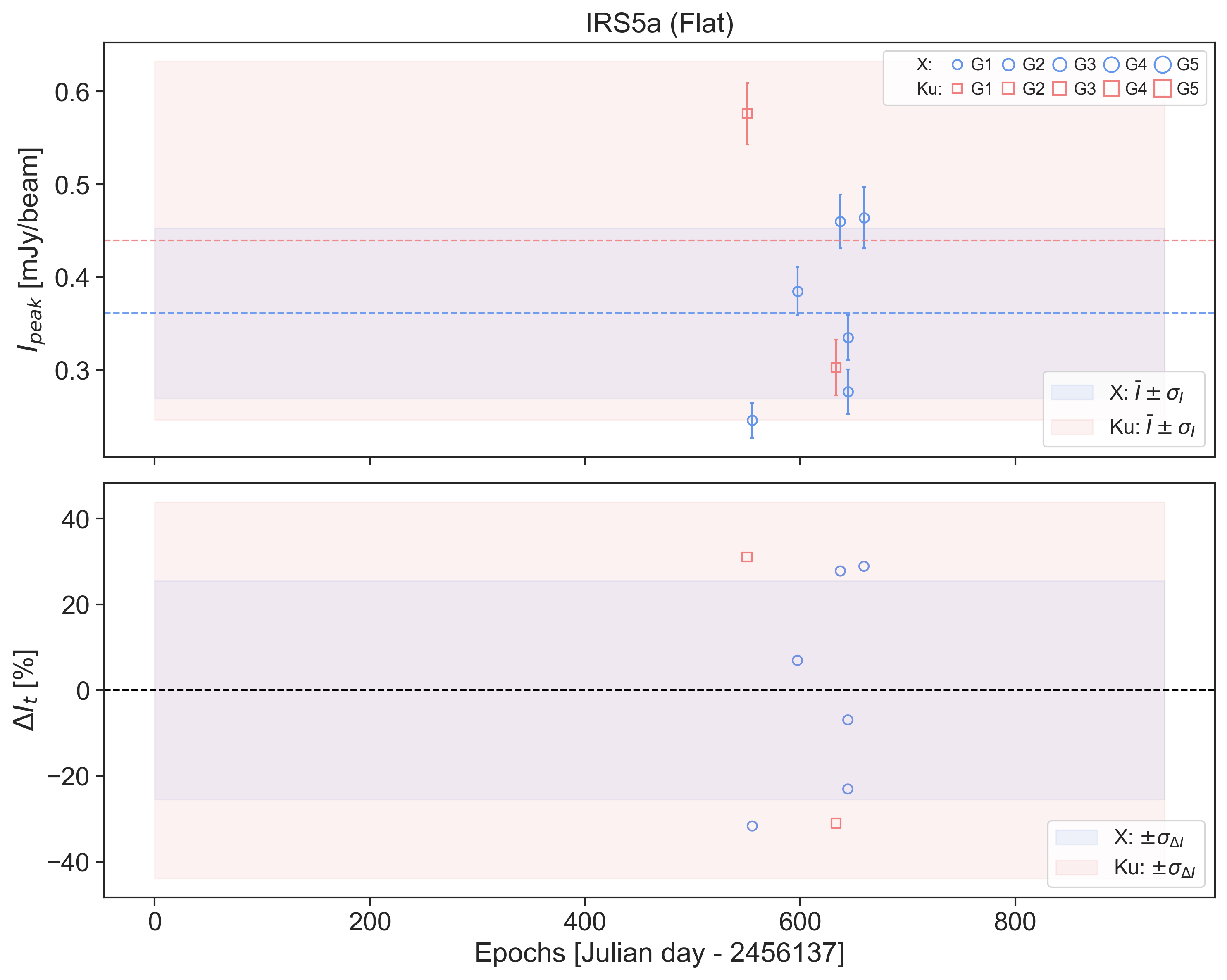}
    \end{subfigure}%
    \begin{subfigure}{0.5\textwidth}
        \centering
        \includegraphics[width=\linewidth]{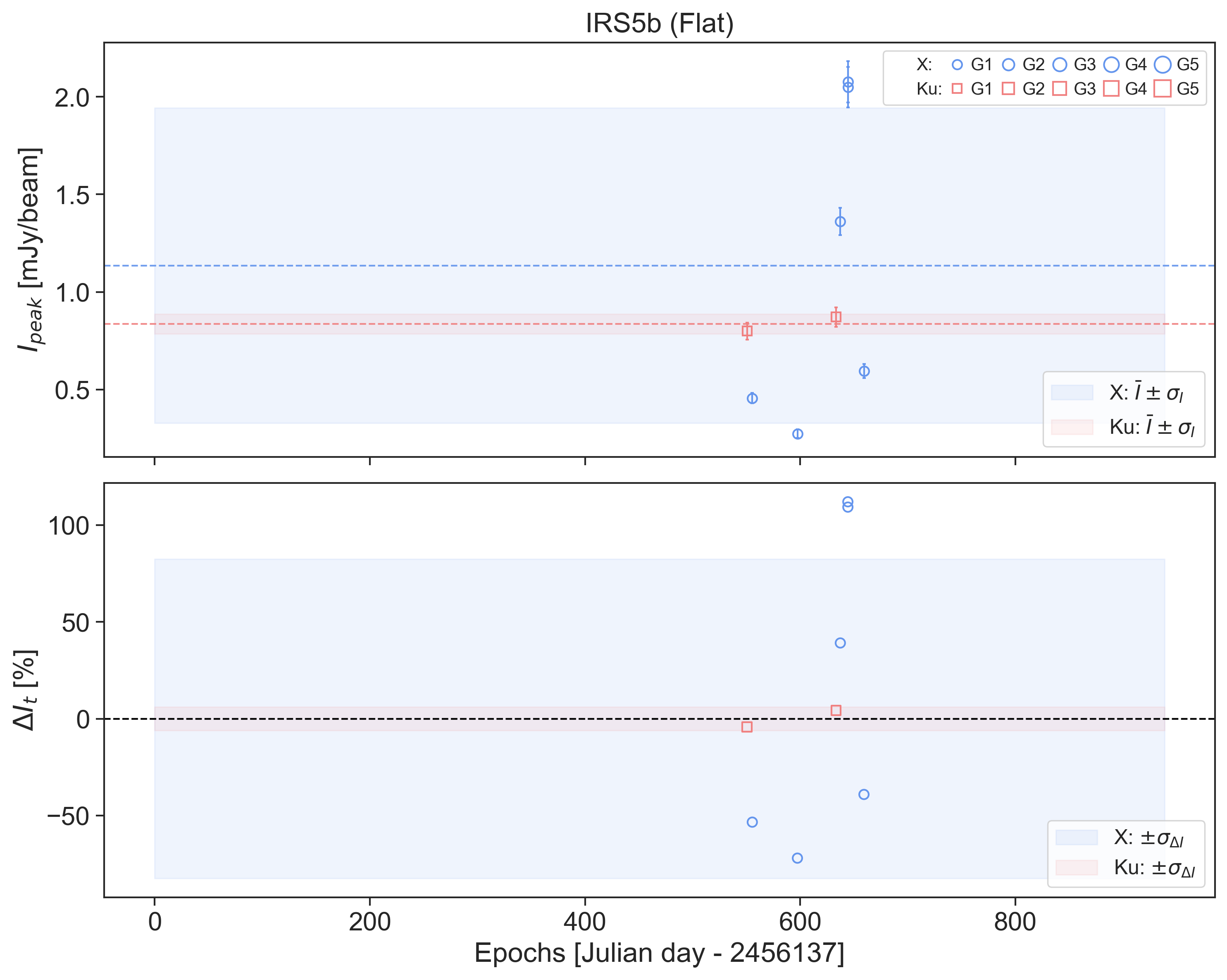}
    \end{subfigure}%
    
    \caption{Light curves of the sources ordered by evolutionary stage (continued). These sources were only detected in the best resolution group G1.}
    \label{fig:panel3}
\end{figure*}

%\section{Tables}


\begin{landscape}
 \begin{table}
  \caption{JVLA Epoch Observations.}
  \label{tab:obs}
  \begin{tabular}{cccccccccc}
    \hline
    Epoch & Day & Time & Frequency & Array & Robust & Resolution & Synthesized Beam & rms noise & Flux/Phase/BP  \\
    ID & & UTC & (GHz) & Configuration & Weighting & Group & (arcsec$\times$arcsec; deg) & ($\mu$Jy/beam) & calibrators  \\
    \hline
    1 & 0 & 2012-03-15 14:21:04 & 9.0 & C & 0.5 & 4 & $9.76\times 3.56; 7.0^\circ$ & 25.9 & 3C286/J1924-2914/J2355+4950 \\
2 & 1 & 2012-03-16 14:27:40 & 9.0 & C & 0.5 & 4 & $9.76\times 3.56; 7.0^\circ$ & 24.6 & 3C286/J1924-2914/J2355+4950 \\
3 & 2 & 2012-03-17 14:13:52 & 9.0 & C & 0.5 & 4 & $9.76\times 3.56; 7.0^\circ$ & 21.1 & 3C286/J1924-2914/J2355+4950 \\
4 & 2 & 2012-03-17 15:09:19 & 9.0 & C & 0.5 & 4 & $9.76\times 3.56; 7.0^\circ$ & 21.2 & 3C286/J1924-2914/J2355+4950 \\
5 & 2 & 2012-03-17 15:39:16 & 9.0 & C & 0.5 & 4 & $9.76\times 3.56; 7.0^\circ$ & 23.6 & 3C286/J1924-2914/J2355+4950 \\
7 & 7 & 2012-03-22 13:54:13 & 9.0 & C & 0.5 & 4 & $9.76\times 3.56; 7.0^\circ$ & 23.9 & 3C286/J1924-2914/J2355+4950 \\
6 & 7 & 2012-03-22 14:27:19 & 9.0 & C & 0.5 & 4 & $9.76\times 3.56; 7.0^\circ$ & 25.1 & 3C286/J1924-2914/J2355+4950 \\
8 & 10 & 2012-03-25 14:15:25 & 9.0 & C & 0.5 & 4 & $9.76\times 3.56; 7.0^\circ$ & 26.2 & 3C286/J1924-2914/J2355+4950 \\
9 & 16 & 2012-03-31 13:49:51 & 9.0 & C & 0.5 & 4 & $9.76\times 3.56; 7.0^\circ$ & 25.4 & 3C286/J1924-2914/J2355+4950 \\
10 & 18 & 2012-04-02 13:06:28 & 9.0 & C & 0.5 & 4 & $9.76\times 3.56; 7.0^\circ$ & 32.5 & 3C48/J1924-2914/J2355+4950 \\
11 & 135 & 2012-07-28 05:56:25 & 9.0 & B & 0.5 & 2 & $2.48\times 2.37; 16.6^\circ$ & 30.2 & 3C48/J1924-2914/J2355+4950 \\
12 & 180 & 2012-09-12 01:45:08 & 9.0 & BnA & 0.5 & 2 & $2.48\times 2.37; 16.6^\circ$ & 31.9 & 3C286/J1924-2914/J2355+4950 \\
13 & 339 & 2013-02-17 16:14:13 & 14.0 & D & 0.0 & 5 & $22.91\times 6.82; -3.2^\circ$ & 49.9 & 3C286/J1924-2914/J2355+4950 \\
14 & 355 & 2013-03-05 13:32:06 & 9.0 & D & 0.5 & -- & $27.31\times 5.44; -18.7^\circ$ & 25.2 & 3C286/J1924-2914/J2355+4950 \\
15 & 407 & 2013-04-26 12:34:48 & 14.0 & D & 0.0 & 5 & $22.91\times 6.82; -3.2^\circ$ & 26.4 & 3C286/J1924-2914/J2355+4950 \\
16 & 409 & 2013-04-28 11:14:39 & 9.0 & D & 0.5 & 5 & $22.91\times 6.82; -3.2^\circ$ & 35.7 & 3C286/J1924-2914/J2355+4950 \\
17 & 445 & 2013-06-03 09:42:25 & 14.0 & DnC & 0.0 & 3 & $4.63\times 1.11; 8.0^\circ$ & 23.6 & 3C286/J1924-2914/J2355+4950 \\
18 & 537 & 2013-09-03 03:55:35 & 14.0 & C & 0.0 & 3 & $4.63\times 1.11; 8.0^\circ$ & 34.4 & 3C286/J1924-2914/J2355+4950 \\
19 & 557 & 2013-09-24 01:13:50 & 9.0 & CnB & 0.5 & 2 & $2.48\times 2.37; 16.6^\circ$ & 29.2 & 3C286/J1924-2914/J2355+4950 \\
20 & 557 & 2013-09-24 01:43:46 & 9.0 & CnB & 0.5 & 2 & $2.48\times 2.37; 16.6^\circ$ & 16.1 & 3C286/J1924-2914/J2355+4950 \\
21 & 558 & 2013-09-25 01:24:46 & 9.0 & CnB & 0.5 & 2 & $2.48\times 2.37; 16.6^\circ$ & 26.3 & 3C286/J1924-2914/J2355+4950 \\
22 & 559 & 2013-09-26 01:34:55 & 9.0 & B & 0.5 & 2 & $2.48\times 2.37; 16.6^\circ$ & 20.1 & 3C286/J1924-2914/J2355+4950 \\
23 & 561 & 2013-09-28 01:58:11 & 9.0 & B & 0.5 & 2 & $2.48\times 2.37; 16.6^\circ$ & 22.5 & 3C286/J1924-2914/J2355+4950 \\
24 & 582 & 2013-10-18 23:35:49 & 9.0 & B & 0.5 & 2 & $2.48\times 2.37; 16.6^\circ$ & 21.2 & 3C286/J1924-2914/J2355+4950 \\
25 & 583 & 2013-10-19 22:47:42 & 9.0 & B & 0.5 & 2 & $2.48\times 2.37; 16.6^\circ$ & 19.9 & 3C286/J1924-2914/J2355+4950 \\
26 & 590 & 2013-10-27 00:53:14 & 14.0 & B & 0.0 & 2 & $2.48\times 2.37; 16.6^\circ$ & 25.9 & 3C286/J1924-2914/J2355+4950 \\
27 & 685 & 2014-01-29 17:34:58 & 14.0 & BnA & 0.0 & 1 & $0.92\times 0.66; 145.1^\circ$ & 13.8 & 3C286/J1924-2914/J2355+4950 \\
28 & 690 & 2014-02-03 16:01:02 & 9.0 & BnA & 0.5 & 1 & $0.92\times 0.66; 145.1^\circ$ & 15.9 & 3C286/J1924-2914/J2355+4950 \\
29 & 732 & 2014-03-17 14:32:24 & 9.0 & A & 0.5 & 1 & $0.92\times 0.66; 145.1^\circ$ & 16.9 & 3C286/J1924-2914/J2355+4950 \\
30 & 768 & 2014-04-22 12:38:47 & 14.0 & A & 0.0 & 1 & $0.92\times 0.66; 145.1^\circ$ & 13.6 & 3C286/J1924-2914/J2355+4950 \\
31 & 772 & 2014-04-26 11:54:18 & 9.0 & A & 0.5 & 1 & $0.92\times 0.66; 145.1^\circ$ & 9.5 & 3C286/J1924-2914/J2355+4950 \\
32 & 779 & 2014-05-03 11:40:53 & 9.0 & A & 0.5 & 1 & $0.92\times 0.66; 145.1^\circ$ & 17.8 & 3C286/J1924-2914/J2355+4950 \\
33 & 779 & 2014-05-03 12:10:50 & 9.0 & A & 0.5 & 1 & $0.92\times 0.66; 145.1^\circ$ & 17.2 & 3C286/J1924-2914/J2355+4950 \\
34 & 794 & 2014-05-18 10:56:52 & 9.0 & A & 0.5 & 1 & $0.92\times 0.66; 145.1^\circ$ & 18.8 & 3C286/J1924-2914/J2355+4950 \\
35 & 843 & 2014-07-06 07:04:16 & 9.0 & D & 0.5 & 5 & $22.91\times 6.82; -3.2^\circ$ & 48.0 & 3C286/J1924-2914/J2355+4950 \\
36 & 848 & 2014-07-11 05:14:05 & 14.0 & D & 0.0 & 5 & $22.91\times 6.82; -3.2^\circ$ & 54.3 & 3C286/J1924-2914/J2355+4950 \\
37 & 877 & 2014-08-09 05:21:25 & 9.0 & D & 0.5 & 5 & $22.91\times 6.82; -3.2^\circ$ & 40.4 & 3C286/J1924-2914/J2355+4950 \\
38 & 900 & 2014-09-01 03:51:54 & 9.0 & D & 0.5 & 5 & $22.91\times 6.82; -3.2^\circ$ & 29.8 & 3C286/J1924-2914/J2355+4950 \\
39 & 1073 & 2015-02-21 16:36:12 & 9.0 & B & 0.5 & 2 & $2.48\times 2.37; 16.6^\circ$ & 14.7 & 3C286/J1924-2914/J2355+4950 \\
%40 & 1101 & 2015-03-21 checar & 9.0 & B & 0.5 & 2 & $2.48\times 2.37; 16.6^\circ$ & 18.3 & 3C286/J1924-2914/J2355+4950 \\

    \hline
  \end{tabular}
 \end{table}
\end{landscape}

%%%%%%%%%%%%%%%%%%%%%%%%%%%%%%%%%%%%%%%%%%%%%%%%%%

% Don't change these lines
\bsp	% typesetting comment
\label{lastpage}
\end{document}